\documentclass[a4paper,12pt]{article}
\usepackage{jheppub}
\usepackage{amsmath}
\usepackage{amssymb}
\usepackage{graphicx}
\usepackage{epsfig}
\usepackage{color}
\usepackage{amsfonts}
\usepackage{adjustbox,lipsum}
\usepackage{hyperref}

\providecommand{\U}[1]{\protect\rule{.1in}{.1in}}
\def\simlt{\stackrel{<}{{}_\sim}}
\def\simgt{\stackrel{>}{{}_\sim}}

\begin{document}
\title{Lepton Flavor Violation in the Inert Scalar Model with Higher Representations}
\author[a]{Talal Ahmed Chowdhury}\author[b]{\& Salah Nasri}
\affiliation[a]{Department of Physics, University of Dhaka, P.O. Box 1000, Dhaka, Bangladesh}
\affiliation[b]{Department of Physics, UAE University, P.O. Box
17551, Al-Ain, United Arab Emirates} \emailAdd{talal@du.ac.bd, snasri@uaeu.ac.ae}

\abstract{
We investigate the lepton flavor violation (LFV) in the inert scalar model with higher
representations. We generalize the inert doublet model with right handed neutrino
by using higher scalar and fermion representation of $SU(2)_{L}$. As the generalized model
and the inert doublet model have the same parameter space, we compare the rates of
$\mu\rightarrow e\gamma$, $\mu\rightarrow ee\overline{e}$ and $\mu-e$ 
conversion in nuclei in the doublet and its immediate extension, the quartet model. 
We show that the corresponding rates are larger 
in the case of higher representation 
compared to the Inert doublet for the same region of 
parameter space. This implies that such extended models 
are more constrained by current LFV bounds and will have better prospects in 
future experiments.}

\maketitle

\section{Introduction}\label{intro}

Neutrino oscillation provides the direct evidence for  lepton flavor violation in
the neutrino sector. Therefore,
one also expects LFV in the charged lepton sector which is yet to be observed.
This is  a generic prediction in most of the neutrino mass models and depending
on the  realization details  of the model, the rates of different LFV processes
can be very different. In this paper, we have focused on radiative neutrino
mass model at one loop proposed in \cite{Ma:2006km}, known as the scotogenic model,
where the scalar content of the model is
the inert doublet. Apart from its role in neutrino mass generation, the inert doublet has been 
extensively studied in the context of dark matter 
\cite{Deshpande:1977rw, LopezHonorez:2006gr, Dolle:2009fn, LopezHonorez:2010tb, 
Agrawal:2008xz, Andreas:2009hj,Nezri:2009jd, Cao:2007rm}, 
mirror model and extra
generation \cite{Martinez:2011ua, Melfo:2011ie}, 
electroweak phase transition \cite{Chowdhury:2011ga, Borah:2012pu, Gil:2012ya, Cline:2013bln, Ahriche:2015mea} and 
collider studies \cite{Lundstrom:2008ai, Dolle:2009ft, Belanger:2015kga}. As the higher scalar representation
is not forbidden by any symmetry in the model, the immediate generalization of the doublet, the quartet with isospin $J=3/2$
was studied in \cite{AbdusSalam:2013eya} to check whether it is viable in providing both light scalar dark matter
and strong electroweak phase transition in the universe. Here we have  incorporated higher scalar representation instead
of the doublet in the scotogenic model and determined  the viable $SU(2)_{L}$ fermion multiplet for generating neutrino mass.
LFV processes in the scotogenic model with inert doublet has been studied in 
\cite{Kubo:2006yx, Sierra:2008wj, Suematsu:2009ww, Adulpravitchai:2009gi, Toma:2013zsa, Vicente:2014wga}
(and references therein). The extension of the scotogenic model has
been addressed in \cite{Ma:2008cu, Law:2013saa}. Also 
larger multiplets have been incorporated in type III seesaw model \cite{Ren:2011mh} and in 
models of radiative neutrino mass generation at higher order with dark matter \cite{Ahriche:2015wha}.

The generalization of scotogenic model with higher $SU(2)_{L}$ 
half-integer representation does not change the parameter set of the
Lagrangian of the inert doublet at the renormalizable level.
Therefore it gives us the  opportunity to investigate  the predictions of LFV processes
for different scalar representations for the same region of parameter space. In particular, we 
compare the  LFV processes for the doublet and the quartet in the light of current experimental bounds and future sensitivities.

There have been many 
great experimental efforts to detect positive LFV signal
in $l_{\alpha}\rightarrow l_{\beta}\gamma$, $l_{\alpha}\rightarrow 3l_{\beta}$ and
$\mu-e$ conversion rate in nuclei. In the case of muon radiative decay,
the MEG collaboration \cite{Adam:2011ch} has put a limit of $\text{Br}(\mu\rightarrow e\gamma)<5.7\times 10^{-13}$
 \cite{Adam:2013mnn} and will have sensitivity of $6\times 10^{-14}$ after acquiring data for three more years 
 \cite{Baldini:2013ke}.
In addition, current bound on branching ratio of lepton flavor violating 3-body decay, 
$\mu\rightarrow ee\overline{e}$ is $1\times 10^{-12}$ set by SINDRUM experiment \cite{Bellgardt:1987du} and Mu3e experiment 
will reach a sensitivity of $10^{-16}$ \cite{Blondel:2013ia}. Furthermore, SINDRUM II experiment has put
current limit on muon to electron ($\mu-e$) conversion rate in Gold (Au) and Titanium (Ti) nucleus
of $7\times 10^{-13}$ \cite{Bertl:2006up} and $4.3\times 10^{-12}$ \cite{Dohmen:1993mp} respectively. The future projects
Mu2e \cite{Glenzinski:2010zz, Bartoszek:2014mya}, DeeMe \cite{Natori:2014yba}, 
COMET \cite{Kuno:2013mha} and PRISM/PRIME \cite{Kuno:2005mm, Barlow:2011zza} 
will improve this bound from $10^{-14}$ to $10^{-18}$. 
For other LFV processes and their experimental bounds, please see Table I of \cite{Toma:2013zsa}.
We have compared the predictions of the LFV processes $\mu\rightarrow e \gamma$, 
$\mu\rightarrow ee\overline{e}$ and
$\mu-e$ conversion rate in Au and Ti for both doublet and quartet scalars 
and our comparison has revealed that the contributions of the quartet in all LFV processes
are larger than those of the doublet for the same region of parameter space. Consequently,
the contribution of higher scalar representation to LFV processes have better experimental prospects. 

The paper is organized as follows. We describe the model in section \ref{zpt}.
In section \ref{LFVprocesses} we present the relevant formulas of
$\mu\rightarrow e \gamma$, $\mu\rightarrow ee\overline{e}$ and $\mu-e$ conversion processes
for the inert doubler and quartet. We present the result in section \ref{resultdiscussion}
and conclude in section \ref{conclusion}. appendix \ref{massspecscalar} contains
the mass spectrum of the inert doublet and quartet in our parametrization. The expressions of
the loop functions are given in appendix \ref{loopappendix}. In appendix \ref{muegZver} 
we collect the Feynman diagrams for $\mu e \gamma$ vertices,
$\mu e Z$ vertices and box diagrams.

\section{The Model}\label{zpt}
Any multiplet charged under $SU(2)_L \times U(1)_Y$ gauge group
is characterized by the quantum numbers $J$ and $Y$, with the electric charge of a   component in  the
multiplet is given by $Q=T_3+Y$. For half-integer
representation $J=n/2$, $T_3$ ranges from $-\frac{n}{2}$ to
$\frac{n}{2}$. So the hypercharge of the multiplet needs to be
$Y=\pm T_3$ for one of the components to have neutral charge. 
For integer representation $n$, similar condition
holds for hypercharge.

The generalized scotogenic model involves one half-integer $SU(2)_{L}$ scalar multiplet
$\Delta$ with hypercharge $Y=1/2$
and three generations of real ($Y=0$) odd dimensional fermionic multiplets, $F_{i}$ ($i=1-3$) charged under
$Z_{2}$ symmetry, $\Delta\rightarrow -\Delta$ and $F_{i}\rightarrow -F_{i}$. 
When the scalar
multiplet is fixed to be $J=n/2$ , $n$ odd, there
are two choices for fermionic multiplet which can give 
$Z_{2}$ even $SU(2)_{L}\times U(1)_{Y}$ invariant Yukawa term with the lepton doublet;
$J=\frac{n-1}{2}$ or $\frac{n+1}{2}$. The  charged lepton sector is augmented by the following terms
\begin{equation}
 {\cal L}\supset -\frac{M_{F_{i}}}{2}\overline{F_{i}^{c}}P_{R}F_{i}+y_{i\alpha}\overline{F}_{i}.
 l_{\alpha}.\Delta+\text{h.c}
 \label{yuk1}
\end{equation}
where the dot represents the proper contractions among $SU(2)$ indices. In the subsequent
analysis we have chosen fermion multiplet to be $J=\frac{n-1}{2}$.

The general Higgs-scalar multiplet potential , symmetric under $Z_2$, can
be written in the following form, 
\begin{eqnarray} 
  \label{potq}
  V_0(\Phi,\Delta)&=&- \mu^2 \Phi^\dagger \Phi + M_0^2 \Delta^\dagger \Delta +
  \lambda_1 (\Phi^\dagger \Phi)^2 + \lambda_2 (\Delta^\dagger \Delta)^2
  +\lambda_3 |\Delta^\dagger T^a \Delta|^2
  +\alpha \Phi^\dagger \Phi \Delta^\dagger \Delta\nonumber\\
  &+&\beta \Phi^\dagger
  \tau^a\Phi \Delta^\dagger T^a \Delta
 +\gamma[ (\Phi^T\epsilon \tau^a\Phi) (\Delta^T
    C T^a \Delta)^\dagger+h.c] 
\end{eqnarray}
Here, $\tau^a$ and $T^{a}$ are the $SU(2)$ generators in fundamental
and $\Delta$'s representation respectively. $C$ is an antisymmetric matrix
analogous to charge conjugation matrix defined as, 
\begin{equation}
  C T^{a} C^{-1}=-T^{aT}
\end{equation} 
Since $C$,  is an  antisymmetric matrix,  it  can only be defined for even 
dimensional space, i.e only for half-integer representation. If the
isospin of the representation is $J$ then $C$ is $(2J+1)\times (2J+1)$
dimensional matrix. The generators are 
normalized in such a way so  
that they satisfy, for fundamental representation, $Tr[\tau^a 
  \tau^b]=\frac{1}{2} \delta^{ab}$ and for other representations,
$Tr(T^{a} T^{b})=D_{2}(\Delta)\delta^{ab}$. Also $T^a T^a=C_{2}(\Delta)$. Here,
$D_{2}(\Delta)$ and $C_{2}(\Delta)$ are Dynkin index and second Casimir
invariant for $\Delta$'s representation. Notice that,
$\gamma$ term is only allowed for representation with
$(J,Y)=(\frac{n}{2},\frac{1}{2})$ and it is essential for the generation of neutrino
mass at one-loop.

The scalar representation with
$(J,Y)=(\frac{n}{2},\frac{1}{2})$ and the fermionic representation with
$(J,Y)=(\frac{n-1}{2},0)$ have the
component fields denoted as $\Delta^{(Q)}$ and $F^{(Q)}$ respectively where $Q$ is the electric
charge. They are written  explicitly as 
\begin{equation}
  \label{hr}
  \bf{\Delta_{\frac{n}{2}}}=\begin{pmatrix}
    \Delta^{(\frac{n+1}{2})}\\
    ...\\
    \Delta^{(0)}\equiv\frac{1}{\sqrt{2}}(S+i\, A)\\
    ...\\
    \Delta^{(-\frac{n-1}{2})}\\
  \end{pmatrix}
\,\textrm{ and }\,
  \bf{F_{\frac{n-1}{2}}}= \begin{pmatrix}
    F^{(\frac{n-1}{2})}\\
    ...\\
    F^{(0)}\\
    ...\\
    F^{(\frac{-n+1}{2})}\\
  \end{pmatrix}
\end{equation}
For the former representation every component represents a unique field
while for the latter there is a redundancy $F^{(-Q)} =
(F^{(Q)})^{*}$.

The choices for real fermion multiplet with the doublet are either $(J,Y)=(0,0)$
or $(1,0)$ and with the quartet, choices are either $(J,Y)=(1,0)$ or $(2,0)$.
Our analysis has focused on the 
following pairs of scalar and fermionic multiplets: $(\Delta_{J=\frac{1}{2}},
{F_{i}}_{J=0})$ and $(\Delta_{J=\frac{3}{2}},
{F_{i}}_{J=1})$. In component fields, the doublet scalar $D$, right handed (RH) neutrino, $N_{R_{i}}$ and
the quartet scalar $\Delta$ and the triplet fermion ${\bf F}_{i}$ are expressed as
\begin{equation} 
  D=\begin{pmatrix}
    C^{+}\\
    D^{0}\equiv \frac{1}{\sqrt{2}}(S+i A)\\
  \end{pmatrix}\,,\,
  N_{R_{i}}\,,\, 
  \Delta=\begin{pmatrix}
    \Delta^{++}\\
    \Delta^{+}\\
    \Delta^{0}\equiv \frac{1}{\sqrt{2}}(S+iA)\\
    \Delta^{'-}
  \end{pmatrix}\,\,
  \text{and}\,\,
  {\bf F}_{i}=\begin{pmatrix} 
    F^{+}\\
    F^{0}\\
    F^{-}
  \end{pmatrix}_{i}
  \label{fieldreps}
 \end{equation}

\subsection{Mass spectra}

We now sketch the general form of mass spectrum for the scalar and fermionic multiplet
which was also presented in \cite{AbdusSalam:2013eya}.
The neutral component of the scalar multiplet ($Y=1/2$) will have $T_{3}$ eigenvalue as
$T_3=-\frac{1}{2}$. Now for the Higgs vacuum expectation value, $\langle
\Phi\rangle=(0,\frac{v}{\sqrt{2}})^{T}$, the term $\langle\Phi^\dagger
\rangle\tau^3\langle\Phi\rangle$ gives $-\frac{v^2}{4}$. So masses for
the neutral components, $S$ and $A$ are splitted by the $\gamma$ term as 
\begin{eqnarray}  \label{nc}
m_{S}^2 &=& M_{0}^2+\frac{1}{2}\left(\alpha+\frac{1}{4}\beta+p(-1)^{p+1}\gamma\right)
v^2\label{masseqscalar}\\
m_{A}^2&=&M_{0}^2+\frac{1}{2}\left(\alpha+\frac{1}{4}\beta-p(-1)^{p+1}\gamma\right)
v^2\label{masseqnpseudo}
\end{eqnarray}
Here, $p=\frac{1}{2}\text{Dim}(\frac{n}{2})=1,2,...$ comes from
$2p\times2p$ $C$ matrix. For the charged component, with $T_3=m$, 
where, $m=n/2,n/2-1,...,-n/2$, we
have, 
\begin{equation}
  m_{(m)}^2=M_{0}^2+\frac{1}{2}\left(\alpha-\frac{1}{2}\beta\,m\right) v^2.
\end{equation}

Moreover, because of the $\gamma$ term, there will be mixing between
components carrying same amount of charge. A component of the
multiplet is denoted as $|J,T_3\rangle$. Components with $|\frac{n}{2},m\rangle$ and 
$|\frac{n}{2},-(m+1)\rangle$ (such that $-m-1\geq -\frac{n}{2}$) will have positive and
negative charge $Q=m+\frac{1}{2}$ respectively. Now
$\langle\Phi\rangle^T\epsilon \tau^a\langle\Phi\rangle$ gives
$\frac{v^2}{2\sqrt{2}}$. Therefore, the mixing matrix between
components with charge $|Q|$ is, 
\begin{equation}
  \label{sc1}
  M^2_{Q}=\begin{pmatrix}
  m^2_{(m)}&\frac{\gamma v^2}{4}\sqrt{\left(\frac{n}{2}-m\right)\left(\frac{n}{2}+m+1\right)}\\
  \\
  \frac{\gamma v^2}{4}\sqrt{\left(\frac{n}{2}-m\right)\left(\frac{n}{2}+m+1\right)}& m^2_{(-m-1)}
  \end{pmatrix}
\end{equation}
And the mass eigenstates are,
\begin{eqnarray}
 \Delta_{1}^{'Q}&=&\cos\theta_{Q}\,\Delta^{Q}_{(m)}+\sin\theta_{Q}\,\Delta_{(-m-1)}^{*Q}\nonumber\\
 \Delta_{2}^{'Q}&=&-\sin\theta_{Q}\,\Delta^{Q}_{(m)}+\cos\theta_{Q}\,\Delta_{(-m-1)}^{*Q}
 \label{egstate}
\end{eqnarray}
where we have
\begin{equation}
 \tan 2\theta_{Q}=\frac{2(M^{2}_{Q})_{12}}{(M^{2}_{Q})_{11}-(M^{2}_{Q})_{22}}
 \label{egstate1}
\end{equation}

Note that the real fermionic multiplet is degenerate at the tree level. However,  there is a small splitting
between the charged and neutral component due to radiative correction which is $O(100\,\text{MeV})$ 
\cite{Cirelli:2005uq}. This splitting is needed in order to treat the  neutral fermion as the dark matter candidate.

\subsection{Neutrino mass generation}
\begin{figure}
 \centerline{\includegraphics[width=5cm]{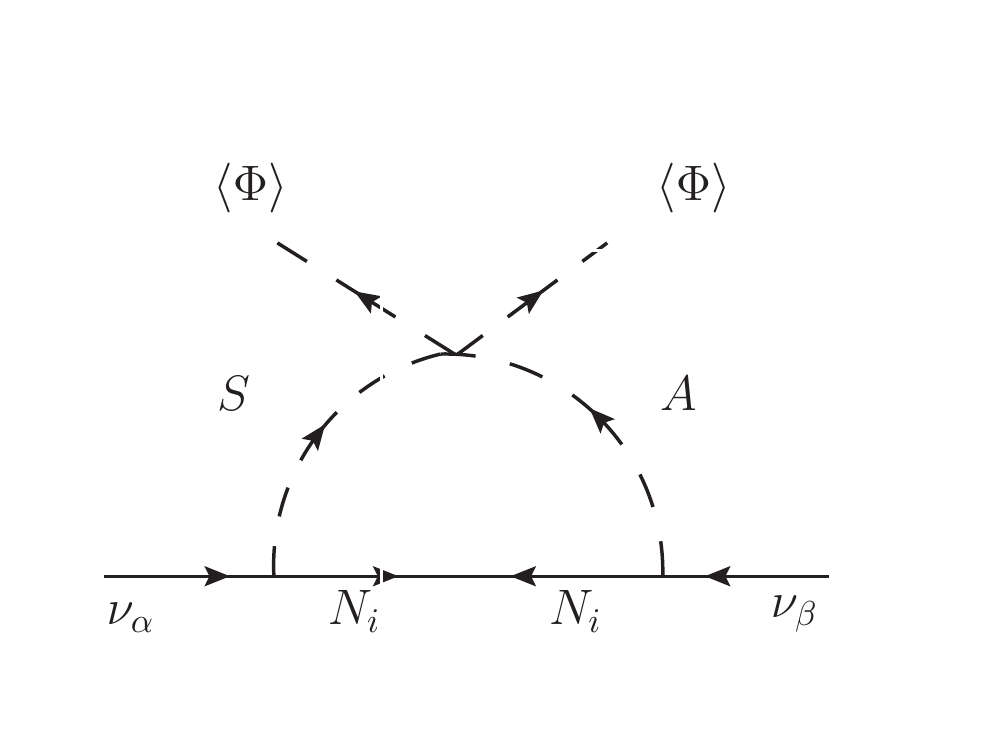}\hspace{0cm}
 \includegraphics[width=5cm]{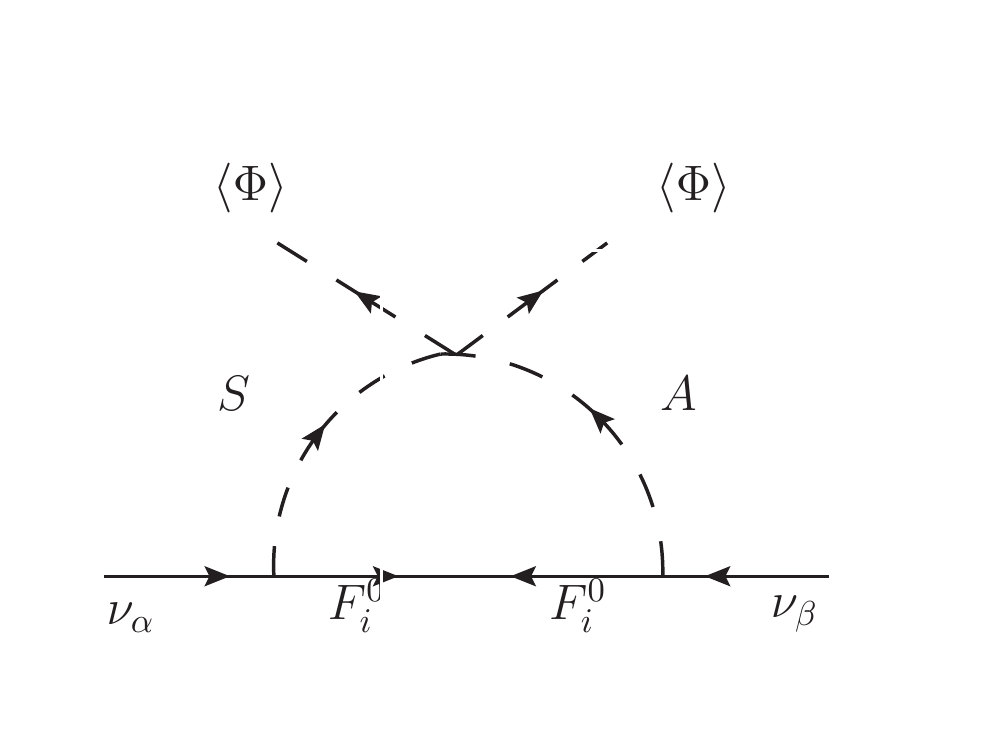}\hspace{0cm}
 \includegraphics[width=5cm]{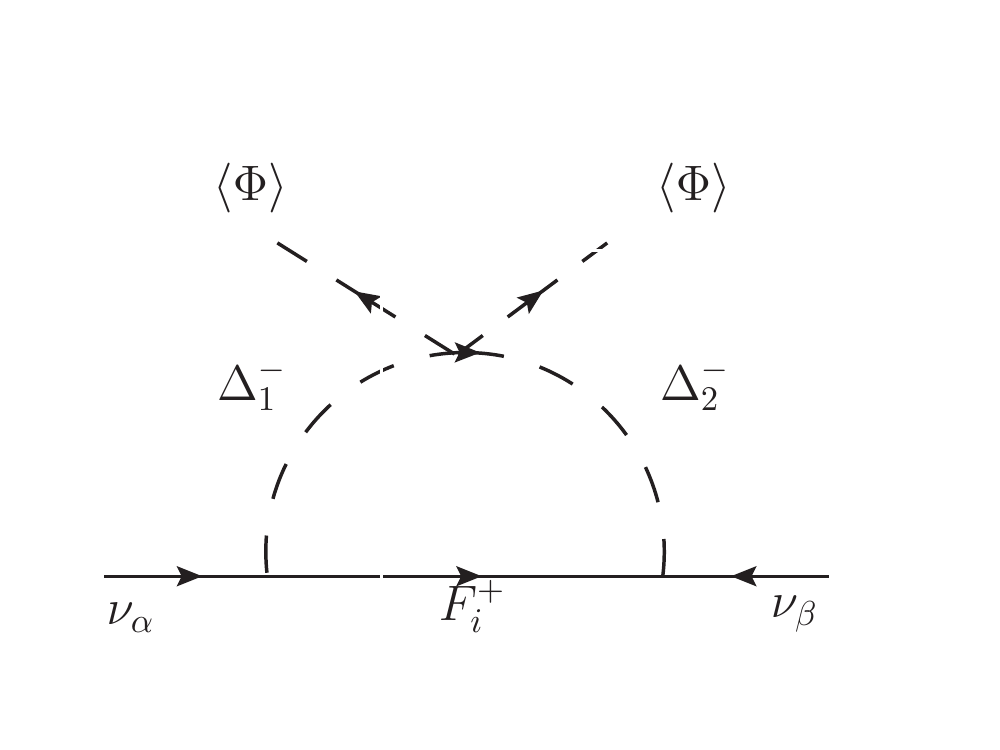}}
 \caption{Neutrino mass generation in the inert doublet (first figure from the left) and the quartet 
 (second and third figures).}
 \label{neutrinomassgen}
\end{figure}

The light neutrino masses are generated at one-loop level as shown in
figure \ref{neutrinomassgen}. The neutrino mass matrix is expressed as
\begin{align}
(m_{\nu})_{\alpha\beta}&=\sum_{i=1}^{3}\frac{y_{\alpha i}y_{i\beta}
 M_{F_{i}}}{16\pi^2}\left\{C^{2}_{\frac{1}{2},0,-\frac{1}{2}}
 \left[\frac{m^2_{S}}{m^{2}_{S}-m^{2}_{F_{i}}}\text{ln}\frac{m^{2}_{S}}{m^{2}_{F_{i}}}
 -\frac{m^2_{A}}{m^{2}_{A}-m^{2}_{F_{i}}}\text{ln}\frac{m^{2}_{A}}{m^{2}_{F_{i}}}\right]\right.\nonumber\\
 &+\left.\sum_{Q\neq0}
 C_{\frac{1}{2},m+\frac{1}{2},m}C_{\frac{1}{2},-m-\frac{1}{2},-m-1}R_{1,m}R_{2,-m-1}
\left[\frac{m^{2}_{Q,1}}{m^{2}_{Q,1}-m^{2}_{F_{i}}}\text{ln}\frac{m^{2}_{Q,1}}{m^{2}_{F_{i}}}-
 \frac{m^{2}_{Q,2}}{m^{2}_{Q,2}-m^{2}_{F_{i}}}\text{ln}\frac{m^{2}_{Q,2}}{m^{2}_{F_{i}}}\right]\right\}\nonumber\\
 &=\left. (y^{T}\Lambda y)_{\alpha\beta}\right.
 \label{nuemass}
\end{align}
Here $C_{m_1,m_2,m_3}$ is the Clebsh-Gordon (CG) coefficient and $m_{1}$, $m_2$ and $m_3$ are the $T_{3}$
eigenvalues of lepton doublet, fermion and scalar multiplet respectively. Non-zero CG
coefficient requires $m_{1}+m_{3}=m_{2}$.
Also $R_{i,m}$ is the
element of the rotation matrix that mixes the two scalar components with same charge $|Q|$ and 
$m^{2}_{Q,i}$ are the corresponding mass eigenvalues. Moreover, $\Lambda_{i}$ is the loop function,
\begin{align}
\Lambda_{i}&=
 \frac{M_{F_{i}}}{16\pi^2}\left\{C^{2}_{\frac{1}{2},0,-\frac{1}{2}}
\left[\frac{m^2_{S}}{m^{2}_{S}-m^{2}_{F_{i}}}\text{ln}\frac{m^{2}_{S}}{m^{2}_{F_{i}}}
 -\frac{m^2_{A}}{m^{2}_{A}-m^{2}_{F_{i}}}\text{ln}\frac{m^{2}_{A}}{m^{2}_{F_{i}}}\right]
 +\sum_{Q\neq0}
 C_{\frac{1}{2},m+\frac{1}{2},m}C_{\frac{1}{2},-m-\frac{1}{2},-m-1}\right.\notag\\
 &\left.R_{1,m}R_{2,-m-1}
 \left[\frac{m^{2}_{Q,1}}{m^{2}_{Q,1}-m^{2}_{F_{i}}}\text{ln}\frac{m^{2}_{Q,1}}{m^{2}_{F_{i}}}-
 \frac{m^{2}_{Q,2}}{m^{2}_{Q,2}-m^{2}_{F_{i}}}\text{ln}\frac{m^{2}_{Q,2}}{m^{2}_{F_{i}}}\right]\right\}
 \label{mass2}
 \end{align}

Therefore the neutrino mass at one loop in the doublet case is given by
\begin{equation}
 (m_{\nu})^{\text{doublet}}_{\alpha\beta}=\sum_{i=1}^{3}\frac{y_{\alpha i}y_{i\beta}
 M_{N_{i}}}{16\pi^2}\left[\frac{m^2_{S}}{m^{2}_{S}-m^{2}_{N_{i}}}\text{ln}\frac{m^{2}_{S}}{m^{2}_{N_{i}}}
 -\frac{m^2_{A}}{m^{2}_{A}-m^{2}_{N_{i}}}\text{ln}\frac{m^{2}_{A}}{m^{2}_{N_{i}}}\right]
 \label{neutrinodoublet}
\end{equation}
where $M_{N_{i}}$ is the mass of the i-th right handed neutrino.
When $m^{2}_{S}\sim m^{2}_{A}\equiv m^{2}_{0}$ then eq. (\ref{neutrinodoublet}) gets simplified
\begin{equation}
 (m_{\nu})^{\text{doublet}}_{\alpha\beta}=\sum_{i=1}^{3}\frac{y_{\alpha i}y_{i\beta}
 \gamma v^2}{16\pi^2 M_{N_{i}}}\left[\frac{m^2_{N_{i}}}{m^{2}_{0}-m^{2}_{N_{i}}}
 +(\frac{m^2_{N_{i}}}{m^{2}_{0}-m^{2}_{N_{i}}})^2\text{ln}\frac{m^{2}_{N_{i}}}{m^{2}_{0}}\right]
 \label{neutrinodoublet1}
\end{equation}

On the other hand, the neutrino mass matrix in the quartet case is given by
\begin{equation}
(M_{\nu})^{\text{quartet}}_{\alpha\beta}=\sum_{i=1}^{2}y_{\alpha i}\Lambda_{i}y_{i\beta}
\label{mass3}
\end{equation}
with the loop factor,
\begin{align}
 \Lambda^{\text{quartet}}_{i}&=\left.\frac{1}{3(4\pi)^2}M_{Fi}\left[\frac{m_{S}^2}{m_{S}^2-M_{Fi}^2}
 \text{ln}\frac{m_{S}^2}{M_{Fi}^2}-\frac{m_{A}^2}{m_{A}^2-M_{Fi}^2}\text{ln}\frac{m_{A}^2}{M_{Fi}^2}\right]\right.\nonumber\\
 &+\left.\frac{1}{6(4\pi)^2}\sin 2\theta M_{Fi}\left[\frac{m_{\Delta^{+}_{1}}^2}{m_{\Delta^{+}_{1}}^2-
 M_{Fi}^2}\text{ln}\frac{m_{\Delta^{+}_{1}}^2}{M_{Fi}^2}
 -\frac{m_{\Delta^{+}_{2}}^2}{m_{\Delta^{+}_{2}}^2-M_{Fi}^2}\text{ln}\frac{m_{\Delta^{+}_{2}}^2}{M_{Fi}^2}\right]\right.
\label{neutrinoquartet}
 \end{align}
Explicit expressions of masses in the inert doublet and quartet models are included in appendix 
\ref{massspecscalar}.

The neutrino mass matrix can be diagonalized as
\begin{equation}
 U^{T}_{PMNS}\,m_{\nu}\,U_{PMNS}\equiv \hat{m}_{\nu}
 \label{mass1}
\end{equation}
where
\begin{equation}
 U_{PMNS}=\begin{pmatrix}
           c_{12}c_{13}&s_{12}c_{13}&s_{13}e^{i\delta}\\
           -s_{12}c_{23}-c_{12}s_{23}s_{13}e^{-i\delta}& c_{12}c_{23}-s_{12}s_{23}s_{13}e^{-i\delta}&s_{23}c_{13}\\
           s_{12}s_{23}-c_{12}c_{23}s_{13}e^{-i\delta}&-c_{12}s_{23}-s_{12}c_{23}s_{13}e^{-i\delta}& c_{23}c_{13}
          \end{pmatrix}
\times \begin{pmatrix}
        1&0&0\\
        0& e^{i\alpha/2}&0\\
        0 & 0 & e^{i\beta/2}
       \end{pmatrix}
\label{upmns}
\end{equation}
Here, $c_{ij}=\cos\theta_{ij}$, $s_{ij}=\sin\theta_{ij}$, $\delta$ is the Dirac phase and $\alpha$, $\beta$
are the Majorana phases.

The Yukawa matrix $y_{i\alpha}$ ($\alpha=e,\mu,\tau$) is expressed using 
the Casas-Ibarra parametrization \cite{Casas:2001sr} so that the
chosen parameter space automatically satisfies the low energy neutrino parameters,
\begin{equation}
 y=\sqrt{\Lambda}^{-1}\,R\,\sqrt{\hat{m}_{\nu}}\,U^{\dagger}_{PMNS}
 \label{casas}
\end{equation}
where $R$ is a complex orthogonal matrix.

\subsection{Perturbativity}

If there are N generations of  right handed fermion multiplet, 
perturbativity of the Yukawa gives the following constraint \cite{Casas:2010wm, Heeck:2012fw}
\begin{equation}
 \text{Tr}(y^\dagger y)=\sum_{i=1}^{3}\sum_{j=1}^{N}|R_{ij}|^{2}\frac{\hat{m}_{\nu_{i}}}{\Lambda_{j}}\simlt O(1)
 \label{yukawa1}
\end{equation}

If $R$ is taken  to be real, the constraint translates into the largest ratio,
$\frac{\hat{m}_{\nu_{i}}}{\Lambda_{j}}\simlt O(1)$, whereas for the general case  when
$R$ is complex, each entry will be bounded as
$|R_{ij}|\simlt \sqrt{\frac{\Lambda_{j}}{3N\hat{m}_{\nu_{i}}}}$.

\section{Lepton flavor violating processes}\label{LFVprocesses}
In this section we have presented the relevant analytical formulas of LFV processes
for the doublet and quartet case. In the standard model due to the GIM suppression
the rate of $\mu\rightarrow e\gamma$ becomes $\sim 10^{-54}$ thus negligible.
On the other hand the presence of heavy right handed neutrino that mixes with
left handed (LH) neutrinos, spoils the GIM suppression
and one could obtain the rate which can be probed by experiment 
\cite{Cheng:1976uq, Cheng:1977nv, Cheng:1980tp, Ma:1980gm, Lim:1981kv, Ilakovac:1994kj, Blum:2007he}.
In inert scalar models, $Z_{2}$ symmetry forbids 
the mixing between LH and RH neutrinos but the enhancements
in the LFV processes are provided by the $C^{\pm}-N_{R_{i}}$ loops in the doublet
and $\Delta-F_{i}$ loops in the quartet model.
We have focused on three LFV processes: $\mu\rightarrow e\gamma$, 
$\mu\rightarrow ee\overline{e}$ and $\mu-e$ conversion in nuclei in this paper as they have the most stringent limits
from the experiments.

\subsection{$\mu\rightarrow e\gamma$}
The branching ratio for $\mu\rightarrow e\gamma$, normalized by 
$\text{Br}(\mu\rightarrow e\overline{\nu_{e}}\nu_{\mu})$, is \cite{Hisano:1995cp, Toma:2013zsa}
\begin{equation}
 \text{Br}(\mu\rightarrow e\gamma)=\frac{3(4\pi)^3\alpha_{em}}{4G_{F}^2}|A_{D}|^2\,\text{Br}(\mu\rightarrow e\nu_{\mu}\overline{\nu_{e}})
 \label{mutoegma}
\end{equation}
where $A_{D}$ is the dipole form factor. The Feynman diagrams of one-loop
contributions by the doublet and quartet to the $\mu e \gamma$ vertex that enters
into the dipole form factor calculation, are given in
figure \ref{muegammavertices}.

The contributions from the doublet is the following,
\begin{equation}
 A^{\text{doublet}}_{D}=\sum_{i=1}^{3}\frac{y^{*}_{e i}y_{i\mu}}{32\pi^2}\frac{1}{m^2_{C}}F^{(n)}(x_{i\sigma})
 \label{addoublet}
\end{equation}
Here $F^{(n)}(x)$ is the loop function given in the appendix \ref{loopappendix} and $x_{i\sigma}=m^{2}_{N_{i}}/m^{2}_{\sigma}$,
where $\sigma=C^{+}$.
On the other hand, the quartet contribution will have two parts
\begin{equation}
 A^{\text{quartet}}_{D}=A^{\text{quartet}}_{D(n)}+A^{\text{quartet}}_{D(c)}
 \label{quartetdipole0}
\end{equation}
where $A^{\text{quartet}}_{D(n)}$ is the contribution of the neutral component 
and  $A^{\text{quartet}}_{D(c)}$ is that of the charged component of the fermion triplet. 
Also, for the notational convenience, we introduce generalized Yukawa coupling
$y_{i\alpha\sigma}=y_{i\alpha}C_{\sigma}$ where $C_{\sigma}$ is the corresponding Clebsh Gordon
coefficient associated with $\sigma$-th component of the quartet. The two contributions are
\begin{equation}
 A^{\text{quartet}}_{D(n)}=\sum_{i=1}^{3}\sum_{\sigma}\frac{y^{*}_{e i\sigma}y_{i\mu\sigma}}{32\pi^2}\frac{1}{m^2_{\sigma}}F^{(n)}(x_{i\sigma})
 \label{adquartet1}
\end{equation}
where $x_{i\sigma}=m^{2}_{F^{0}_{i}}/m^{2}_{\sigma}$, $\sigma=\Delta^{+}_{1},\,\Delta^{+}_{2}$.
And
\begin{equation}
 A^{\text{quartet}}_{D(c)}=-\sum_{i=1}^{3}\sum_{\sigma}\frac{y^{*}_{e i\sigma}y_{i\mu\sigma}}{32\pi^2}\frac{1}{m^2_{\sigma}}F^{(c)}(x_{i\sigma})
 \label{adquartet1}
\end{equation}
where $x_{i\sigma}=m^{2}_{F^{\pm}_{i}}/m^{2}_{\sigma}$,  and $\sigma=\Delta^{++},\,S,\,A$.

\subsection{$\mu\rightarrow ee\overline{e}$}
Now we turn to $\mu\rightarrow ee\overline{e}$ decay. The branching ratio is given as 
\cite{Hisano:1995cp, Arganda:2005ji, Toma:2013zsa}
\begin{align}
 \text{Br}(\mu\rightarrow ee\overline{e})&=\frac{3(4\pi)^2\alpha_{em}^2}{8G_{F}^2}
 \left[|A_{ND}|^2+|A_{D}|^2\left(\frac{16}{3}\text{ln}\frac{m_{\mu}}{m_{e}}-\frac{22}{3}\right)+\frac{1}{6}|B|^2\right.\notag\\
 &+\left.\frac{1}{3}(2|F^{L}_{Z}|^2+F^{R}_{Z}|^2)+\left(-2A_{ND}A^*_{D}+\frac{1}{3}A_{ND}B^*-\frac{2}{3}A_{D}B^*+\text{h.c}\right)\right]\notag\\
 &\times \text{Br}(\mu\rightarrow e\overline{\nu_{e}}\nu_{\mu})
 \label{mutoeee}
\end{align}
where $A_{D}$ and $A_{ND}$ are the dipole and 
non-dipole contribution from the photonic penguin diagrams respectively. Also
$B$ represents the contribution from the box diagrams.
Moreover, $F^{L}_{Z}$ and $F^{R}_{Z}$ are given as
\begin{equation}
 F^{L}_{Z}=\frac{F_{Z}g^{l}_{L}}{g^2m_{Z}^2\sin^2\theta_{W}}\,\,\,,\,\,\,F^{R}_{Z}=\frac{F_{Z}g^{l}_{R}}{g^2m_{Z}^2\sin^2\theta_{W}}
 \label{zcontrib}
\end{equation}
Here, $F_{Z}$ is the Z-penguin contribution and $g^{l}_{L}$ and $g^{l}_{R}$ are the Z-boson coupling
to the LH and RH charged leptons respectively. In this model, Higgs penguin contribution
will be suppressed by the small electron Yukawa coupling, and therefore we have only considered
the photon penguin, Z-boson penguin and box diagrams. 

\subsubsection{$\gamma$-penguin contribution}
First let us consider contributions from the photon penguin diagrams. In this case
the $\gamma$ line of $\mu e \gamma$ vertex given in figure \ref{muegammavertices} will have
$\overline{e}e$  attached to it. 
The photonic non-dipole contribution, $A_{ND}$ for the doublet is in the 
following
\begin{equation}
 A^{\text{doublet}}_{ND}=\sum_{i=1}^{3}\frac{y^{*}_{e i}y_{i\mu}}{96\pi^2}\frac{1}{m^2_{C}}G^{(n)}(x_{i\sigma})
\end{equation}

The photonic non-dipole contribution, for the case of the quartet, will again have two parts,
\begin{equation}
 A^{\text{quartet}}_{ND}=A^{\text{quartet}}_{ND(n)}+A^{\text{quartet}}_{ND(c)}
 \label{quartetnondipole0}
\end{equation}
Here $A^{\text{quartet}}_{ND(n)}$ is the contribution of the neutral component 
and $A^{\text{quartet}}_{ND(c)}$ is the contribution of 
the charged component of the fermion triplet.
\begin{equation}
 A^{\text{quartet}}_{ND(n)}=\sum_{i=1}^{3}\sum_{\sigma = \Delta^{+}_{1},\,\Delta^{+}_{2}}\frac{y^{*}_{e i\sigma}y_{i\mu\sigma}}{96\pi^2}\frac{1}{m^2_{\sigma}}G^{(n)}(x_{i\sigma})
 \label{adquartet1}
\end{equation}
where again $x_{i\sigma}=m^{2}_{F^{0}_{i}}/m^{2}_{\sigma}$.
And the charged component of fermion triplet contributes as follows,
\begin{equation}
 A^{\text{quartet}}_{ND(c)}=-\sum_{i=1}^{3}\sum_{\sigma}\frac{y^{*}_{e i\sigma}y_{i\mu\sigma}}{96\pi^2}\frac{1}{m^2_{\sigma}}G^{(c)}(x_{i\sigma})
 \label{adquartet1}
\end{equation}
with $x_{i\sigma}=m^{2}_{F^{\pm}_{i}}/m^{2}_{\sigma}$, and $\sigma = \Delta^{++},\,S,\,A$. 
The loop functions $F^{(n)}(x)$, $F^{(c)}(x)$, $G^{(n)}(x)$ and $G^{(c)}(x)$ 
are given in the appendix \ref{loopappendix}.

\subsubsection{Z-penguin contribution}
Now we focus on the Z-penguin diagram. The Feynman diagrams of one-loop contributions
from the doublet and the quartet to the $\mu e Z$ vertex are 
presented in figure \ref{muezvertices}. In Z-penguin diagram, the $Z$ line of 
$\mu e Z$ vertex will have $\overline{e}e$ line attached to it.
For the doublet, the contribution is given by the neutral fermion.
Following the formulas given in 
\cite{Arganda:2005ji, Abada:2014kba, Arganda:2014lya}\footnote{\cite{Arganda:2005ji} contained
a mistake in the calculation of Z-penguin diagram which was pointed out in \cite{Krauss:2013gya}.
Subsequently, correct results were presented in \cite{Abada:2014kba} and \cite{Arganda:2014lya}.
Moreover, $C_{00}$ of \cite{Abada:2014kba} and $C_{24}$ of \cite{Arganda:2014lya} only differ
by an overall minus sign.}
\begin{equation}
 F^{\text{doublet}}_{Z(n)}=-\frac{1}{16\pi^2}\sum_{i=1}^{3}y^{*}_{e i}y_{i\mu}\left[2\,g_{ZC^{+}C^{-}}\,
 C_{24}(m_{N_{i}},m_{C},m_{C})+g^{l}_{L}B_{1}(m_{N_{i}},m_{C})\right]
 \label{doubletZ}
\end{equation}
Here, $g_{ZC^{+}C^{-}}$ is the Z boson coupling to $C^{\pm}$ of the doublet and $g^{l}_{L}$
is the Z boson coupling to LH charged leptons given by
\begin{equation}
 g^{l}_{L}=\frac{g}{\cos\theta_{W}}\left(-\frac{1}{2}+\sin^2\theta_{W}\right)
 \label{Zlepton}
\end{equation}

On the other hand, the quartet contribution is 
\begin{equation}
 F^{\text{quartet}}_{Z}=F^{\text{quartet}}_{Z(n)}+F^{\text{quartet}}_{Z(c)}
 \label{quartetZ0}
\end{equation}
where the neutral fermion of the triplet contributes as
\begin{align}
 F^{\text{quartet}}_{Z(n)}& =-\frac{1}{16\pi^2}\sum_{i=1}^{3}\sum_{\sigma_{1},\sigma_{2}}\left[2\,y^{*}_{e i\sigma_{1}}y_{i\mu\sigma_{2}}\,g_{Z\sigma_{1}\sigma_{2}}\,
 C_{24}(m_{F^{0}_{i}},m_{\sigma_{1}},m_{\sigma_{2}})\right.\notag\\
 & \left. +y^{*}_{e i\sigma_{1}}y_{i\mu\sigma_{1}}g^{l}_{L}B_{1}(m_{F^{0}_{i}},m_{\sigma_{1}})\right]
\label{quartetZ1}
 \end{align}
where  $\sigma_{1,2}\in \{\Delta^{+}_{1},\Delta^{+}_{2}\}$ and $g_{Z\sigma_{1}\sigma_{2}}$ is the
Z boson coupling to $\sigma_{1}$ and $\sigma_{2}$ scalars of the quartet.
The charged fermion of the triplet has the following contribution
\begin{align}
 F^{\text{quartet}}_{Z(c)} & =-\frac{1}{16\pi^2}\sum_{i=1}^{3}\sum_{\sigma_{1},\sigma_{2}}\left\{y^{*}_{e i\sigma_{1}}y_{i\mu\sigma_{1}}\,g_{ZF^{\pm}_{i}\overline{F^{\pm}_{i}}}\,%
 \left[\left(2C_{24}(m_{\sigma_{1}},m_{F^{\pm}_{i}},m_{F^{\pm}_{i}})+\frac{1}{2}\right)\right.\right. \notag \\
 & +\left. m^{2}_{F^{\pm}_{i}}%
 C_{0}(m_{\sigma_{1}},m_{F^{\pm}_{i}},m_{F^{\pm}_{i}})\right]%
 +2\,y^{*}_{e i\sigma_{1}}y_{i\mu\sigma_{2}}\,g_{Z\sigma_{1}\sigma_{2}}\,%
 C_{24}(m_{F^{\pm}_{i}},m_{\sigma_{1}},m_{\sigma_{2}}) \notag \\
 &\left. + y^{*}_{e i\sigma_{1}}y_{i\mu\sigma_{1}}g^{l}_{L}B_{1}(m_{F^{\pm}_{i}},m_{\sigma_{1}})\right\}
 \label{quartetZ2}
\end{align}
Here $\sigma_{1}$ and $\sigma_{2}$ range over the $S,\,A,\ \Delta^{++}$, and  $g_{ZF^{\pm}_{i}\overline{F^{\pm}_{i}}}$ is the coupling
of Z boson to charged fermions. Moreover, $B_{1}$, $C_{0}$ and $C_{24}$ are the loop functions,
adopted from \cite{Arganda:2005ji, Arganda:2014lya, Abada:2014kba}, presented in the appendix \ref{loopappendix}.
As $B_{1}$ and $C_{24}$ arise from divergent loop integrals, for large $M$,
\begin{equation}
 C_{24}(M,m,m)\rightarrow \frac{1}{4}\text{ln}\frac{M^2}{\mu^2}\,\,,\,\,B_{1}\rightarrow \frac{1}{2}\text{ln}\frac{M^2}{\mu^2}
 \label{decoup}
\end{equation}
Therefore the combination $2 xC_{24}+yB_{1}$ in Z-penguin contribution eq. (\ref{quartetZ1})
or in eq. (\ref{quartetZ2}) is vanishing at very large mass $M$ when there are
specific relations set by group theoretical requirements in vertex factors $x$ and $y$.

\subsubsection{Box contribution}
Lastly the box contribution for the doublet case, presented in figure \ref{boxfigs},
is \cite{Arganda:2005ji}
\begin{equation}
 e^2\,B^{\text{doublet}}_{(n)}=\frac{1}{16\pi^2}\sum_{i,j=1}^{3}\left[\frac{\tilde{D}_{0}}{2}y^{*}_{ei}y_{i\mu}y^{*}_{ej}y_{je}
 +D_{0}m_{N_{i}}m_{N_{j}}y^{*}_{ei}y^{*}_{ei}y_{j\mu}y_{je}\right]
\label{boxdoublet}
\end{equation}
where, $\tilde{D}_{0}=\tilde{D}_{0}(m_{N_{i}},m_{N_{j}},m_{C},m_{C})$ and
$D_{0}=D_{0}(m_{N_{i}},m_{N_{j}},m_{C},m_{C})$ are  loop functions given in the appendix B.

For the quartet case, the contribution of the box diagram can be written as 
\begin{equation}
 B^{\text{quartet}}=B^{\text{quartet}}_{(n)}+B^{\text{quartet}}_{(c)}
 \label{boxquartet0}
\end{equation}
with $B^{\text{quartet}}_{(n)}$ is the contribution due to the neutral fermions and it is given by
\begin{equation}
 e^2\,B^{\text{quartet}}_{(n)}=\frac{1}{16\pi^2}\sum_{i,j=1}^{3}\sum_{\sigma_{1},\sigma_{2}}\left[\frac{\tilde{D}_{0}}{2}y^{*}_{ei\sigma_{1}}y_{i\mu\sigma_{2}}y^{*}_{ej\sigma_{2}}y_{je\sigma_{1}}
 +D_{0}m_{F^{0}_{i}}m_{F^{0}_{j}}y^{*}_{ei\sigma_{1}}y^{*}_{ei\sigma_{2}}y_{j\mu\sigma_{2}}y_{je\sigma_{1}}\right]
\label{boxquartet1}
\end{equation}
where, $\tilde{D}_{0}=\tilde{D}_{0}(m_{F^{0}_{i}},m_{F^{0}_{j}},m_{\sigma_{1}},m_{\sigma_{2}})$ and
$D_{0}=D_{0}(m_{F^{0}_{i}},m_{F^{0}_{j}},m_{\sigma_{1}},m_{\sigma_{2}})$.
Here, $\sigma_{1,2}$ ranges over $\Delta^{+}_{1}$ and $\Delta^{+}_{2}$.

The term $B^{\text{quartet}}_{(c)} $ corresponds to  the contribution of the  charged fermions and it reads
\begin{equation}
 e^2\,B^{\text{quartet}}_{(c)}=\frac{1}{16\pi^2}\sum_{i,j=1}^{3}\sum_{\sigma_{1},\sigma_{2}}\frac{\tilde{D}_{0}}{2}y^{*}_{ei\sigma_{1}}
 y_{i\mu\sigma_{2}}y^{*}_{ej\sigma_{2}}y_{je\sigma_{1}}
\end{equation}
Here, $\tilde{D}_{0}=\tilde{D}_{0}(m_{F^{\pm}_{i}},m_{F^{\pm}_{j}},m_{\sigma_{1}},m_{\sigma_{2}})$ and 
$\sigma_{1,2}$ ranges over $\Delta^{++},\,S,\,A$.

\subsection{$\mu-e$ conversion in nuclei}
The conversion rate, normalized by the muon capture rate is 
\cite{Kitano:2002mt, Arganda:2007jw, Toma:2013zsa, Crivellin:2014cta}
\begin{align}
 \text{CR}(\mu-e,\text{Nucleus})&=\frac{p_{e}E_{e}m^3_{\mu}G_{F}^2\alpha_{em}^3Z_{eff}^{4}F_{p}^2}{8\pi^2 Z\,\Gamma_{\text{capt}}}
 \left\{|(Z+N)(g^{(0)}_{LV}+g^{(0)}_{LS})+(Z-N)(g^{(1)}_{LV}+g^{(1)}_{LS})|^{2}\right.\notag\\
 &+\left.|(Z+N)(g^{(0)}_{RV}+g^{(0)}_{RS})+(Z-N)(g^{(1)}_{RV}+g^{(1)}_{RS})|^{2}\right\}
 \label{mueconv}
\end{align}
Here, $Z$ and $N$ are the number of protons and neutrons in the nucleus, $Z_{eff}$ is the effective atomic
charge, $F_{p}$ is the nuclear matrix element and $\Gamma_{\text{capt}}$ represents
the total muon capture rate. $p_{e}$ and $E_{e}$ are the momentum and energy of the electron (taken as 
$\sim m_{\mu}$ in the numerical evaluation). $g^{(0)}_{XK}$ and $g^{(1)}_{XK}$ ($X=L,R$ and $K=V,S$)
in the above expression are given as
\begin{eqnarray}
 g^{(0)}_{XK}=\frac{1}{2}\sum_{q=u,d,s}(g_{XK(q)}G^{(q,p)}_{K}+g_{XK(q)}G^{(q,n)}_{K})\nonumber\\
 g^{(1)}_{XK}=\frac{1}{2}\sum_{q=u,d,s}(g_{XK(q)}G^{(q,p)}_{K}-g_{XK(q)}G^{(q,n)}_{K})
 \label{nuclear1}
\end{eqnarray}
$g_{XK(q)}$ are the couplings in the effective Lagrangian describing $\mu-e$ conversion,
\begin{equation}
 {\cal L}_{eff}=-\frac{G_{F}}{\sqrt{2}}\sum_{q}\left\{[g_{LS(q)}\overline{e}_{L}\mu_{R}+g_{RS(q)}\overline{e}_{R}\mu_{L}]\overline{q}q+
 [g_{LV(q)}\overline{e}_{L}\gamma^{\mu}\mu_{L}+g_{RV(q)}\overline{e}_{R}\gamma^{\mu}\mu_{R}]\overline{q}\gamma_{\mu}q\right\}
\end{equation}
$G^{(q,p)},\, G^{(q,n)}$ are the numerical factors that arise when quark matrix elements
are replaced by the nucleon matrix elements,
\begin{equation}
 \langle p|\overline{q}\Gamma_{K}q|p\rangle=G^{(q,p)}_{K}\overline{p}\Gamma_{K}p\,\,,\,\,
 \langle n|\overline{q}\Gamma_{K}q|n\rangle=G^{(q,n)}_{K}\overline{n}\Gamma_{K}n
 \label{nuclear2}
\end{equation}
For the inert scalar model, the $\mu-e$ conversion rate receives the $\gamma$, Z and Higgs penguin
contributions. In $\gamma$ and Z penguin diagrams, $\overline{q}q$ (q=u,d,s) line is attached to
$\gamma$ line of $\mu e \gamma$ vertex and Z boson line of $\mu e Z$ vertex respectively. It
doesn't receive any box contribution because there is no coupling
between inert scalars and quarks because of the $Z_{2}$ symmetry. Moreover, Higgs penguin
contribution is small compared to $\gamma$ and Z penguin diagrams because of small Yukawa couplings thus
neglected in our numerical analysis. The relevant effective coupling for the conversion
in the inert scalar model is
\begin{eqnarray}
 g_{LV(q)}&=&g^{\gamma}_{LV(q)}+g^{Z}_{LV(q)}\nonumber\\
 g_{RV(q)}&=&g_{LV(q)}|_{L\leftrightarrow R}\nonumber\\
 g_{LS(q)}&\approx& 0\,\,\,,\,\,\,g_{RS(q)}\approx 0\nonumber
\end{eqnarray}
The relevant couplings are
\begin{eqnarray}
 g^{\gamma}_{LV(q)}&=&\frac{\sqrt{2}}{G_{F}}e^2Q_{q}(A_{ND}-A_{D})\label{nuclear31}\\
 g^{Z}_{LV(q)}&=&-\frac{\sqrt{2}}{G_{F}}\frac{g^{q}_{L}+g^{q}_{R}}{2}\frac{F_{Z}}{m_{Z}^2}
 \label{nuclear32}
\end{eqnarray}
Here $Q_{q}$ is the electric charge of the quarks and Z boson couplings to the quarks are
\begin{equation}
 g^{q}_{L}=\frac{g}{\cos\theta_{W}}(T^{q}_{3}-Q_{q}\sin^2\theta_{W})\,\,,\,\,
 g^{q}_{R}=-\frac{g}{\cos\theta_{W}}Q_{q}\sin^2\theta_{W}
 \label{nuclear4}
\end{equation}
Also the relevant numerical factors for nucleon matrix elements are
\begin{equation}
 G^{(u,p)}_{V}=G^{(d,n)}_{V}=2\,\,,\,\,G^{(d,p)}_{V}=G^{(u,n)}_{V}=1
 \label{nuclear5}
\end{equation}

\section{Results and Discussion}\label{resultdiscussion}

In this section we have presented our numerical results and discussed the phenomenological implications of those results
for larger scalar multiplets.  But before presenting the results, we
have listed all the constraints regarding dark matter and collider searches
so that our analysis can focus on parameter space for where both inert doublet and quartet models are viable.

There are two possible dark matter (DM) candidates in the inert scalar models. 
In the doublet model they are
the lightest right handed neutrino, $N_{1}$ and the lightest neutral scalar, 
$S$ of the doublet. On the other hand, in the quartet
model the neutral component of the lightest fermion triplet, $F_{1}^{0}$
and the lightest neutral scalar, $S$ of the quartet can play the dark matter role.
In both cases fermionic and scalar DM give rise to different phenomenology. In this preliminary study of
comparing different LFV rates in inert scalar models, we have chosen the scalar as the DM particle and 
used the constraints associated with it in our analysis.

\subsection{Constraints and parameter space}\label{constranitssec}

\subsubsection{Collider constraints}\label{colliderconst}

For the doublet scalar, the collider searches have put the following mass 
constraints, $m_{C^{+}}\simgt 100$ GeV, $m_{S}\simgt 65-80$ GeV and $m_{A}\simgt 140$ GeV
\cite{Lundstrom:2008ai, Dolle:2009ft, Gustafsson:2012aj, Aoki:2013lhm, Belanger:2015kga}.
Although there hasn't been any collider studies on the quartet, one can recast the constraints of the doublet case
onto the quartet. As the quartet scalar has the cascade decay channel, we can expect multilepton final states
along with missing transverse energy similar to doublet. Therefore, the mass constraints for quartet, compatible
with bounds on electroweak precision observable \cite{Beringer:1900zz}, are
$m_{\Delta_{1,2}}, m_{\Delta^{++}}\simgt 100$ GeV, $m_{S} \simgt 65-80$ GeV and $m_{A} \simgt 140$ GeV.
Considering $S$ as the DM also set the mass hierarchy in quartet components:
$m_{S}<m_{\Delta_{1}^{+}}<m_{\Delta^{++}}<m_{\Delta_{2}^{+}}<m_{A}$. 
In contrast, the scalar masses in the TeV scale for both doublet and quartet are fairly unconstrained.

In the case of fermions, the masses of RH neutrino in the doublet case are not constrained by current collider
data. In contrast, fermion triplet of the quartet case, having
gauge interaction, will have an accessible collider signature. In \cite{Aad:2013yna} the mass of the 
charged component of the triplet is excluded up to 270 GeV with 8 TeV 20.3 $\text{fb}^{-1}$ LHC data.
Moreover, in \cite{Cirelli:2014dsa} 
it was shown that the projected reach for 14 TeV collider with 3 $\text{ab}^{-1}$ 
luminosity (High luminosity LHC phase)
would be $M_{F}\simlt 500$ GeV, for (future) 100 TeV pp collider with 3 $\text{ab}^{-1}$ luminosity 
in mono-jet searches,
$M_{F}\simlt 1.3$ TeV, and with 30 $\text{ab}^{-1}$ luminosity, $M_{F}\simlt 1.7$ TeV.

\subsubsection{DM Constraints}\label{darkmatter}

The dark matter density of the universe measured by Planck collaboration is 
$\Omega_{DM}h^2=0.1196\pm 0.0031\,(68\%\,\text{CL})$ \cite{Ade:2013zuv}.
In the inert scalar model, there are two viable mass region of scalar DM. They are the low mass region
($m_{S}<m_{W}$) and the high mass region ($m_{S}\gg m_{W}$). The low mass DM region of doublet model
has been extensively studied. In addition, same region for DM in the quartet 
was addressed in \cite{AbdusSalam:2013eya}
where it was shown that it is harder to achieve low mass dark matter with correct 
relic density compared to the 
doublet because, for most
of the parameter space,
bounds on electroweak $T$ parameter sets the mass of single charged component, $\Delta_{1}^{+}$ 
close to the DM mass and therefore it is not only in tension with collider bounds but 
also opens up coannihilation channel and leads
to a sub-dominant DM in the universe. 

In the high mass region of the doublet, as shown in \cite{Hambye:2009pw}, the DM mass starts
from a lower bound of $m_{0}=534\pm 25$ GeV (where the thermal freeze-out only happens through the gauge interaction) 
to $20$ TeV if the higgs-scalar coupling, $\lambda_{S}\simlt 2\pi$. 
The maximal mass splitting compatible with correct relic density, are
\begin{equation}
|m_{A}-m_{S}|\simlt 16.9\,\text{GeV},\,\,\, |m_{C^{+}}-m_{S}|\simlt 14.6\,\text{GeV}
\label{doubletmasssplit}
\end{equation}
when $m_{S}\sim O(5\,\text{TeV})$.

\begin{figure}[h!]
 \centerline{\includegraphics[width=7.5cm]{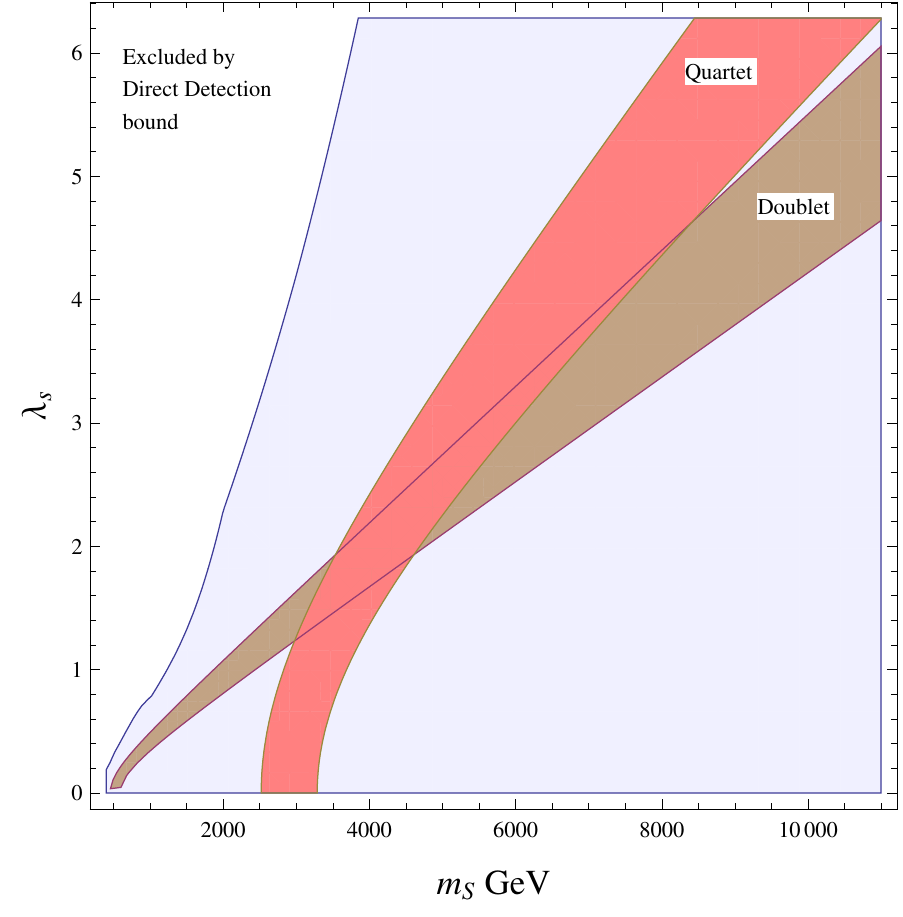}\hspace{0mm}
 \includegraphics[width=7.5cm]{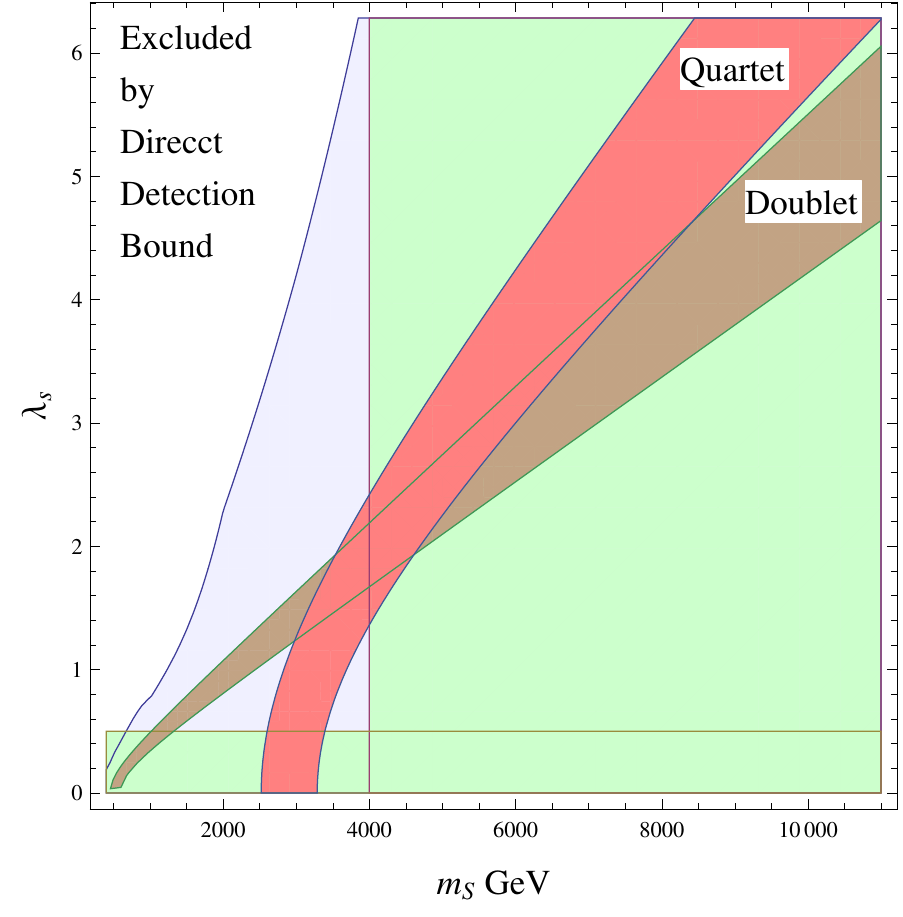}}
 \caption{Correlation between the mass of the DM, $m_{S}$ and the effective coupling
 between the Higgs and the DM, $\lambda_{S}$ for the doublet and quartet case. Here, the white
 region is excluded by the direct
 detection bound from the LUX collaboration \cite{Akerib:2013tjd}. The left figure
 represents the correlation without taking into account the Sommerfeld enhancement
 in the thermal freeze-out. In the right figure, for the green shaded region,
 Sommerfeld enhancement is not negligible.}
 \label{allowerdrelic}
\end{figure}

In the case of high mass region for the quartet, we have used
FeynRules \cite{Alloul:2013bka} to generate the model files for MicrOMEGAS \cite{Belanger:2013oya}
and have found out that the DM mass starts
from a  lower bound of $2.46$ TeV (freeze out only through gauge interaction)\footnote{without
considering the Sommerfeld enhancement}
to upper bound of $14$ TeV
set again by $\lambda_{S}\simlt 2\pi$ bound. 
In this case, the mass splitting between the DM and other components are 
\begin{eqnarray}
|m_{A}-m_{S}|&\simlt& 16\,\text{GeV},\,\,|m_{\Delta_{2}^{+}}-m_{S}|\simlt 14\,\text{GeV}\nonumber\\
|m_{\Delta^{++}}-m_{S}|&\simlt& 12\,\text{GeV},\,\,|m_{\Delta_{1}^{+}}-m_{S}|\simlt 1\,\text{GeV}
\label{quartetmasssplit}
\end{eqnarray}
when $m_{S}\sim O(5\,\text{TeV})$. figure \ref{allowerdrelic} presents the $m_{S}-\lambda_{S}$ plane with
allowed region for both doublet and quartet scalar DM by the relic density and direct detection bound
\cite{Akerib:2013tjd}. Here, 
$\lambda_{S}$ is effective coupling of $S$ to Higgs field as can be seen in eq. (\ref{masseqscalar}).
From figure \ref{allowerdrelic}, we can see that there is an overlapping region on the plane
where doublet and quartet DM satisfy the constraints simultaneously. 

The $\gamma$ coupling which controls the mass splitting between scalar (DM) and 
pseudoscalar component, has the range $\gamma\in [10^{-9},2.7]$ for the doublet and $\gamma\in[10^{-9},1.36]$
to be consistent with the relic density. But it gets another constraint from 
bounds on DM inelastic scattering with nuclei. If the typical velocity of a DM particle, $\chi$ 
is $\beta_{\chi} c\sim 220\,\text{km}/\text{sec}$,
the inelastic scattering is kinematically forbidden if the splitting $\Delta_{\chi}$
between DM and the next to lightest
component is larger,
\begin{equation}
 \Delta_{\chi}>\frac{\beta_{\chi}^2 m_{\chi}M_{\text{nucleus}}}{2(m_{\chi}+M_{\text{nucleus}})}\nonumber
\end{equation}
Therefore one would require, $\gamma\simgt 10^{-5}$ to kinematically forbid the inelastic scattering of scalar DM
with O(TeV) mass. As the inelastic scattering is mediated by the exchange of
Z boson and the scattering cross section is in the order of $10^{-40}-10^{-39}\,\text{cm}^2$, which is
much larger than the direct detection bounds,
the allowed range of $\gamma$ for doublet and quartet DM are
$\gamma\in [10^{-5},2.7]$ and $\gamma\in[10^{-5},1.36]$, respectively.

\subsubsection{Gamma ray constraints and Sommerfeld enhancement}\label{sommerfeld}
Compared to the collider searches and DM direct detection experiments, indirect detection can 
set limits on the inert scalar DM at the TeV mass range because of a
certain enhancement in the annihilation cross sections. 

At small relative velocity, two particles interacting via a long range force receive
non-perturbative enhancement in the interaction cross section which is known as
Sommerfeld enhancement \cite{sommerfeldref}. 
When the mass of the DM is much larger than the mass of W and Z bosons, the electroweak
interaction effectively behaves like a long range force, thus pair annihilation cross
sections of the DM also receive Sommerfeld enhancements as pointed in
\cite{Hisano:2003ec, Hisano:2004ds, Hisano:2005ec}. At present, as the 
relative velocity of DM is about
$10^{-3}$, Sommerfeld enhancement significantly boosts the indirect
detection signals, specially the gamma rays produced from the DM annihilation and put stringent constraint on the DM in the light of 
the experimental observations. In fact it was shown for the case of 
wino dark matter \cite{Fan:2013faa, Cohen:2013ama} and 
minimal DM models (5-plet fermion and 7-plet scalar
with zero hypercharge) \cite{Cirelli:2015bda, Garcia-Cely:2015dda, Aoki:2015nza}
(and references therein)
that they are highly constrained to be the dominant DM of
the universe by the experimental limits on gamma ray spectrum due to
the Sommerfeld enhancement in the pair annihilation cross section.

Having electroweak charge, the heavy DM component of the 
inert scalar multiplet is also expected to have
enhancements in both weak and scalar interactions. Although the
full treatment of Sommerfeld enhancement for inert scalar model is
beyond the scope of this work, following \cite{ArkaniHamed:2008qn, Slatyer:2009vg},
we introduce the dimensionless
parameters to curve out the regions of the parameter space where the
enhancement takes place and where the enhancement is negligible. The parameters
are, $\epsilon_{v_{\text{DM}}}=(v_{\text{DM}}/c)/\alpha$, $\epsilon_{\phi}=(m_{\phi}/m_{\text{DM}})/\alpha$
and $\epsilon_{\delta}=\sqrt{2\delta/m_{\text{DM}}}/\alpha$. Here $v_{\text{DM}}$ is the relative
velocity of the DM particle, $m_{\phi}$ is the mass of the gauge boson carrying the force,
$\delta$ is the mass splitting between the DM and the next to lightest charged component
of the multiplet and $\alpha$ is the coupling constant of the relevant interaction.
It was shown in \cite{Slatyer:2009vg} that the Sommerfeld enhancement is relevant 
if $\epsilon_{v_{\text{DM}}},\,\epsilon_{\phi},\,\epsilon_{\delta}\simlt 1$. On the other hand,
it is negligible for the region of 
parameter space where any of $\epsilon_{v_{\text{DM}}},\,\epsilon_{\phi},\,\epsilon_{\delta}> 1$.

\begin{figure}
 \centerline{\includegraphics[width=7.5cm]{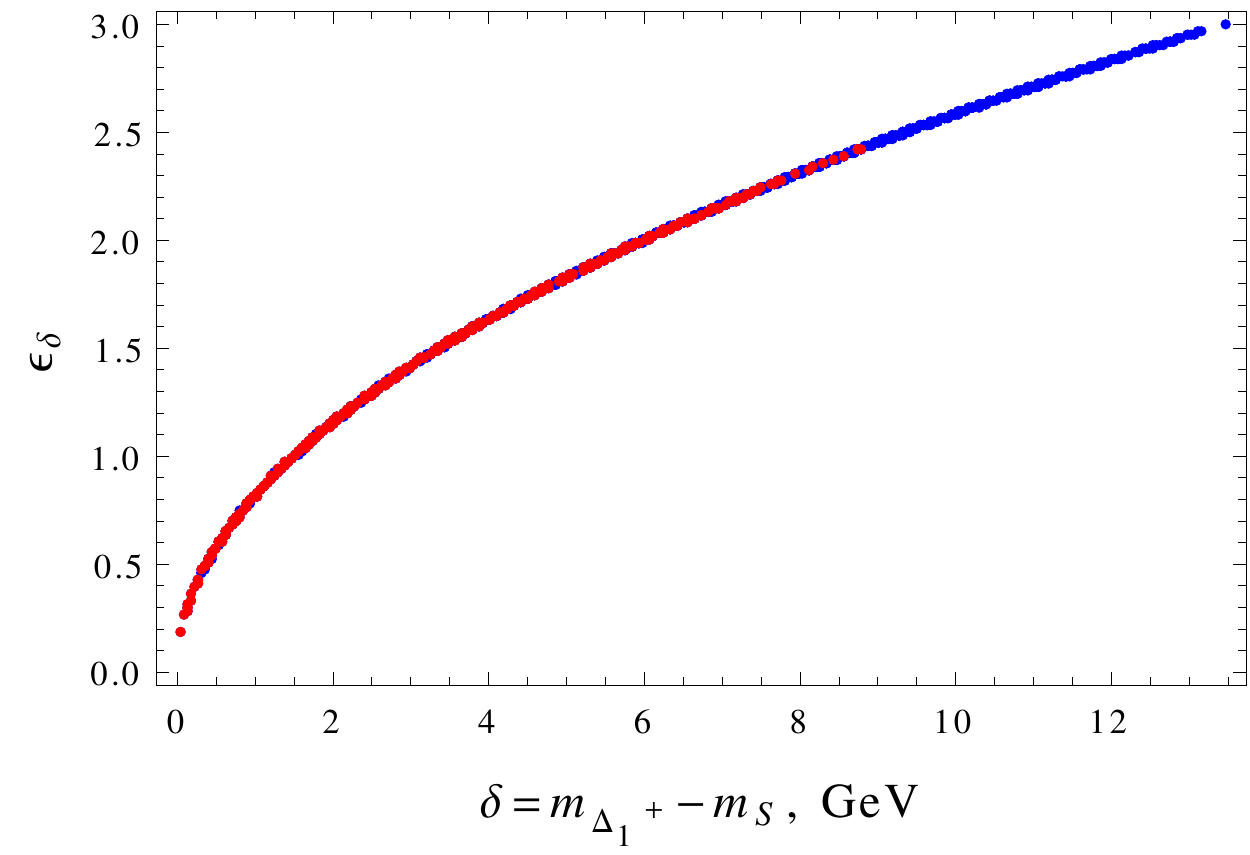}}
 \caption{$\epsilon_{\delta}$ vs $\delta=m_{\Delta^{+}_{1}}-m_{S}$ for DM mass, $m_{S}=3000$ GeV in the
 quartet.
 Here, blue points are allowed by stability conditions on the scalar potential and perturbative
 limits on scalar couplings. Red points are allowed by the bounds on electroweak precision observables.}
 \label{allowedepsilon}
\end{figure}

In the case of the minimal DM models,
the processes contributing to the gamma spectrum from DM annihilation are, DM DM $\rightarrow W^{+}W^{-},ZZ$ where the decay and fragmentation of 
W and Z pairs produce secondary photons and
DM DM $\rightarrow \gamma\gamma,\gamma Z$ producing line spectrum of mono energetic photons.
The Sommerfeld enhancement takes place when the DM-DM two particle state changes 
into $\text{DM}^{+}\text{DM}^{-}$ two particle state,
where $\text{DM}^{\pm}$ is the next to lightest charged state, 
by exchanging W boson and subsequently charged states annihilate.
For the minimal DM case, the DM and next to lightest charged state is almost 
degenerate (only  loop induced mass splitting of the $O(100)$ MeV), so
$\epsilon_{\delta}<1$ for $\alpha_{w}=1/30$ and TeV scale DM and one can have
Sommerfeld enhanced annihilation cross section. On the other hand, for the inert scalar
models, the following terms in the scalar potential
\begin{equation}
 V\supset \beta \Phi^\dagger
  \tau^a\Phi \Delta^\dagger T^a \Delta
 +\gamma[ (\Phi^T\epsilon \tau^a\Phi) (\Delta^T
    C T^a \Delta)^\dagger+h.c]
    \label{scalarmasssplit}
\end{equation}
can split the DM component and other charged component of the multiplet after electroweak
symmetry breaking. For example, for quartet, when $m_{S}=3$ TeV and $\delta=m_{\Delta^{+}_{1}}
-m_{S}=1.5$ GeV, $\epsilon_{\delta}$ is $1.001$. In addition, from figure \ref{allowedepsilon},
we can see that
the bounds on electroweak precision observables allow maximum mass splitting to be $8.78$ GeV and
corresponding $\epsilon_{\delta}$ is $2.46$. Therefore for such 
mass splitting, according to \cite{Slatyer:2009vg}, the
Sommerfeld enhancement can be negligible in the inert scalar models.

Moreover, Sommerfeld enhancement also affects the thermal freeze-out of the minimal DM
as pointed out in \cite{Cirelli:2007xd, Cirelli:2009uv}.
Such enhancement is also expected in the case of inert scalar DM. But if the 
thermal freeze-out happens after the electroweak phase transition, one can
introduce enough mass splitting so that $\epsilon_{\delta}> 1$. In fact,
$\delta=1.5$ GeV is compatible with the observed DM relic density of the universe with
$m_{S}=3$ TeV for both doublet and quartet scalar DM. 
On the other hand, if freeze-out temperature, $T_{F}$ is larger than the
critical temperature of electroweak phase transition, $T_{\text{PT}}$,
the thermal freeze-out takes place before 
the electroweak phase transition and 
there will not be any mass splitting to suppress the enhancement. Therefore
thermal DM scenario of inert scalar DM will be different than that of the broken phase.
But the value of the critical temperature of the electroweak phase transition depends on the model,
order of the transition and its dynamics (see for example \cite{Quiros:1999jp, Morrissey:2012db})
. For this reason, 
we consider the range, $T_{\text{PT}}=100-200$ GeV for
the transition temperature. Now if $x_{f}=M_{\text{DM}}/T_{F}\sim 20$,
we see that for $m_{\text{DM}}>4$ TeV, freeze out takes place in the unbroken
phase and will involve Sommerfeld enhanced annihilation cross-sections.

On the other hand, for $M_{\text{DM}}<4$ TeV, the DM freezes out in the broken phase. 
So one can
introduce the enough mass splitting, $\delta\sim 1.5$ GeV between the DM and next to lightest
charged state to suppress the enhancement in the annihilation cross sections.

For the inert scalar multiplets, apart from gauge interactions, 
the DM interacting via higgs exchange is also expected to have
enhancement. In this case, the Yukawa potential $V_{\text{sc}}$ experienced by the DM is 
\begin{equation}
 V_{\text{sc}}(r)
=\alpha_{\text{sc}}\frac{e^{-m_{h}r}}{r}\,\,\,\text{with}\,\,\,
\alpha_{\text{sc}}=\frac{\lambda_{S}^2}{4\pi} \frac{v^2}{m_{S}^2} 
\label{scalaryukawa}
\end{equation}
For example, if $m_{S}=1$ TeV and $\lambda_{S}=\pi$, $\alpha_{\text{sc}}=0.047$ so
$\epsilon_{\phi}=2.6$ for the Higgs exchange, therefore the enhancement is generally not important
for scalar interaction with the DM mass at TeV range.

In summary, although the DM with mass at TeV range in the inert scalar model is
expected to have Sommerfeld enhancement in the gauge interactions and can have
significantly enhanced indirect detection signal, there is a small common region
of parameter space for doublet and quartet as seen from figure \ref{allowerdrelic} (right) 
where one can have enough mass splitting to suppress
the Sommerfeld enhancement in the inert scalar models and such mass splitting is
compatible with the observed DM relic density. Therefore in the subsequent analysis,
we only focus that small region of parameter space with benchmark point,
$m_{S}=3$ TeV and $\delta=1.5$ GeV and have left the complete analysis
of Sommerfeld enhancement in the inert quartet case for future work \cite{futureDMcase}.

\subsubsection{Scalar coupling and LFV rates with scalar DM}\label{gammascalarDM}
\begin{figure}[h!]
 \centerline{\includegraphics[width=7.5cm]{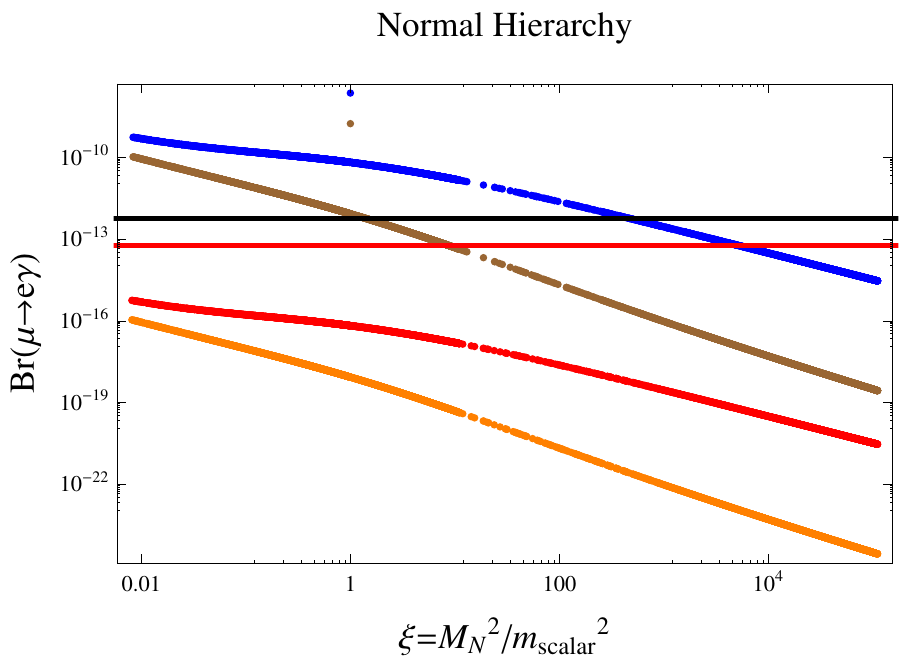}\hspace{0cm}
 \includegraphics[width=7.5cm]{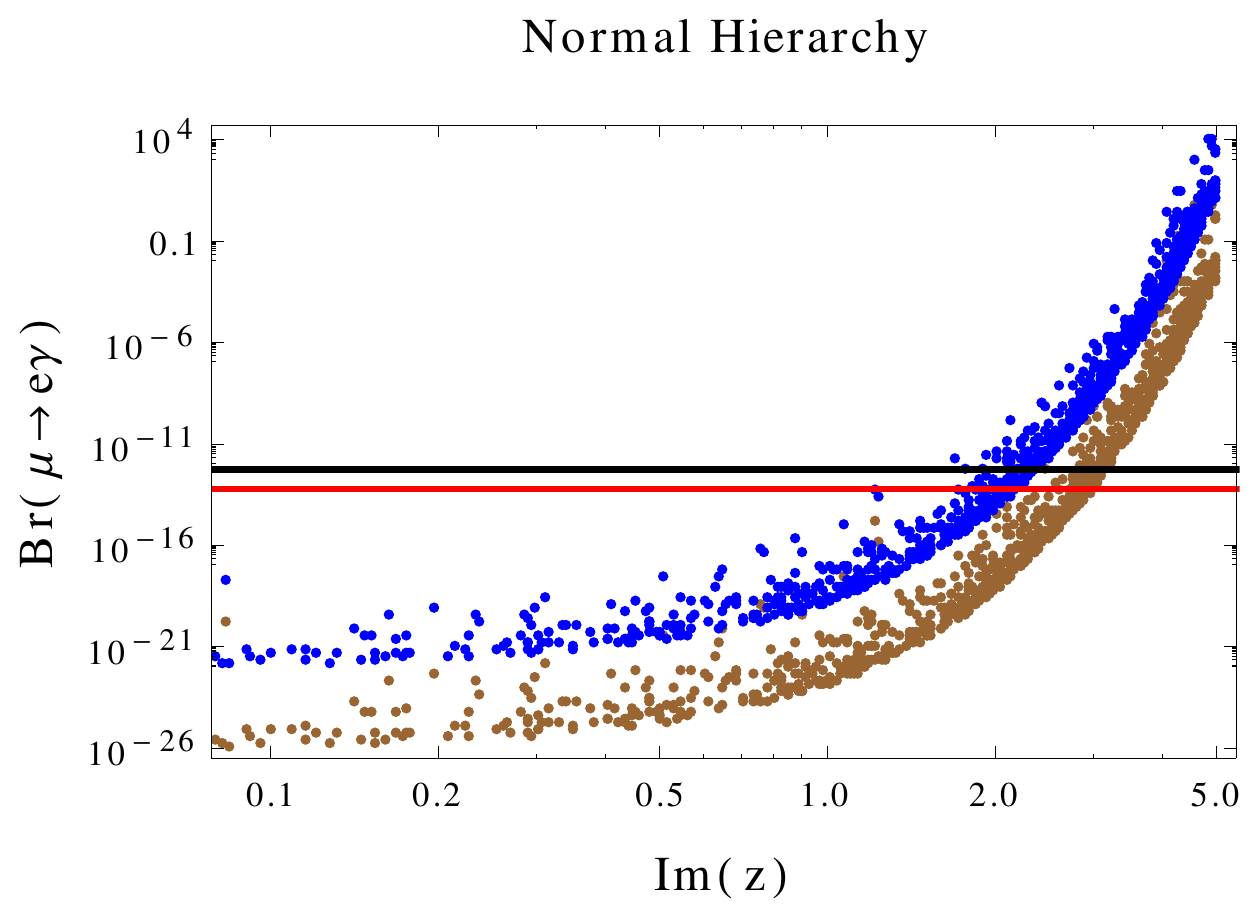}}
 \caption{Left figure presents the dependence of the rate of LFV processes on the $\gamma$ when
 the $R$ is a real matrix. Here we have considered only $\text{Br}(\mu\rightarrow e\gamma)$
 for illustration. The brown and blue represents the rate in the doublet and quartet cases
 respectively for $\gamma=10^{-9}$. On the other hand, the orange and red points represents the rate
 in the doublet and quartet cases respectively for $\gamma=10^{-5}$. Right figure presents the
 correlation of the rate in doublet (brown points) and quartet (blue points)
 with imaginary part of the complex angle, Im(z), when we consider complex $R$
 matrix. Here The scalar mass is fixed at $m_{\text{scalar}}=3000$ GeV and $\gamma=10^{-5}$.
 The black horizontal line is the current bound $5.7\times 10^{-13}$ and red line is projected
bound $6\times 10^{-14}$.}
\label{gammacomparison}
\end{figure}

There is a correlation between the $\gamma$ coupling of the scalar sector 
and the rate of LFV processes when $R$ in eq. (\ref{yuk1})
is a real orthogonal matrix.
As we can see from eq. (\ref{neutrinodoublet1})
and eq. (\ref{neutrinoquartet}) that the smaller value of $\gamma$ leads to smaller value of the loop factor $\Lambda_{i}$
and thus neutrino mass. This in turn increases the Yukawa coupling, as in eq. (\ref{casas}), and becomes
inconsistent with perturbativity bound eq. (\ref{yukawa1}) when $\gamma$ is very small.
On the other hand, large value of $\gamma$ implies 
larger separation in $m_{S}$ and $m_{A}$ and also in $m_{\Delta^{+}_{1}}$ and $m_{\Delta^{+}_{2}}$, thus
larger value of $\Lambda_{i}$
and in this case
the value of Yukawa coupling is reduced. In figure \ref{gammacomparison} (left),
We have illustrated this by comparing $\text{Br}(\mu\rightarrow e\gamma)$ for
$\gamma=10^{-9}$ and $10^{-5}$ respectively. We can see that for $\gamma=10^{-5}$, the rate has
become out of reach for current and future 
experiments. Therefore in the case of real $R$ matrix,
$\gamma\sim O(10^{-9})$ leads to appreciable LFV rates. However we have seen in Sec. \ref{darkmatter} that
as one would require, $\gamma\simgt 10^{-5}$ to kinematically forbid the inelastic scattering of scalar DM
with O(TeV) mass so considering only real $R$ will lead to negligible rates of LFV processes.

On the other hand, in the case of complex $R$, 
such correlation between $\gamma$ and the rates of LFV processes is not straightforward
because the size of Yukawa coupling also depends on the imaginary part of the complex angles in $R$.
For simplicity, we have added an imaginary part, Im(z), in three angles of $R$ and 
in figure \ref{gammacomparison} (right),  we can see that, despite having $\gamma=10^{-5}$, 
$\text{Br}(\mu\rightarrow
e\gamma)$ become comparable to the current bound with increasing values of Im(z). Again
perturbativity of the Yukawa coupling typically put upper bound on Im(z) of $O(3-5)$.
Therefore, one can have viable scalar DM in both doublet and quartet models where $\xi>1$ with appreciable LFV rates
by tuning the Imaginary part of complex angles in $R$.

\subsubsection{Viable parameter space}\label{viableparametersp}
The parameter space for the model consists of 
$\{M_{0},\alpha,\beta,\gamma\}$ of the scalar sector 
and $\{M_{N(F)},y_{i\alpha}\}$ of the fermionic sector. Here $M_{N}$ and $M_{F}$ 
are the masses of RH neutrino and real fermion triplet 
(as the components of the triplet are degenerate at tree level) respectively.

The focus of this preliminary study is the comparison among different LFV rates in
doublet and quartet model with scalar DM. At first, from figure \ref{allowerdrelic},  as an exemplary point, 
we have chosen the mass of scalar DM to be $m_{S}=3$ TeV 
with $\lambda_{S}=1.3$ in the $m_{S}-\lambda_{S}$ plane
so that scalar DM is viable both in doublet and quartet model. 
Moreover, $\gamma$ is set to be $10^{-5}$ to be consistent with bounds from
DM direct detection. As the
components of the scalar multiplet are almost degenerate apart from the
very small splitting induced by non zero $\gamma$. Therefore we set the average mass 
of the scalar components at $m_{\text{scalar}}=3$ TeV.

There are two sets of fermion mass range we have considered in our analysis. 
For the comparison of LFV rates with the variation of fermion masses both in doublet and quartet model,
we have evaluate them in two sets, namely,
i) where $\xi=M^2_{N(F)}/m^2_{\text{scalar}}<1$ so that the scalar component ceases to be the DM and
ii) where $\xi>1$ where the scalar component is the DM.
We have varied the masses of RH neutrinos and the fermion triplet
within the range, $M_{N(F)}\in (270\,\text{GeV},\,30\,\text{TeV})$ which encompasses both sets mentioned
above. $270$ GeV is taken as the lower limit of fermion mass as triplet fermion is excluded
up to that mass in collider searches. Also such range is considered to see how the LFV rates vary with the mass of the fermion in addition to the
DM aspects of inert scalar model.

We have used the experimental values of low energy neutrino parameters, 
$U_{\text{PMNS}}$, $\Delta m^2_{\text{solar}}$
and $\Delta^2_{\text{atm}}$ as the input in eq. (\ref{yuk1}) for
Yukawa couplings. For both normal and inverted hierarchies, we could only vary
the lowest neutrino mass, $m_{\nu_{1}}$, the Dirac phase, $\delta$ and Majorana phases, 
$\alpha_{\nu},\, \beta_{\nu}$ and three complex angles, $z_{1},\,z_{2},\,z_{3}$ of, $R$.
In our numerical analysis,as an simplification, the lowest neutrino
mass is set to $m_{\nu}=1$ meV, $\delta\in [0, 2\pi]$, $\alpha=\beta=0$ and common imaginary part
in $z_{i}=\theta_{i}+i\,\text{Im}(z_{i})$, Im(z) with the range $(0,5)$.

Summarizing, our input parameters in the numerical scans are 
$\{M_{0},\alpha,\beta,\gamma,M_{N}=M_{F}=\tilde{M},m_{\nu_{1}},\delta,\alpha_{\nu},\beta_{\nu},\theta_{1}
,\theta_{2},\theta_{3},\text{Im}(z)\}$ satisfying all the constraints mentioned above.
Therefore, we can compare the LFV rates in both models for common viable
point in the parameter space.

\subsection{LFV processes}\label{LFVprocessessection}

In the inert scalar models with scalar DM in the high mass regime, 
there is no direct correlation between the 
Yukawa couplings and DM properties. Also we have seen that the real 
matrix $R$ and $\gamma\simgt 10^{-5}$ (scalar 
DM direct detection constraint) give rise to small Yukawa couplings which in turn lead to LFV rates
beyond the reach of current and future experiments as seen in figure \ref{gammacomparison} (left). But
the size of the Yukawa coupling can be enhanced by varying the imaginary part of complex angles of $R$ without
substantially affecting the phenomenology of the scalar DM and despite having $\gamma\simgt 10^{-5}$, we
can easily obtain the LFV rates within the experimental range.

So first we have compared the rates of $\mu\rightarrow e\gamma$, $\mu\rightarrow ee\overline{e}$ and
$\mu-e$ conversion rate with $\gamma=10^{-5}$ and the real $R$ matrix by varying the fermion masses for
the doublet and quartet models. Then
we vary Im(z) within its constrained limits and determine the region allowed by current and future bounds
on the rates of these three LFV processes for both doublet and quartet cases. 
Also when $\xi>1$, we have our scalar dark matter in both doublet and quartet models.

\subsubsection{$\text{Br}(\mu\rightarrow e \gamma)$}\label{muegammasec}

\begin{figure}[h!]
\centerline{\includegraphics[width=7.5cm]{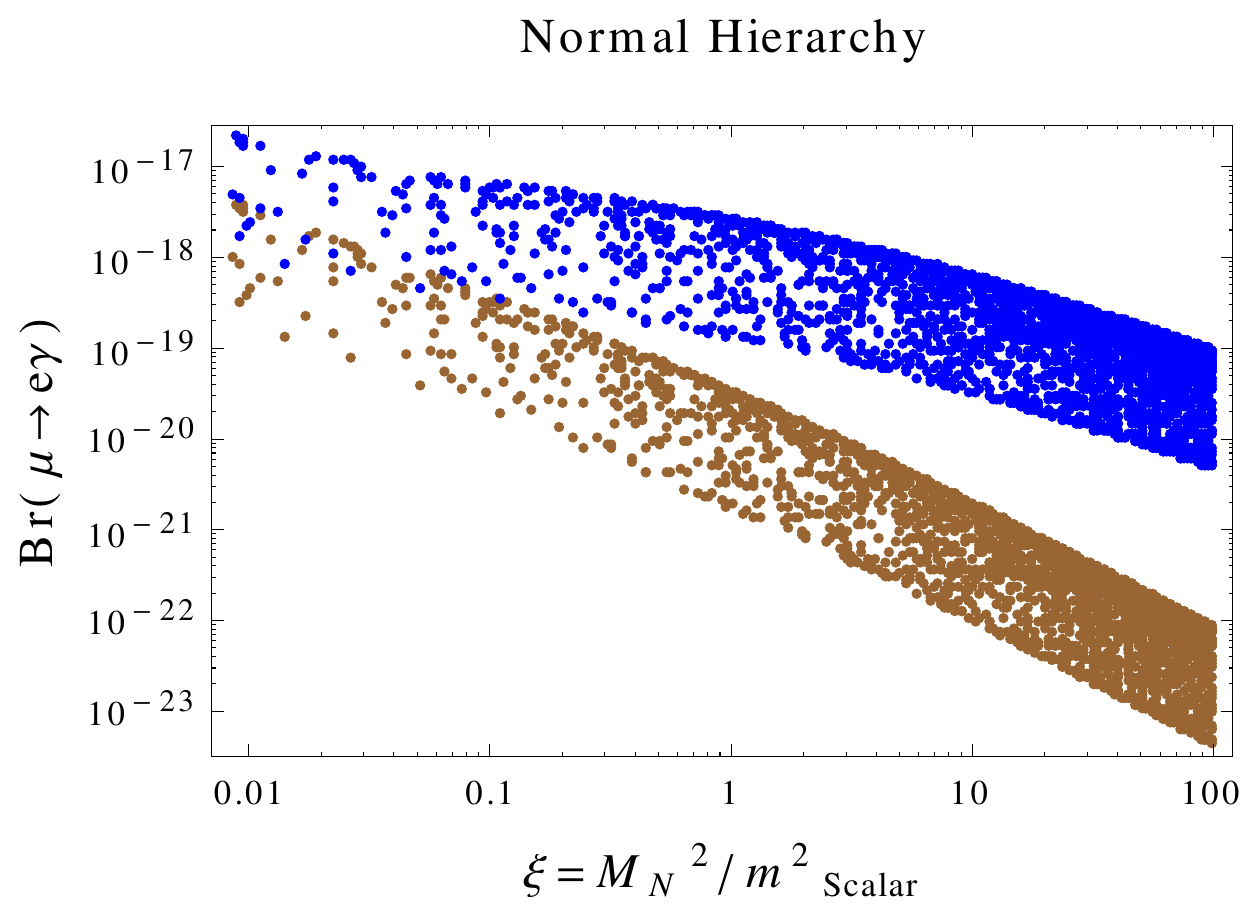}\hspace{0 cm} 
\includegraphics[width=7.5cm]{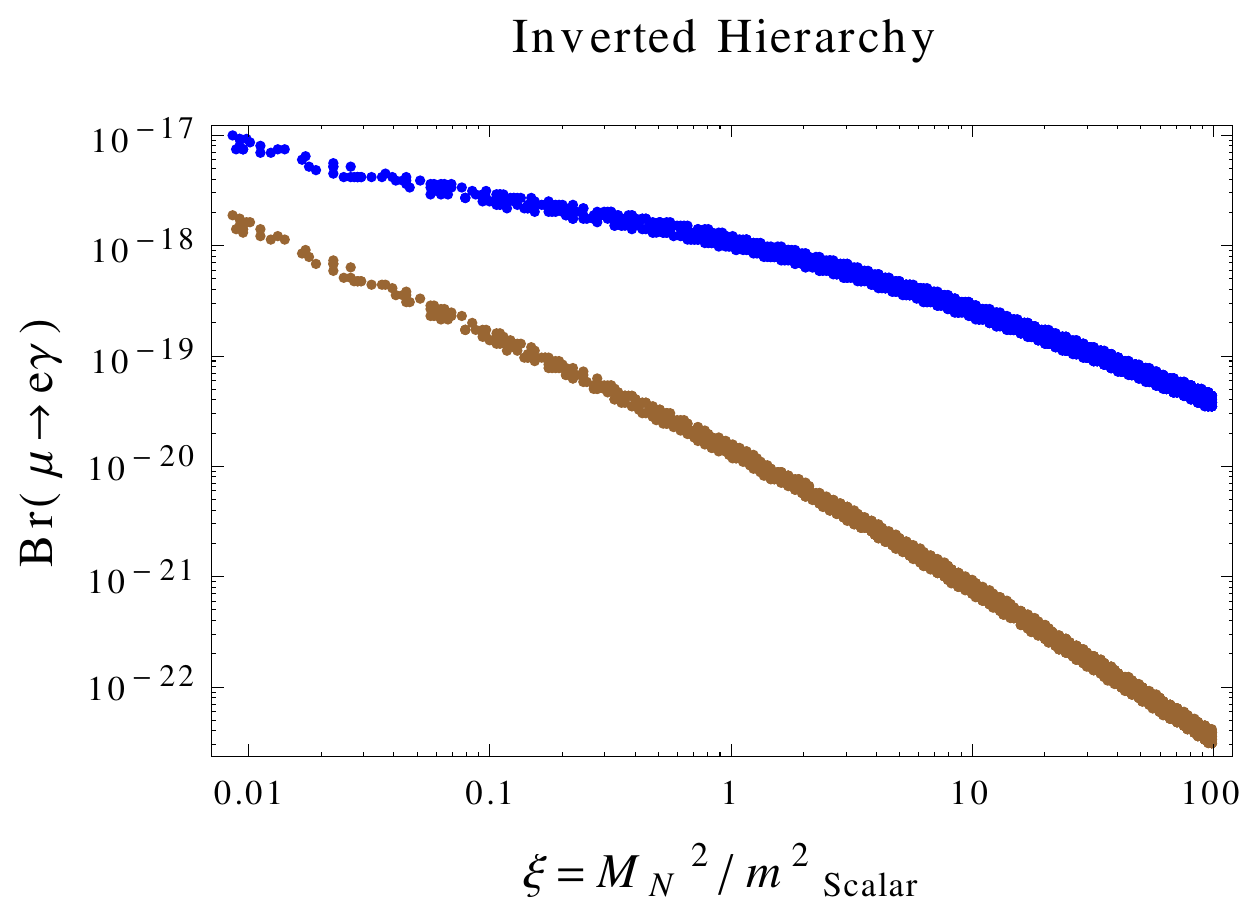}} 
\caption{Correlation between $\xi=M^2_{N(F)}/m^2_{\text{scalar}}$ and $\text{Br}(\mu\rightarrow e\gamma)$
for doublet (brown points) and quartet (blue points) with normal (left fig.) and inverted (right fig.)
hierarchy for light neutrino mass. Here we have taken $M_{N(F)}$ to be degenerate, random Dirac phase
$\delta$ and random real matrix $R$. Also we have set Majorana phases $\alpha_{\nu}$ and $\beta_{\nu}$ to be zero
in this case. The scalar mass is fixed at
$m_{\text{scalar}}=3000$ GeV. Also $\gamma=10^{-5}$ and light neutrino mass, $m_{\nu_{1}}=1$ meV.
}
\label{Brmuegamma}
\end{figure}

Due to the excellent bound put by the MEG collaboration \cite{Adam:2011ch, Adam:2013mnn}, $\mu\rightarrow e\gamma$
is one of the most well studied LFV processes. figure \ref{Brmuegamma} shows the comparison of this process between
the doublet (brown points) and the quartet (blue points) scalar. 
We can see that the quartet
contribution to $\mu\rightarrow e\gamma$ is larger than that of the doublet. For the same parameter point,
in the quartet case, additional charged and neutral scalar ($\Delta_{1}^{\pm}$, $\Delta_{2}^{\pm}$,
$\Delta^{\pm\pm}$, $S$ and $A$) and fermion states ($F^{0}_{i}$ and $F^{\pm}_{i}$) enter in the loop
compared to single charged scalar ($C^{\pm}$) and neutral fermion state ($N_{i}$) in the doublet case 
and as the contributions of
extra states are additive, the rate has increased in the quartet case than that of the doublet. From
figure \ref{Brmuegamma} we can see that $\text{Br}(\nu\rightarrow e \gamma)$ is larger for the quartet
than the doublet for both $\xi<1$ and $\xi>1$ (where the doublet and quartet scalars are the DM).

\subsubsection{$\text{Br}(\mu\rightarrow e e\overline{e})$}\label{mueeesection}

\begin{figure}[h!]
\centerline{\includegraphics[width=7.5cm]{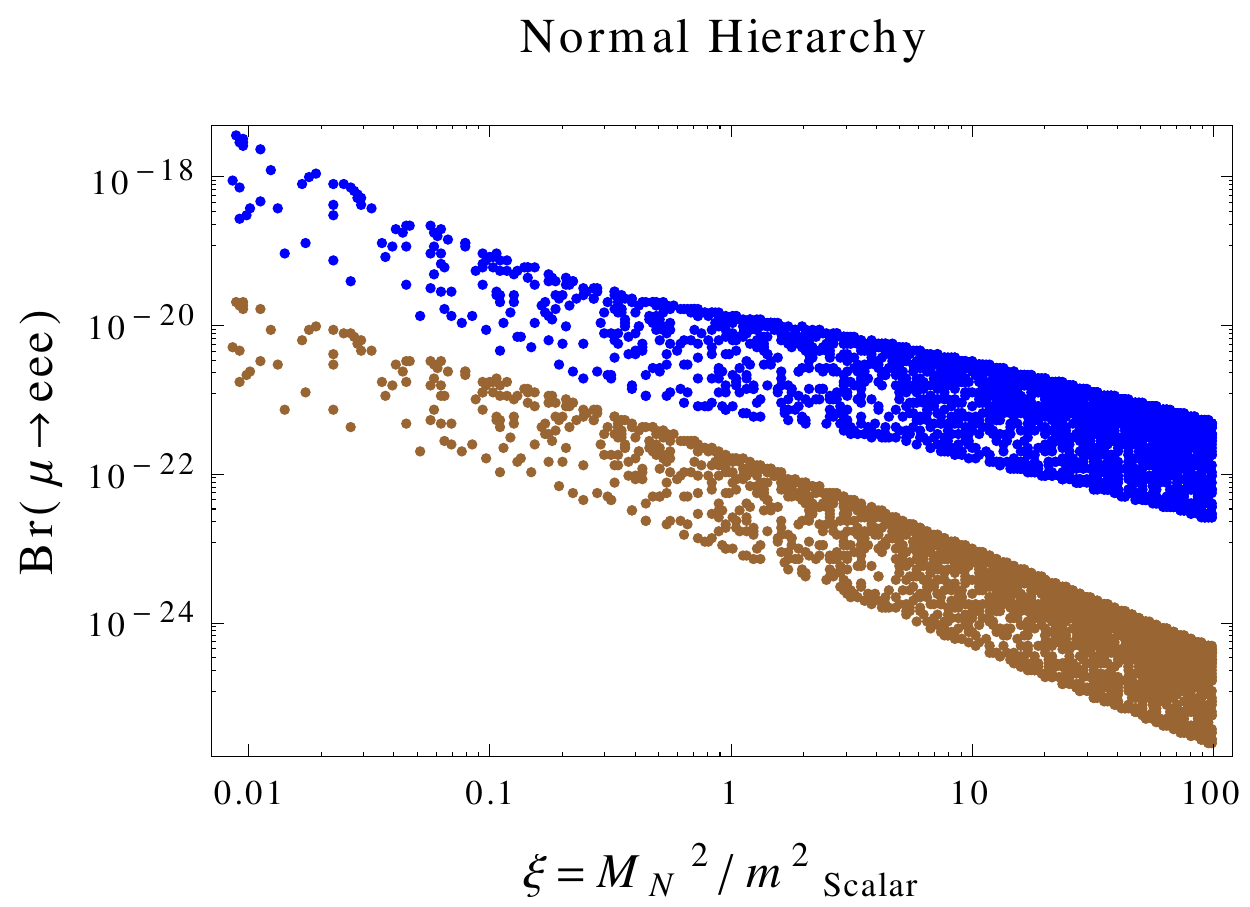}\hspace{0 cm} 
\includegraphics[width=7.5cm]{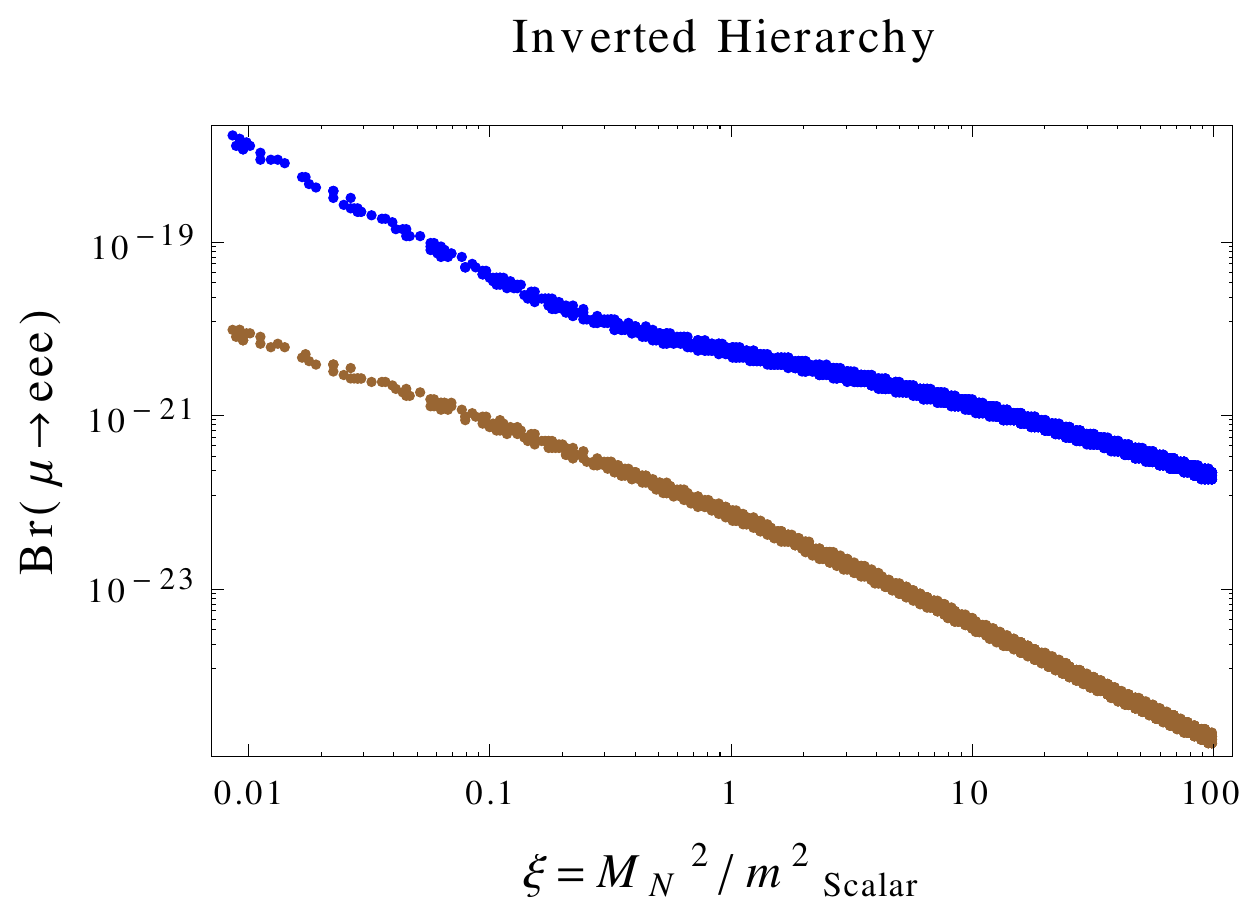}} 
\caption{Correlation between $\xi=M^2_{N(F)}/m^2_{\text{scalar}}$ and $\text{Br}(\mu\rightarrow ee\overline{e})$
for doublet (brown points) and quartet (blue points) with normal (left fig.) and inverted (right fig.)
hierarchy for light neutrino mass. Here we have taken same input parameters as in $\text{Br}(\mu\rightarrow
e\gamma)$.
}
\label{Brmueee}
\end{figure}

In $\mu\rightarrow e e \overline{e}$, the dominant contributions are coming from $\gamma$-penguin and Box diagrams.
The Higgs penguin diagram is suppressed by the small electron Yukawa coupling. The Z penguin contribution is small
because of the cancellation that takes place between $C_{24}$ and $B_{1}$ terms in eq. (\ref{doubletZ}) and also
between the same terms in eq. (\ref{quartetZ1}) when $m_{\sigma_{1}}=m_{\sigma_{2}}$. Moreover, similar cancellation
takes place between the first two lines and third line of eq. (\ref{quartetZ2})
due to the specific relations among the couplings in front of the vertices.
Therefore Z penguin contribution is also small in $\mu\rightarrow e e \overline{e}$ for both inert doublet and quartet
case. Also note that the Z contribution in the quartet case is relatively bigger than that in the doublet because 
in the quartet $m_{\sigma_{1}}$ and $m_{\sigma_{2}}$ are not exactly equal when $\sigma_{1}\neq \sigma_{2}$. Hence
one receives larger Z-penguin contribution in the quartet compared to the doublet. Still this contribution
is numerically not significant if we compare it with $\gamma$ penguin diagram or box diagram contributions.
From figure \ref{Brmueee}, we can see that 
$\text{Br}(\mu\rightarrow e e \overline{e})$ is larger for quartet (blue points) compared to the doublet
(brown points) for both $\xi<1$ and $\xi>1$ cases.

\subsubsection{$\mu-e$ conversion rate}\label{mueconvsection}

\begin{figure}[h!]
\centerline{\includegraphics[width=7.5cm]{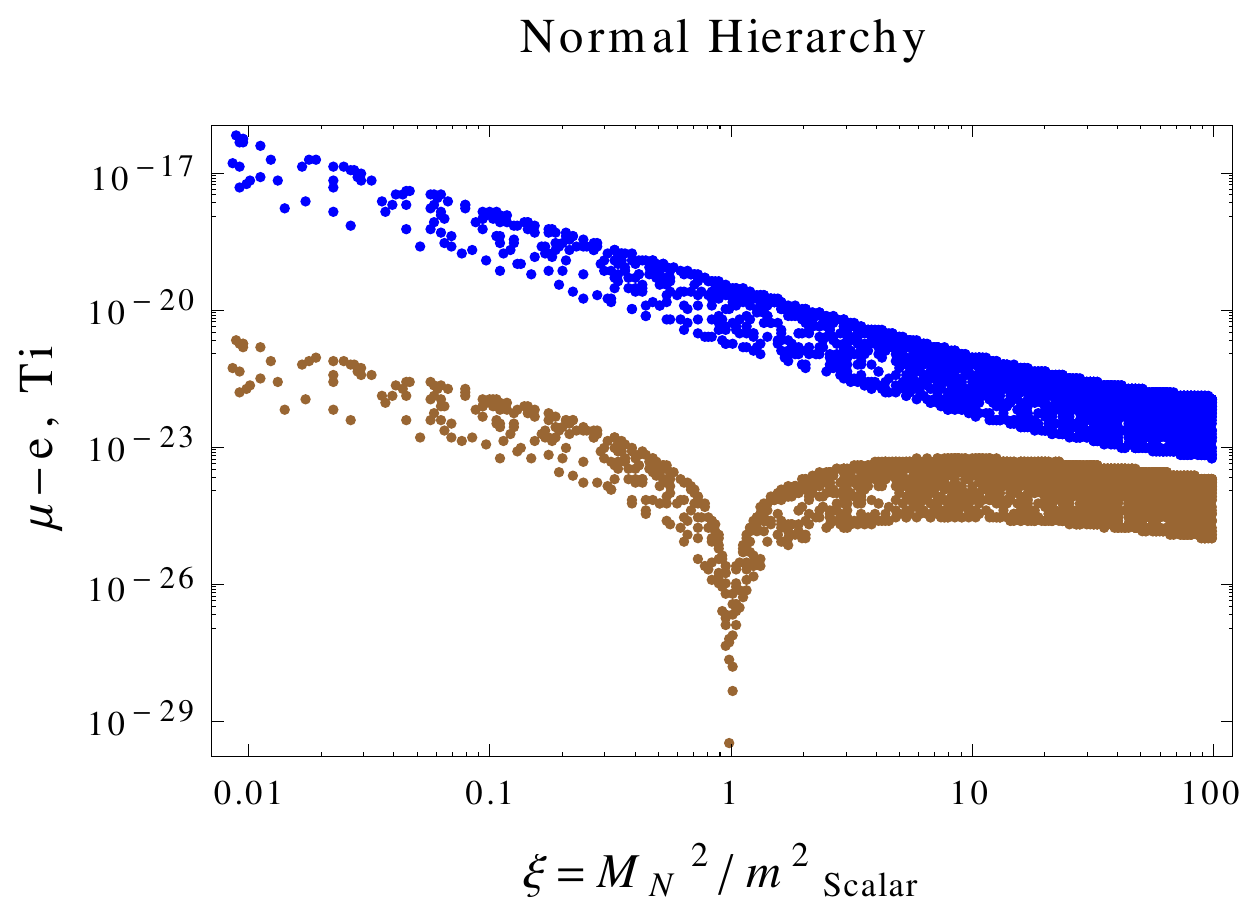}\hspace{0 cm} 
\includegraphics[width=7.5cm]{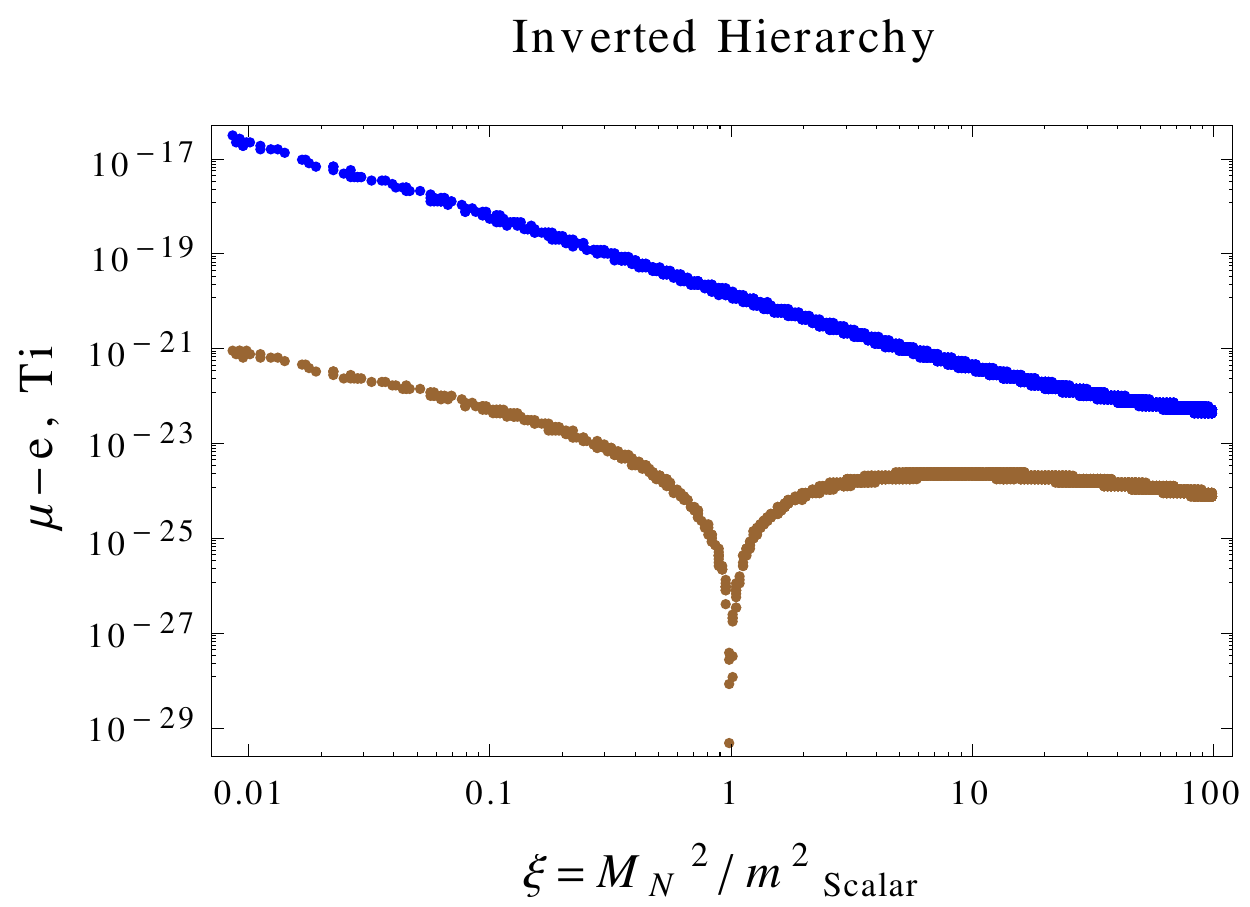}} 
\caption{Correlation between $\xi=M^2_{N(F)}/m^2_{\text{scalar}}$ and $\mu-e$ conversion rate for Ti nucleus.
for doublet (brown points) and quartet (blue points) with normal (left fig.) and inverted (right fig.)
hierarchy for light neutrino mass. Here we have taken same input parameters as in $\text{Br}(\mu\rightarrow
e\gamma)$.
}
\label{mueconveTifig}
\end{figure}

Another prominent LFV process currently under investigation is the $\mu-e$ conversion in nuclei. Here
we have calculated the $\mu-e$ conversion rate for Ti and Au nuclei in the inert model with doublet
and quartet. From figure \ref{mueconveTifig}, we can see that the $\mu-e$ conversion rate is larger for
the quartet (blue points) compared to the doublet (brown points). The dip occurs in the doublet
contribution at $\xi=1$ because at that value, the dipole contribution $A^{\text{doublet}}_{D}$
and the non-dipole contribution $A^{\text{doublet}}_{ND}$ are equal as they are coming from single
$\gamma$ penguin diagram involving charged scalar $C^{\pm}$ and neutral fermion $N_{i}$ and 
eq. (\ref{nuclear31}) indicates that the effective coupling is zero for doublet at that point.
On the other hand, for quartet case $A^{\text{quartet}}_{D}$ and $A^{\text{quartet}}_{ND}$ at $\xi=1$
are different because more than one charged scalar contribute to the $\gamma$ penguin diagrams. Again
we can see from figure \ref{mueconveTifig} that the conversion rate is larger for the quartet than that for
the doublet for both $\xi<1$ and $\xi>1$ cases. 
We have not included the figure for $\mu-e$ conversion rate in 
Au nuclei as it is similar to figure \ref{mueconveTifig}.

\subsubsection{LFV rates in the doublet and quartet}\label{doubletvsquartet}
As expected, the LFV rates seen in figure \ref{Brmuegamma}, \ref{Brmueee}
and \ref{mueconveTifig} are very small for real $R$ and $\gamma=10^{-5}$. The rates will reduce even more
if we increase $\gamma$. Still the rates are larger for the quartet compared to the doublet
for $\xi<1$ and $\xi>1$ case where scalar is treated as the DM candidate. Now 
we increase the value of Im(z) and calculate the LFV rates
with increasing values of $\tilde{M}$.
\begin{figure}[h!]
 \centerline{\includegraphics[width=7.5cm]{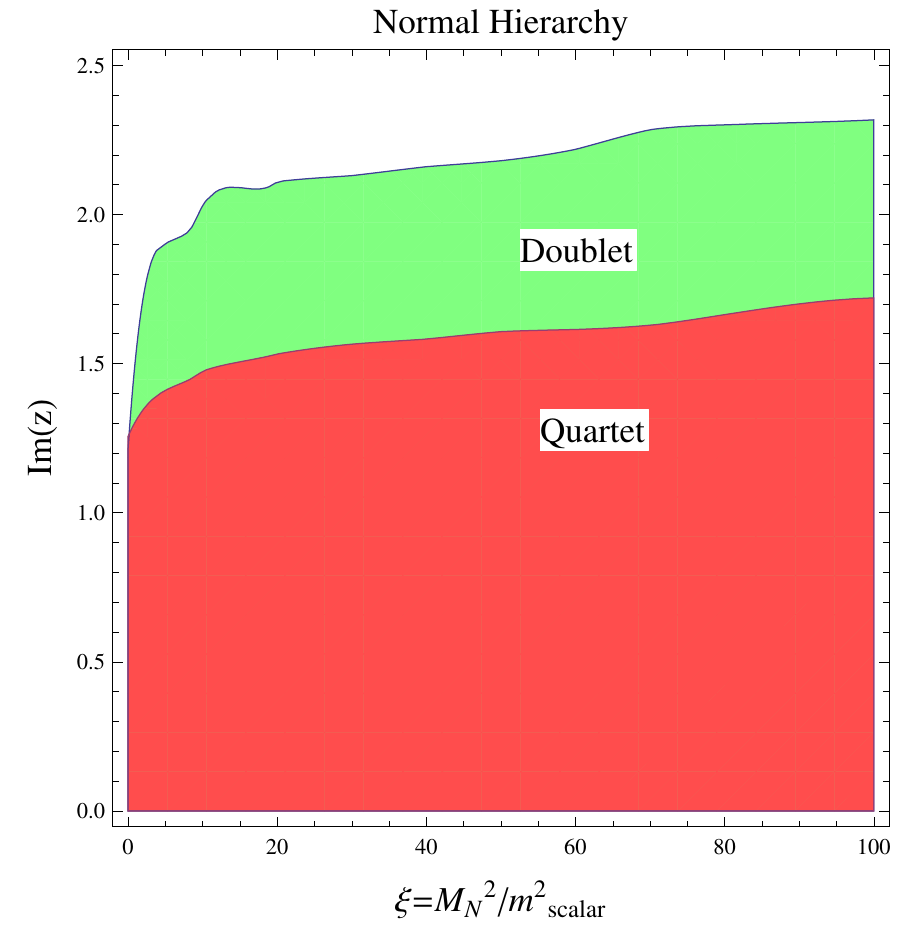}\hspace{0cm}
 \includegraphics[width=7.5cm]{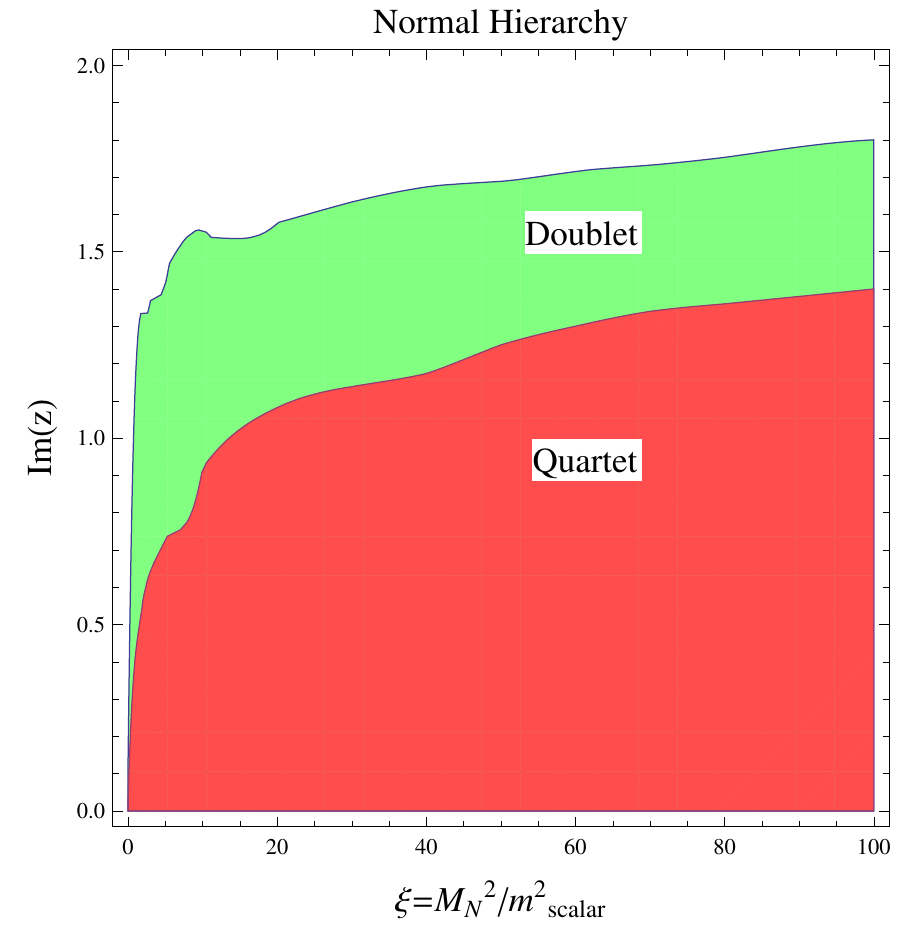}}
 \caption{The $\xi-\text{Im(z)}$ plane for degenerate $M_{N(F)}$, random Dirac phase
$\delta$, zero Majorana phases $\alpha_{\nu}=\beta_{\nu}=0$ and light neutrino mass, $m_{\nu_{1}}=1$ meV.
The scalar mass is $m_{\text{scalar}}=3000$ GeV with $\gamma=10^{-5}$. In (left), the current bounds are
imposed:
$\text{Br}(\mu\rightarrow e\gamma)\simlt 5.9\times 10^{-13}$, $\text{Br}(\mu\rightarrow ee\overline{e})
\simlt 1\times 10^{-12}$ and $\mu-e$ conversion rate for Ti $\simlt 4.3\times 10^{-12}$. In (right),
the future sensitivity are considered: $\text{Br}(\mu\rightarrow e\gamma)\simlt 6.4\times 10^{-14}$,
$\text{Br}(\mu\rightarrow ee\overline{e})
\simlt 1\times 10^{-16}$ and $\mu-e$ conversion rate for Ti $\simlt 10^{-18}$.}
\label{LFVconstraints}
\end{figure}

From figure \ref{LFVconstraints}, we can see that LFV rates in the quartet are more constrained than those
in the doublet for common parameter space satisfying all the restrictions of Sec. \ref{constranitssec}.
The allowed regions on $\xi$-Im(z) plane are reduced further for both doublet and quartet models if
one imposes the sensitivity of future lepton flavor violating experiments. 
The case for inverted hierarchy shows similar pattern so we have only
presented results regarding normal hierarchy.

\section{Conclusions}\label{conclusion}

The scotogenic model is a well studied neutrino mass model and lepton flavor violation
is one of its important phenomenological aspects. In this study we present the comparison
among different LFV processes in the inert doublet and the quartet model, taking into account
the current experimental limits and future sensitivity. There are two possible dark matter candidates
in the inert scalar models: scalar and fermionic DM. In this study we have considered scalar DM
and evaluated LFV rates for common parameter space subjected to collider bounds, DM constraints 
for doublet and quartet model and low-energy neutrino parameters. 
Our results are summarized as follows

\begin{itemize}
 \item $\text{Br}(\mu\rightarrow e \gamma)$,
$\text{Br}(\mu\rightarrow e e \overline{e})$ and $\mu-e$ conversion rates in
nuclei in the quartet model are larger than those in the doublet model for the same parameter space
as seen from figure \ref{Brmuegamma}, \ref{Brmueee} and \ref{mueconveTifig}.
In the case of higher scalar representation more particles enter into the loops and their
contributions are additive in the LFV processes. Therefore we can have larger rates of
different LFV processes compared to the lower scalar representation. From figure \ref{LFVconstraints},
we can see that, LFV processes in higher scalar representation 
are more constrained by the current and near-future experiments.
In addition, this phenomenological result is complementary to the 
appearance of low scale Landau pole for higher representations 
\cite{AbdusSalam:2013eya, DiLuzio:2015oha, Hamada:2015bra}.

\item There is no significant deviation from figure \ref{Brmuegamma}-\ref{LFVconstraints}
for non-degenerate right handed neutrinos and real fermion triplets.
In the case of large hierarchy, $m_{N_{3}}>>m_{N_{1,2}}$, the dominant contribution
comes from only the lightest generation.
\end{itemize}

We would like to emphasize here that the conclusion of our preliminary study is applicable 
to the inert scalar models 
where scalar DM is considered. But there is much room for an improved analysis. For example,
in the case of fermionic DM, the DM constraints will be different and will
have different viable parameter set for the LFV rate comparison. Also one needs to
study the DM properties and viability of a common parameter space where $\xi\sim 1$. 
Therefore, further quantitative analysis of the fermionic DM aspects 
in the quartet model will be presented in a future publication \cite{futureDMcase}.
Furthermore, similar analysis can
be carried out for $\tau\rightarrow \mu\gamma$, $\tau\rightarrow ee\overline{e}$, 
$\tau\rightarrow \mu\mu\overline{\mu}$ in the inert scalar models to probe the flavor structure
of the Yukawa sector and to have better constraints on the higher scalar
representation in the light of experimental limits.

\section*{Acknowledgements}

We would like to thank Avelino  Vicente  for stimulating discussion. 
T.A.C is grateful to Fernando Quevedo, Bobby Acharya and the HECAP section of 
ICTP for the support and the hospitality where the initial part of this work has been carried out.
We are also indebted to the Referee for the constructive report, for which, the result and presentation
of our study has been substantially improved.

\appendix

\section{Scalar masses}\label{massspecscalar}

\subsection{Inert Doublet}
The mass spectrum for the inert doublet in our parametrization eq. (\ref{potq}) is,
\begin{eqnarray}
m^2_{S}&=&M^2_{0}+\frac{1}{2}\left(\alpha+\frac{1}{4}\beta+\gamma\right)v^2\nonumber\\
m^2_{A}&=&M^2_{0}+\frac{1}{2}\left(\alpha+\frac{1}{4}\beta-\gamma\right)v^2\nonumber\\
m^2_{C}&=&M^2_{0}+\frac{1}{2}\left(\alpha+\frac{1}{4}\beta\right)v^2
\label{doubetmass}
\end{eqnarray}

\subsection{Inert Quartet}
In the inert quartet case, the $\gamma$ term, apart from splitting $S$ and $A$, 
also mixes two single charged components of the
quartet. According to eq. (\ref{sc1}), the mass matrix for single
charged fields in $(\Delta^{+},\Delta^{'+})$ basis is 
\begin{equation}
  M^2_{+}=\begin{pmatrix}
  M_0^2+\frac{1}{2}(\alpha-\frac{1}{4}\beta)v^2& \frac{\sqrt{3}}{2}\gamma v^2\\
  \frac{\sqrt{3}}{2}\gamma v^2& M_0^2+\frac{1}{2}(\alpha+\frac{3}{4}\beta)v^2
  \end{pmatrix}
\end{equation}
Diagonalizing the mass matrix, we have mass eigenstates for single
charged fields, $\Delta_1^+=\Delta^{+}\cos\theta +\Delta^{'+}\sin\theta$,
$\Delta_2^+=-\Delta^{+}\sin\theta +\Delta^{'+}\cos\theta$ with
$\tan2\theta=-\frac{2\sqrt{3}\gamma}{\beta}$. 

Therefore the mass spectrum of the quartet is
\begin{eqnarray}
\label{qrmass1}
m^2_{S(A)}&=&M^2_{0}+\frac{1}{2}\left(\alpha+\frac{1}{4}\beta\mp 2\gamma\right)v^2\nonumber\\
m^2_{\Delta^{++}}&=&M^2_{0}+\frac{1}{2}\left(\alpha-\frac{3}{4}\beta\right)v^2\nonumber\\
m^2_{\Delta^{+}_{1}(\Delta^{+}_{2})}&=&M^2_{0}+\frac{1}{2}\left(\alpha+\frac{1}{2}\beta \mp \frac{1}{2}\sqrt{\beta^2+12\gamma^2}\right)v^2
\label{massquartet}
\end{eqnarray}

Because of the mixing between two single charged states, the mass
relation is 
\begin{equation}
  m_{S}^2+m_{A}^2=m_{\Delta_{1}^{+}}^2+m_{\Delta_{2}^{+}}^2
\end{equation}

\section{Loop functions}\label{loopappendix}

The loop functions relevant for the dipole and non-dipole form factors from $\mu e \gamma$ vertex are
\begin{eqnarray}
 F^{(n)}(x)&=&\frac{1-6x+3 x^2+2 x^3-6 x^2\text{ln}x}{6(1-x)^{4}}\label{dipoleneu}\\
 F^{(c)}(x)&=&\frac{2+3x-6x^2+x^3+6x\text{ln}x}{6(1-x)^{4}}\label{dipolechar}\\
 G^{(n)}(x)&=&\frac{2-9x+18x^2-11x^3+6x^3\text{ln}x}{6(1-x)^4}\label{nondipoleneu}\\
 G^{(c)}(x)&=&\frac{16-45x+36x^2-7x^3+6(2-3x)\text{ln}x}{6(1-x)^4}\label{nondipolechar}
\end{eqnarray}

In the following we collect the Passarino-Veltman loop functions.
\begin{equation}
 B_{1}(m_{1},m_2)=-\frac{1}{2}-\frac{m_1^{4}-m_{2}^{4}+2m_{1}^4\text{ln}\frac{m_{2}^2}{m_{1}^{2}}}{4(m_{1}^2-m_{2}^2)^2}
 +\frac{1}{2}\text{ln}\frac{m_{2}^{2}}{\mu^{2}}
 \label{bfunction}
\end{equation}

\begin{equation}
 C_{0}(m_{1},m_{2},m_{3})=\frac{m_{2}^{2}(m_{1}^2-m_{3}^2)\text{ln}\frac{m_{2}^2}{m_{1}^2}-(m_{1}^{2}-m_{2}^{2})m_{3}^2
 \text{ln}\frac{m_{3}^{2}}{m_{1}^{2}}}{(m_{1}^2-m_{2}^{2})(m_{1}^{2}-m_{3}^{2})(m_{2}^{2}-m_{3}^{2})}
\end{equation}

\begin{align}
 C_{24}(m_{1},m_{2},m_{3})&=\frac{1}{8(m_{1}^2-m_{2}^{2})(m_{1}^{2}-m_{3}^{2})(m_{2}^{2}-m_{3}^{2})}\left[-2(m_{1}^{2}+m_{2}^{2})
 m_{3}^{4}\,\text{ln}\frac{m_{3}^{2}}{m_{1}^{2}}
 -(m_{3}^2-m_{1}^{2})\right.\notag\\
 &\left.\left(2m_{2}^{4}\,\text{ln}\frac{m_{2}^{2}}{m_{1}^{2}}
 +(m_{1}^{2}-m_{2}^{2})(m_{2}^{2}-m_{3}^{2})
 \left(2\,\text{ln}\frac{m_{1}^2}{\mu^{2}}-3\right)\right)\right]
\end{align}

\begin{eqnarray}
 \tilde{D}_{0}(m_1,m_2,m_3,m_4)&=&\frac{m_{2}^{4}\,\text{ln}\frac{m_{2}^{2}}{m_{1}^{2}}}
 {(m_{2}^{2}-m_{1}^{2})(m_{2}^{2}-m_{3}^{2})(m_{2}^{2}-m_{4}^{2})}-\frac{m_{3}^{4}\,\text{ln}\frac{m_{3}^{2}}{m_{1}^{2}}}
 {(m_{3}^{2}-m_{1}^{2})(m_{3}^{2}-m_{2}^{2})(m_{3}^{2}-m_{4}^{2})}\nonumber\\
 &-&\frac{m_{4}^{4}\,\text{ln}\frac{m_{4}^{2}}{m_{1}^{2}}}
 {(m_{4}^{2}-m_{1}^{2})(m_{4}^{2}-m_{2}^{2})(m_{4}^{2}-m_{3}^{2})}
 \label{dtilde}
\end{eqnarray}

\begin{eqnarray}
 D_{0}(m_1,m_2,m_3,m_4)&=&\frac{m_{2}^{2}\,\text{ln}\frac{m_{2}^{2}}{m_{1}^{2}}}
 {(m_{2}^{2}-m_{1}^{2})(m_{2}^{2}-m_{3}^{2})(m_{2}^{2}-m_{4}^{2})}-\frac{m_{3}^{2}\,\text{ln}\frac{m_{3}^{2}}{m_{1}^{2}}}
 {(m_{3}^{2}-m_{1}^{2})(m_{3}^{2}-m_{2}^{2})(m_{3}^{2}-m_{4}^{2})}\nonumber\\
 &-&\frac{m_{4}^{2}\,\text{ln}\frac{m_{4}^{2}}{m_{1}^{2}}}
 {(m_{4}^{2}-m_{1}^{2})(m_{4}^{2}-m_{2}^{2})(m_{4}^{2}-m_{3}^{2})}
 \label{dzero}
\end{eqnarray}
\newpage
\section{$\mu e\gamma$ vertex, $\mu e Z$ vertex and box diagrams}\label{muegZver}

\subsection{$\mu e \gamma$ vertex}\label{muegammafeyn}
\begin{figure}[h!]
\vspace{-1cm}
 \centerline{
 \includegraphics[width=4cm]{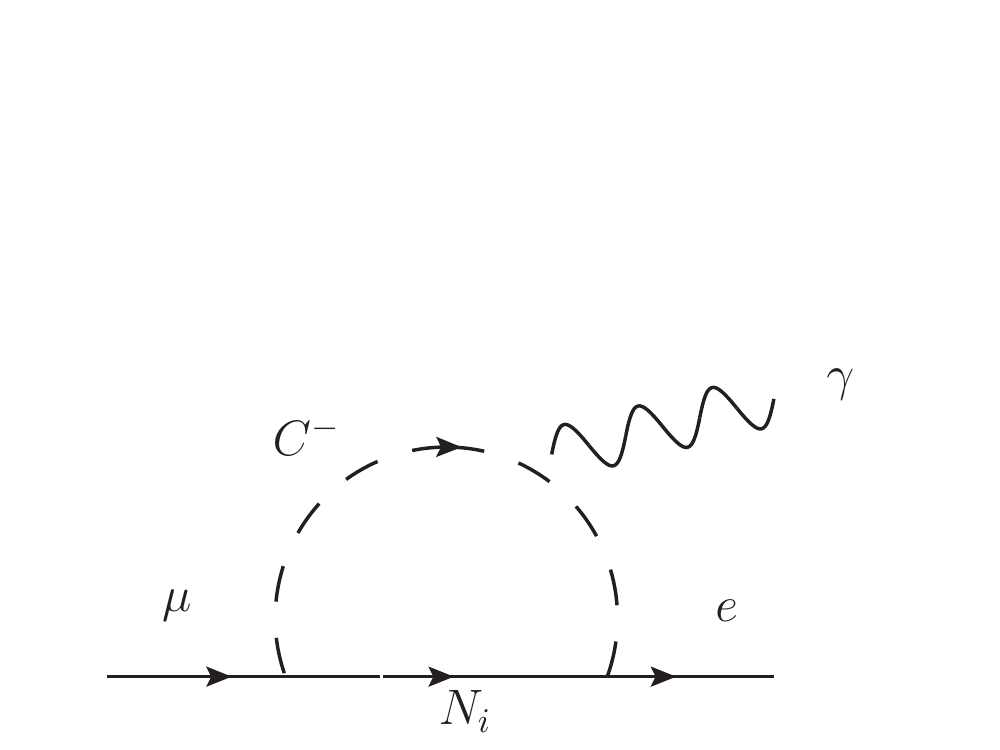}\hspace{0cm}
 \includegraphics[width=4cm]{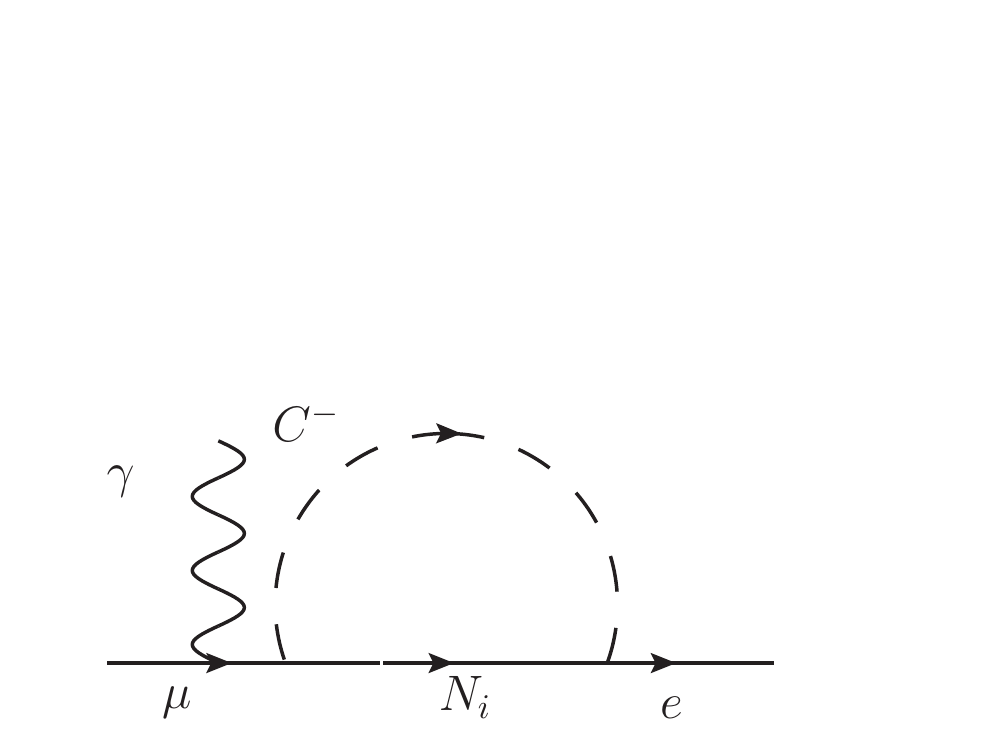}\hspace{0cm}
 \includegraphics[width=4cm]{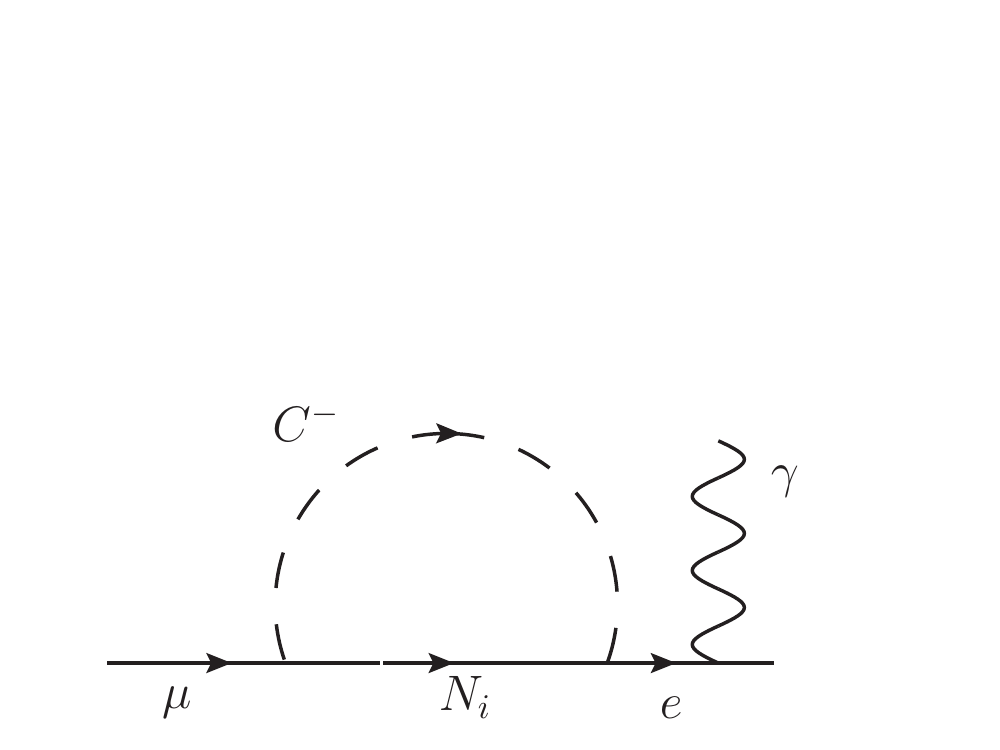}}\vspace{-1.2cm}
 \centerline{\includegraphics[width=4cm]{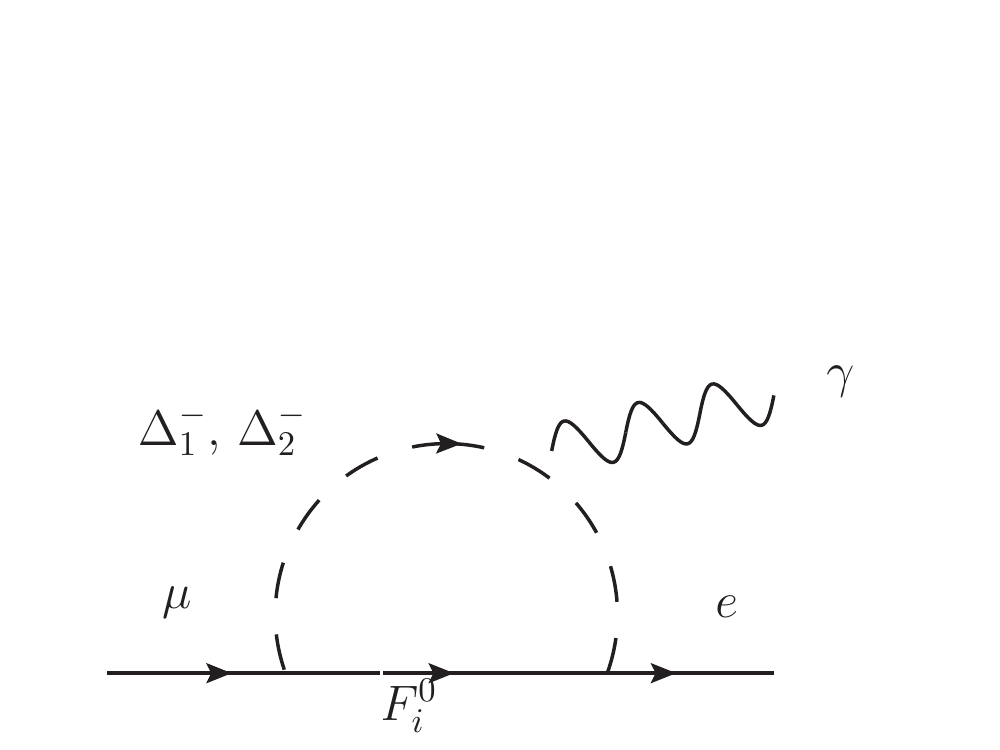}\hspace{0cm}
 \includegraphics[width=4cm]{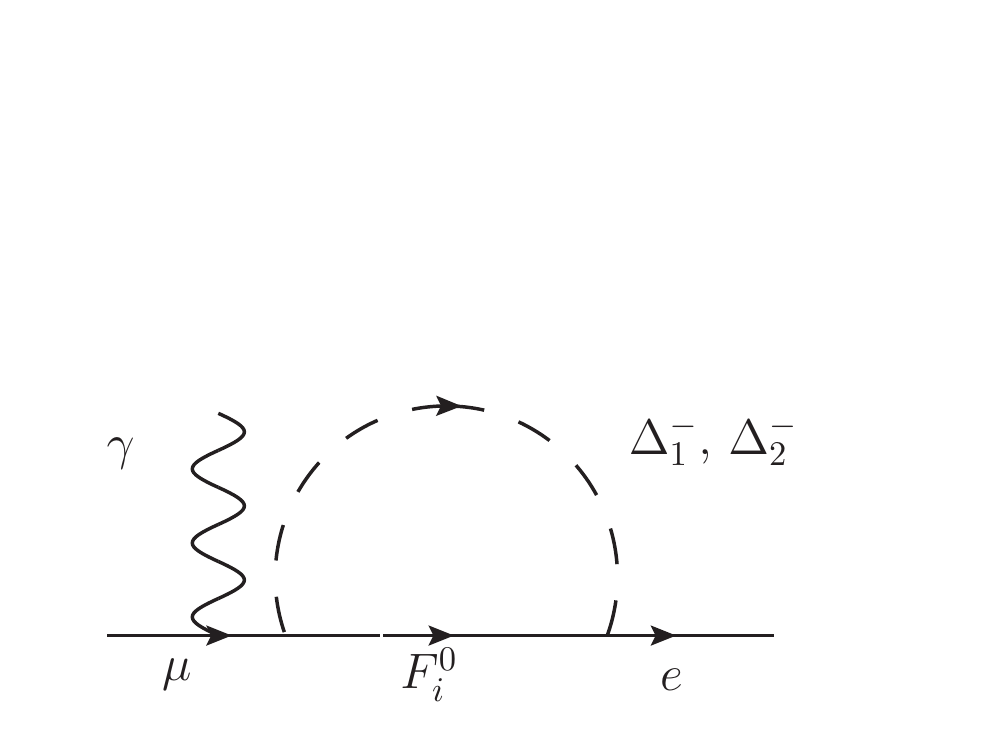}\hspace{0cm}
 \includegraphics[width=4cm]{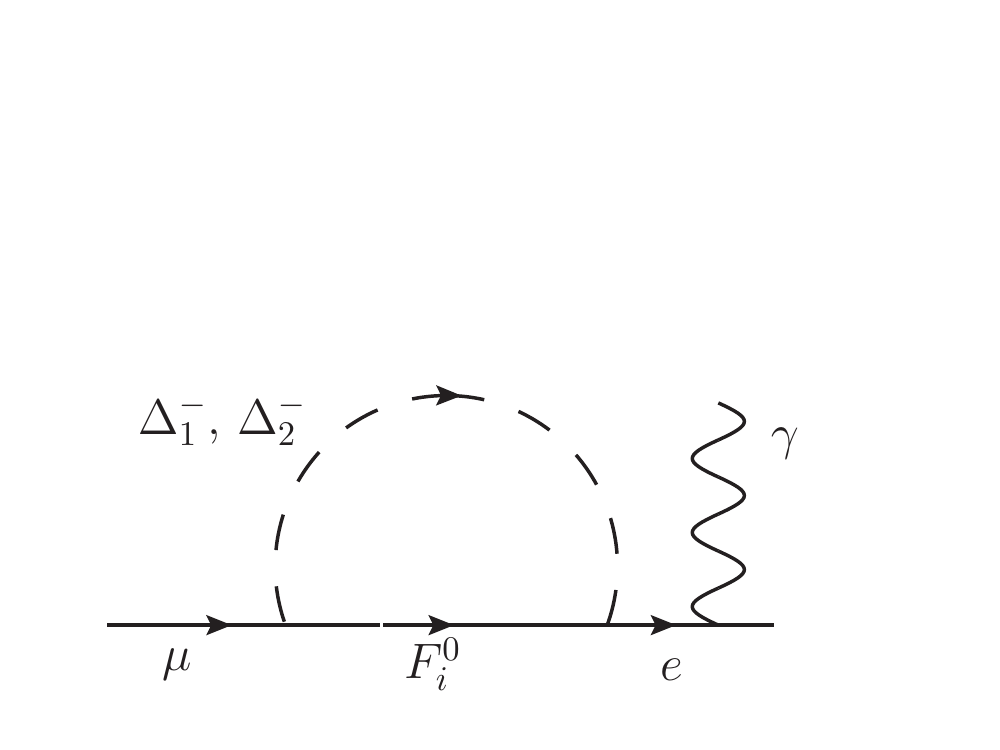}}\vspace{-0.75cm}
 \centerline{\includegraphics[width=4cm]{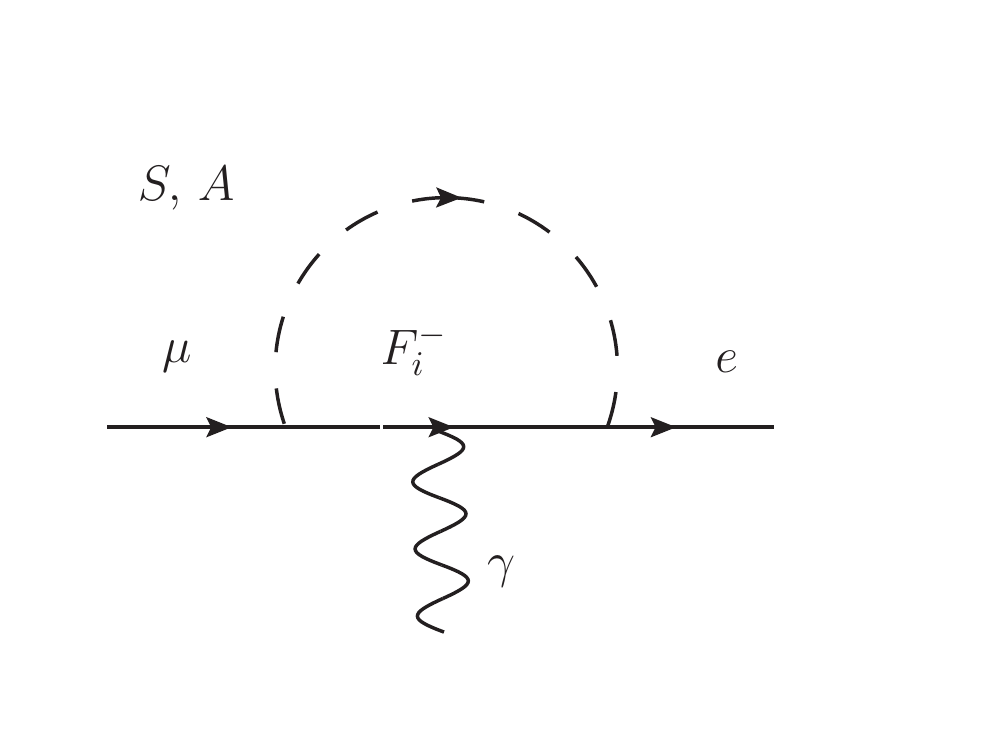}\hspace{0cm}
 \includegraphics[width=4cm]{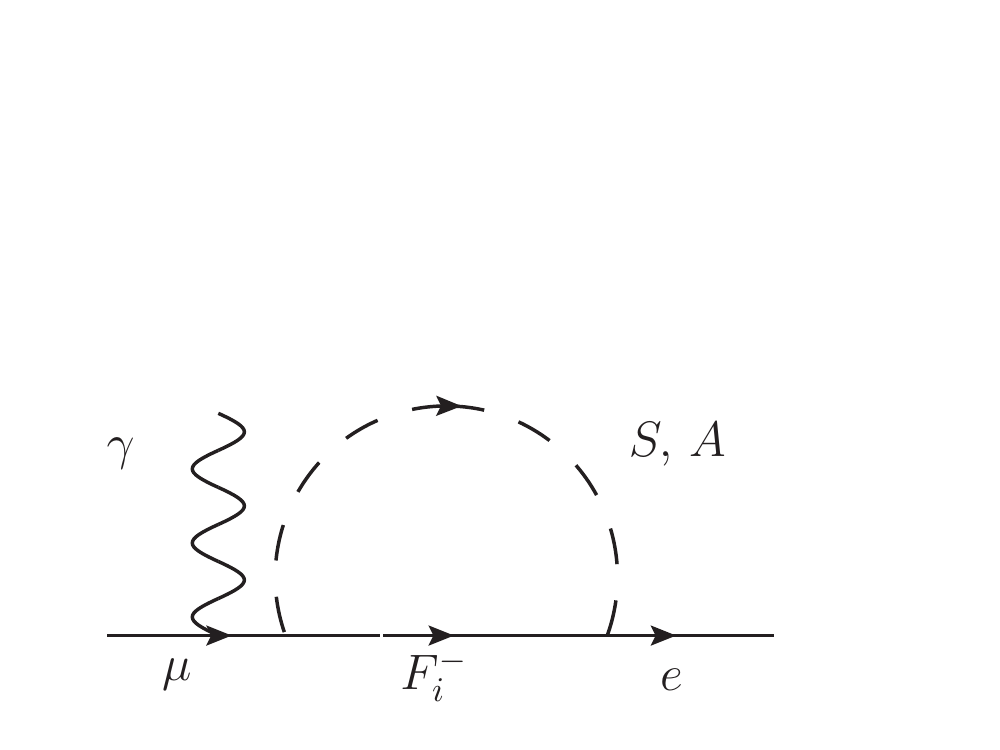}\hspace{0cm}
 \includegraphics[width=4cm]{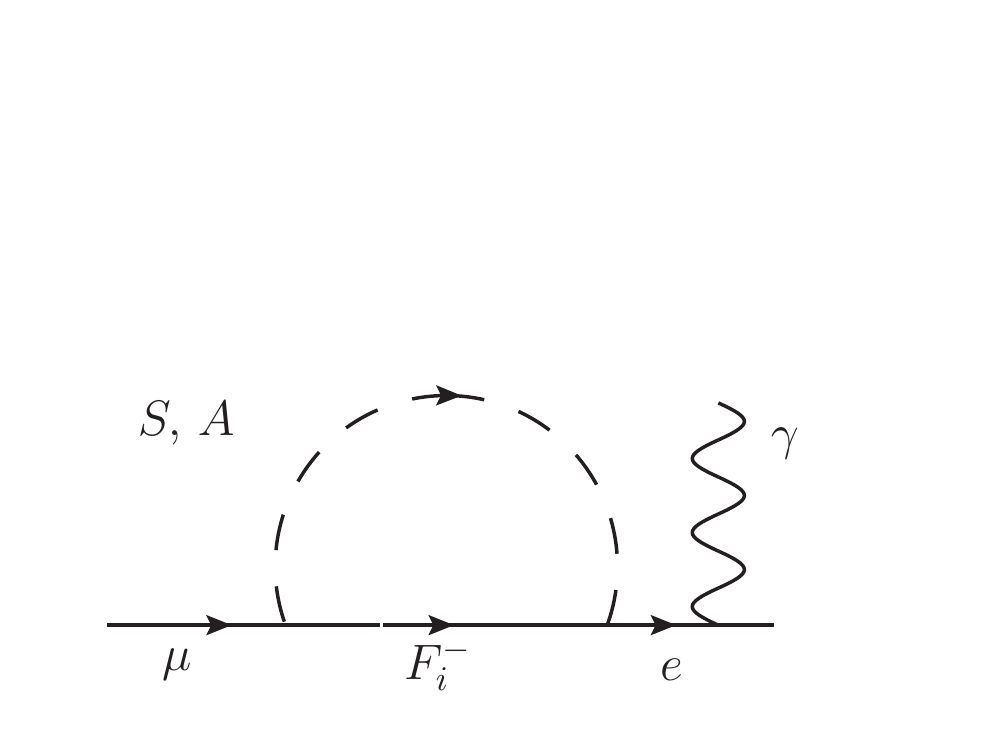}}\vspace{-1.10cm}
 \centerline{\includegraphics[width=4cm]{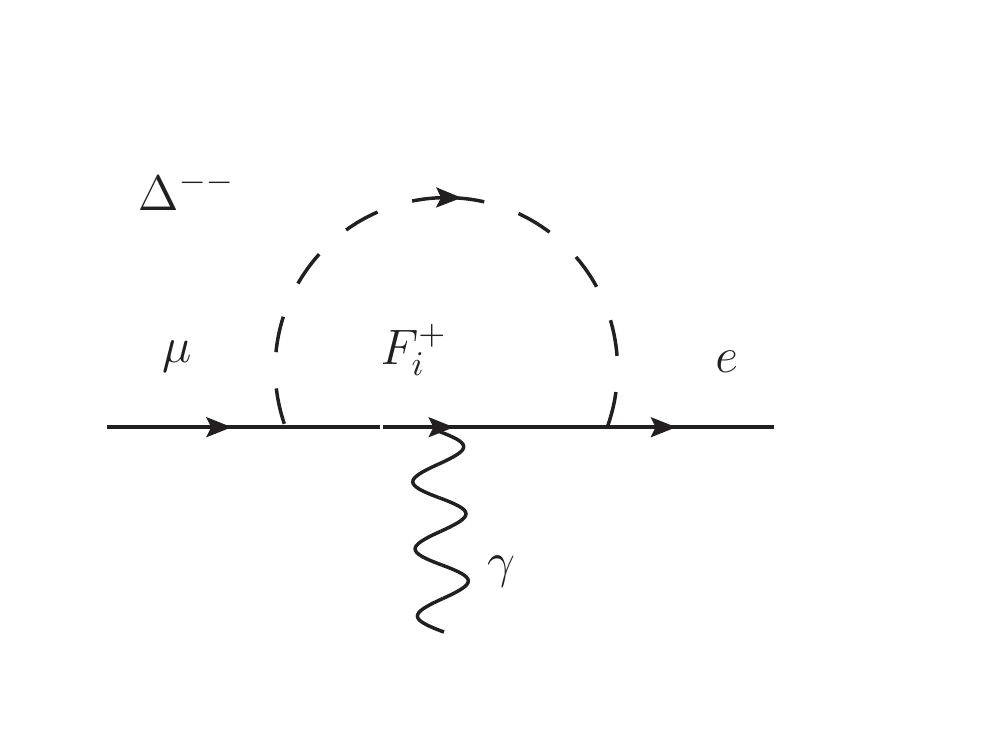}\hspace{0cm}
 \includegraphics[width=4cm]{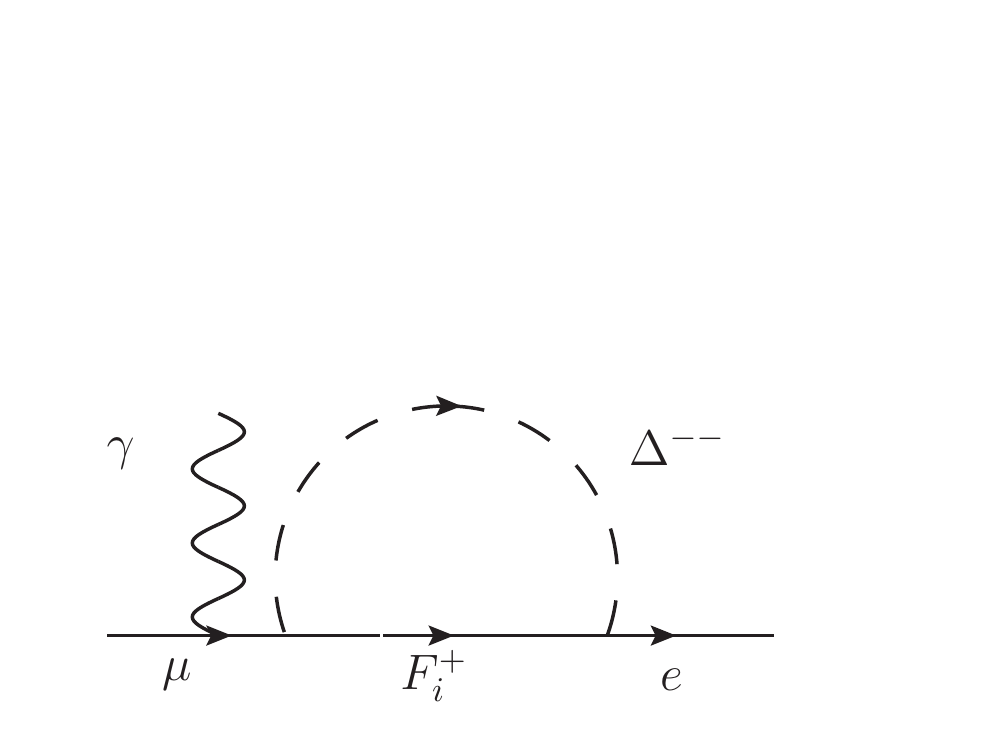}\hspace{0cm}
 \includegraphics[width=4cm]{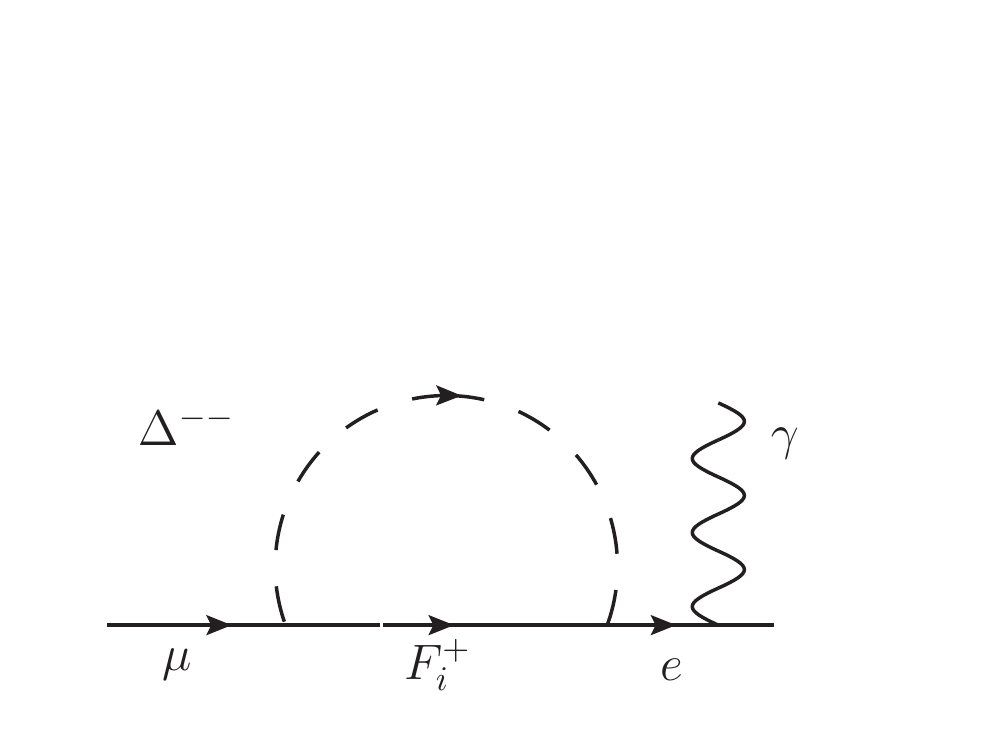}}
 \vspace{-0.5cm}
 \caption{$\mu e\gamma$ vertex and the self energy diagrams of the
 external fermions for the doublet (first row) and the quartet cases
 (second to fourth rows).}
 \label{muegammavertices}
\end{figure}

Here we present in figure \ref{muegammavertices}  the Feynman diagrams of one-loop contributions of the doublet and quartet to
the $\mu e \gamma$ vertex.

\subsection{$\mu e Z$ vertex}\label{muezfeyn}
\begin{figure}[h!]
 \centerline{\includegraphics[width=4cm]{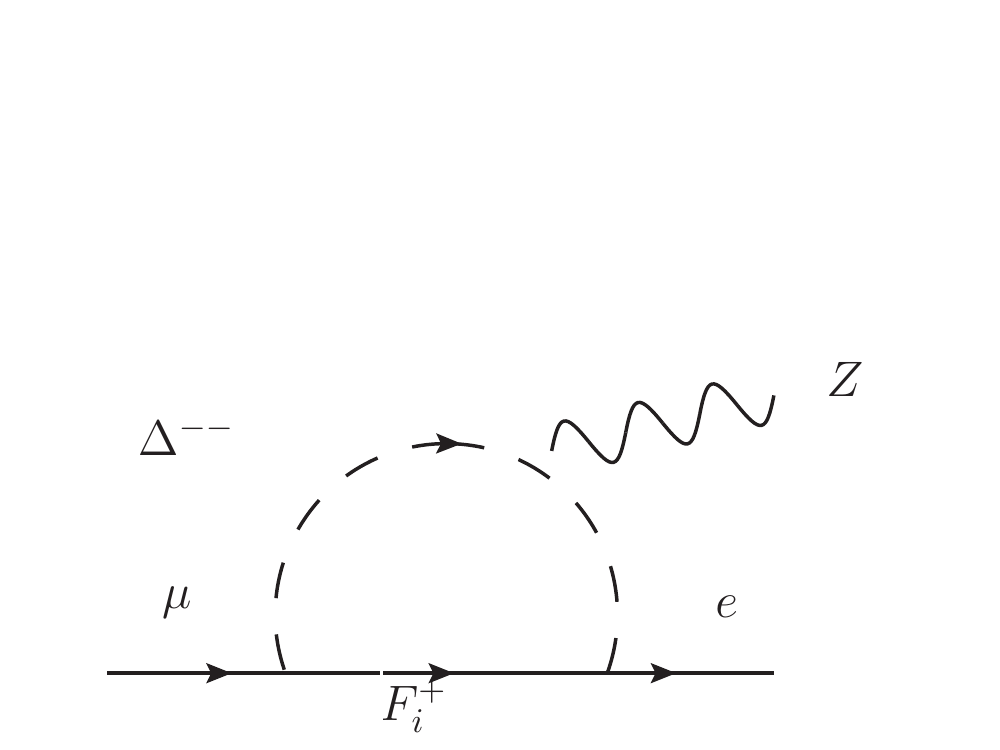}\hspace{-1cm}
 \includegraphics[width=4cm]{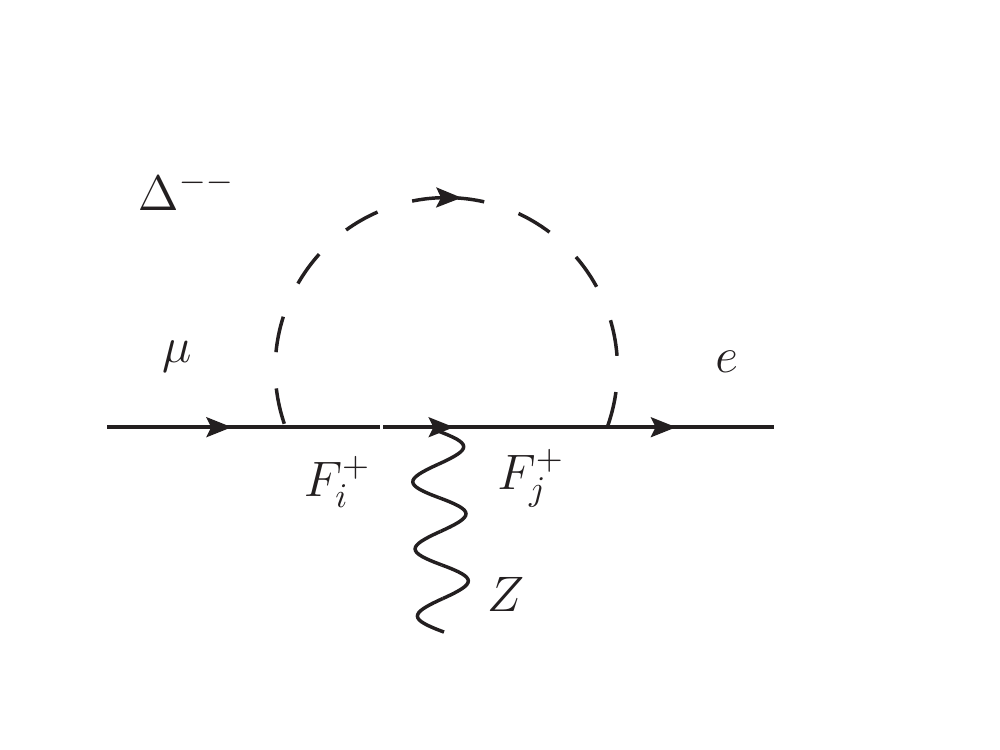}\hspace{-1cm}
 \includegraphics[width=4cm]{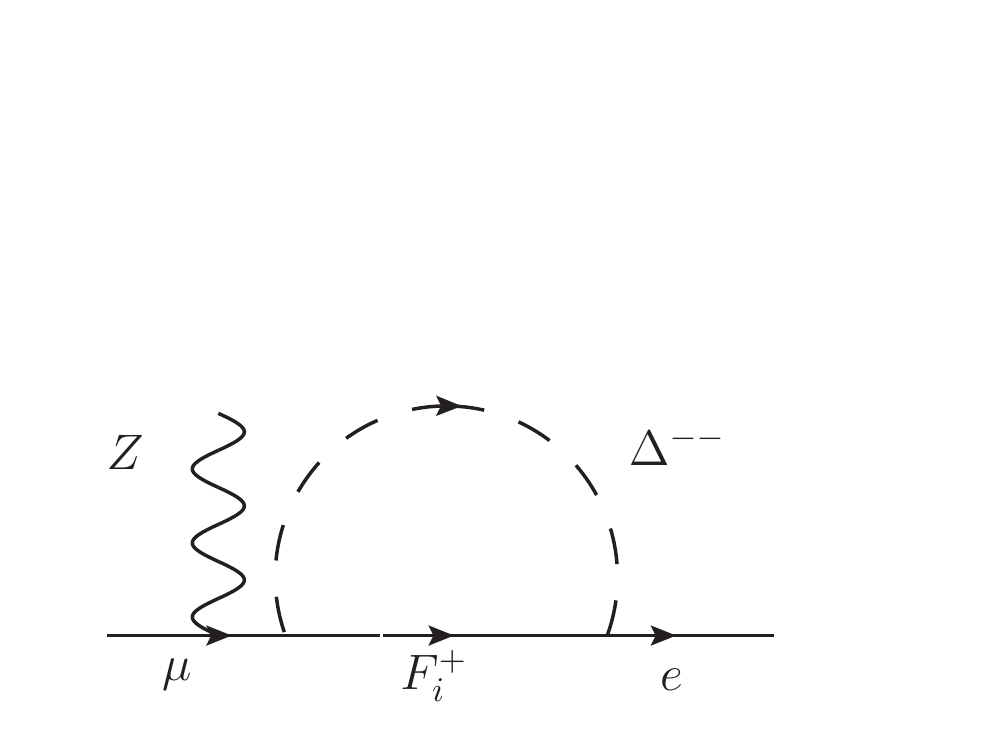}\hspace{-1cm}
 \includegraphics[width=4cm]{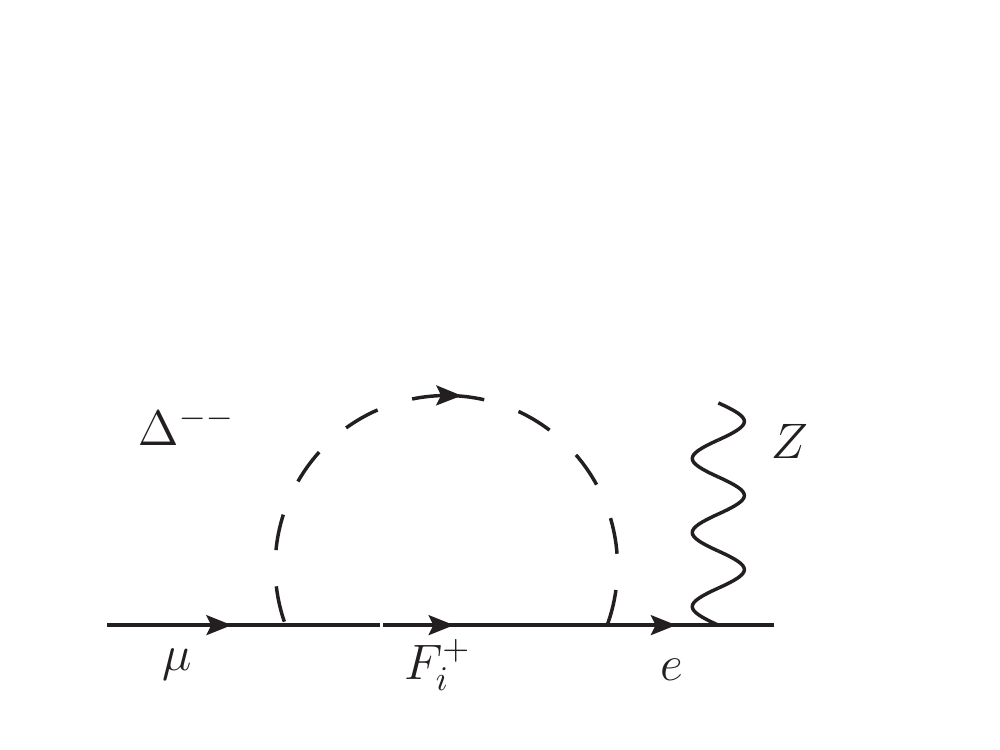}}
 \vspace{-0.25cm}
 \caption{$\mu e Z$ vertex and the self energy diagrams of the external fermions for the 
 doublet (first row) and the quartet cases (second to fourth
 rows).}
 \label{muezvertices}
\end{figure}

We present in figure \ref{muezvertices} the Feynman diagrams of one-loop contributions of the doublet and the quartet to
the $\mu e Z$ vertex.
\begin{figure}[h!]
\vspace{-1cm}
 \centerline{\includegraphics[width=4cm]{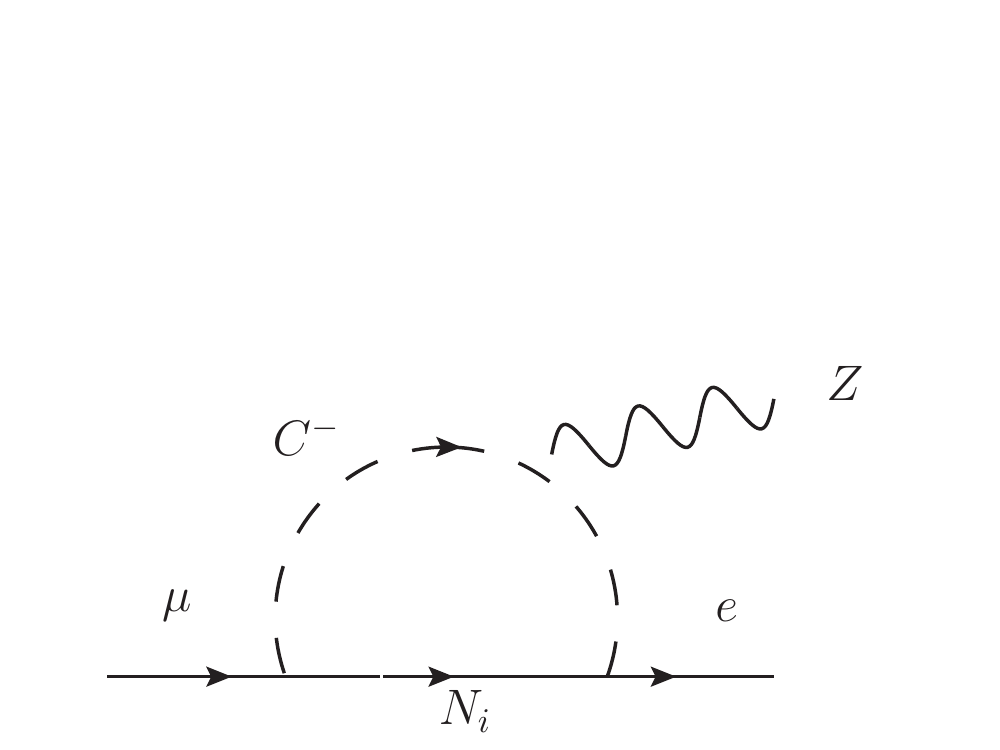}\hspace{0cm}
 \includegraphics[width=4cm]{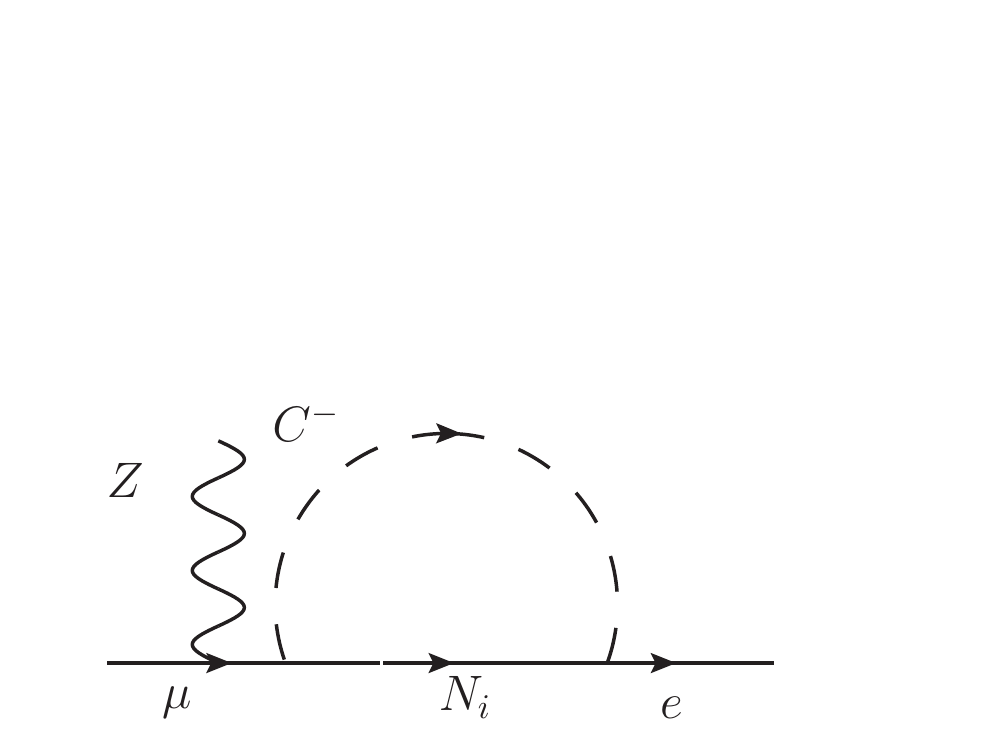}\hspace{0cm}
 \includegraphics[width=4cm]{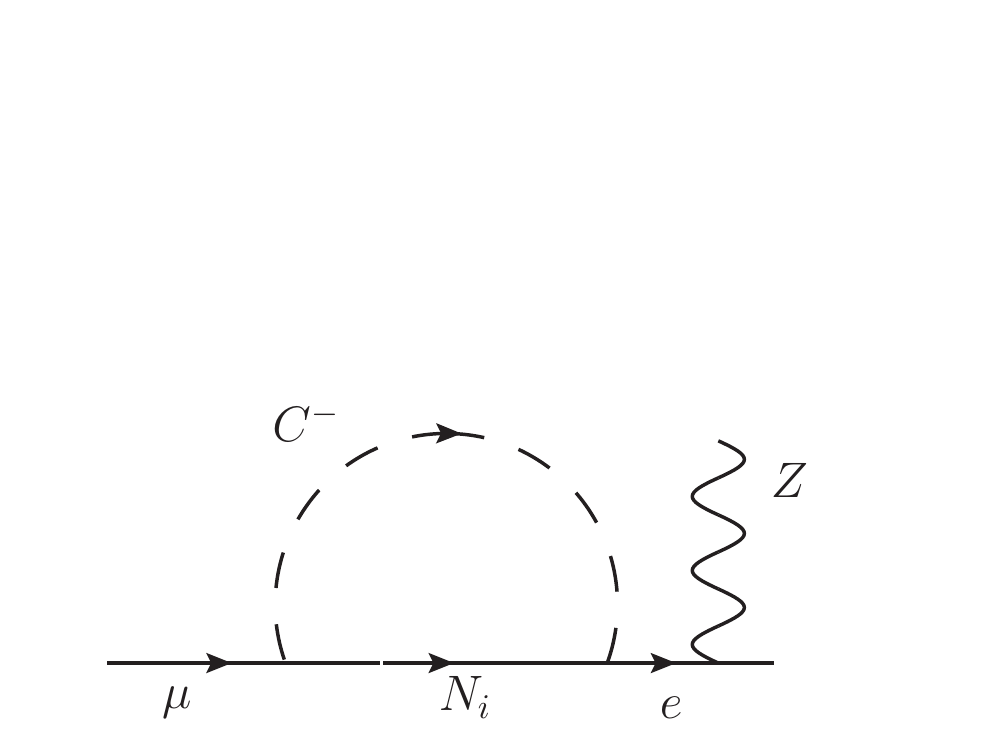}}\vspace{-1cm}
 \centerline{\includegraphics[width=4cm]{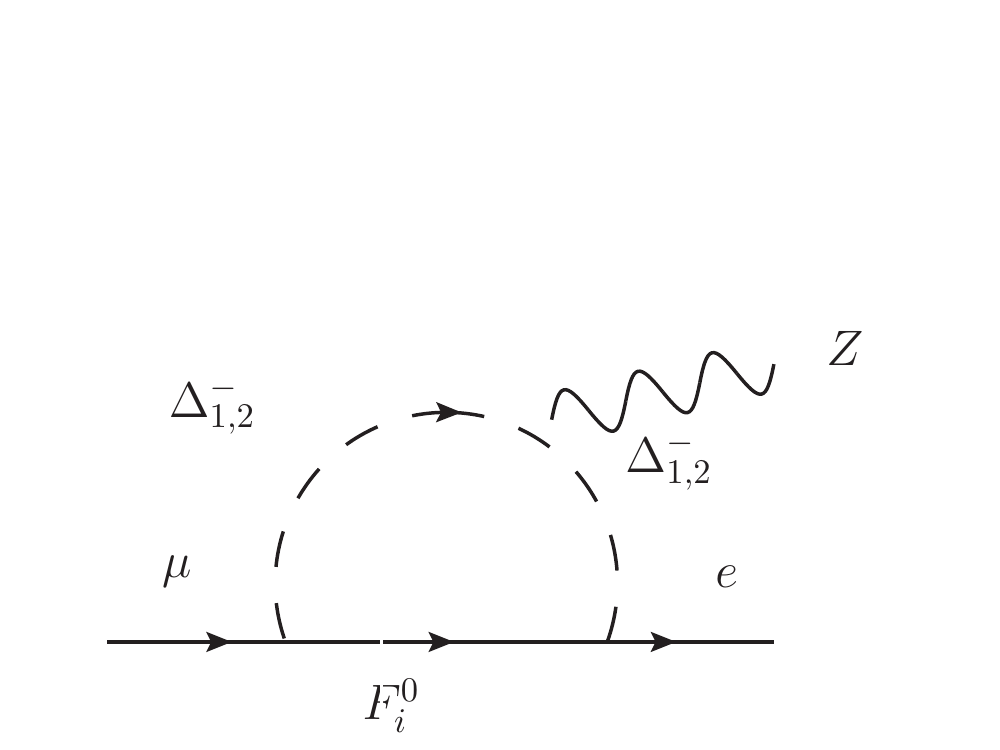}\hspace{0cm}
 \includegraphics[width=4cm]{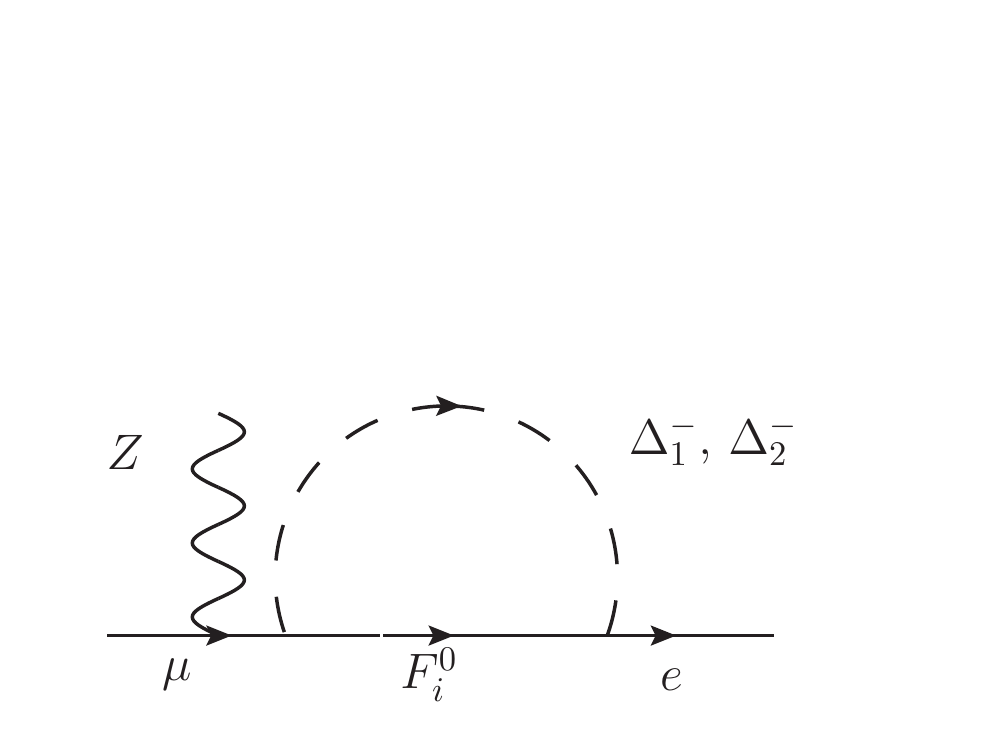}\hspace{0cm}
 \includegraphics[width=4cm]{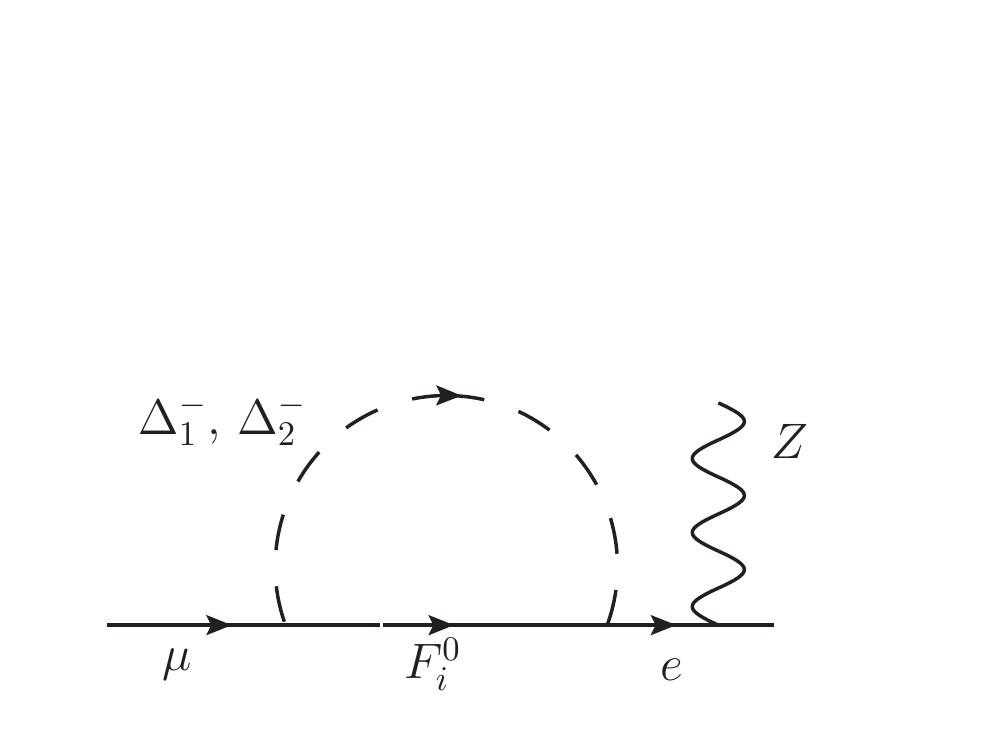}}\vspace{0cm}
 \centerline{\includegraphics[width=4cm]{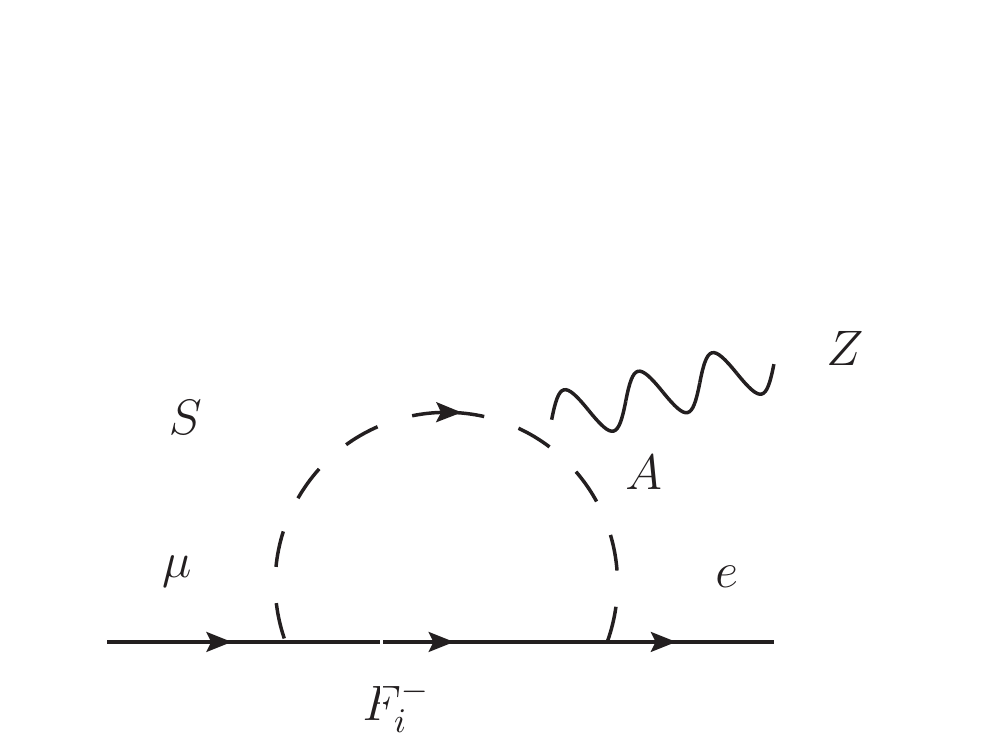}\hspace{-1cm}
 \includegraphics[width=4cm]{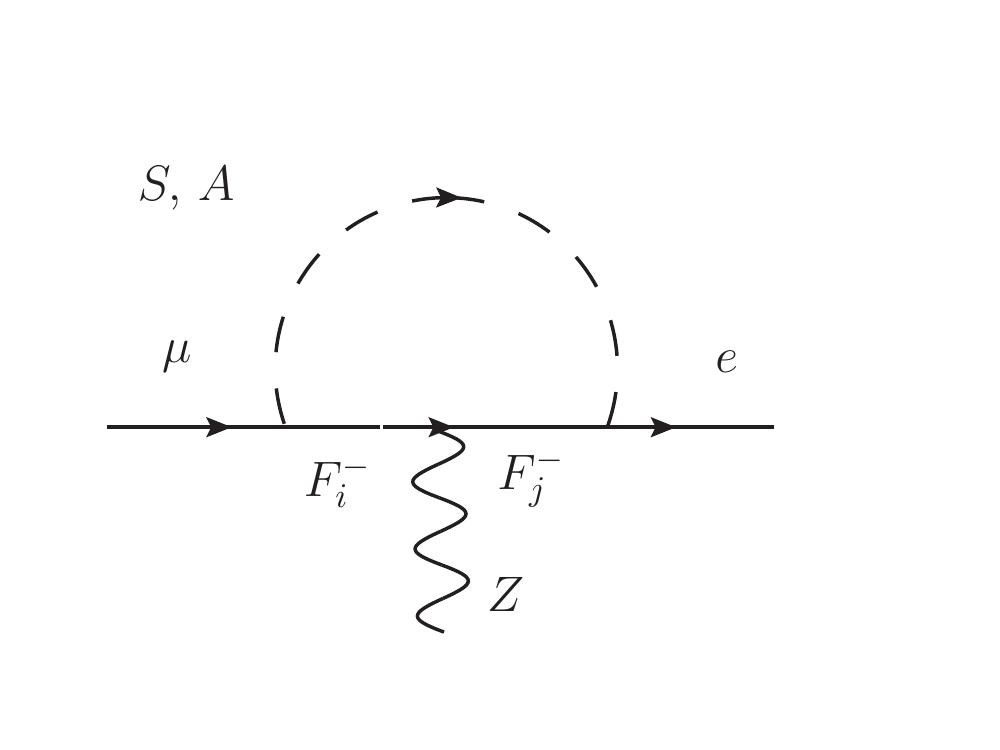}\hspace{-1cm}
 \includegraphics[width=4cm]{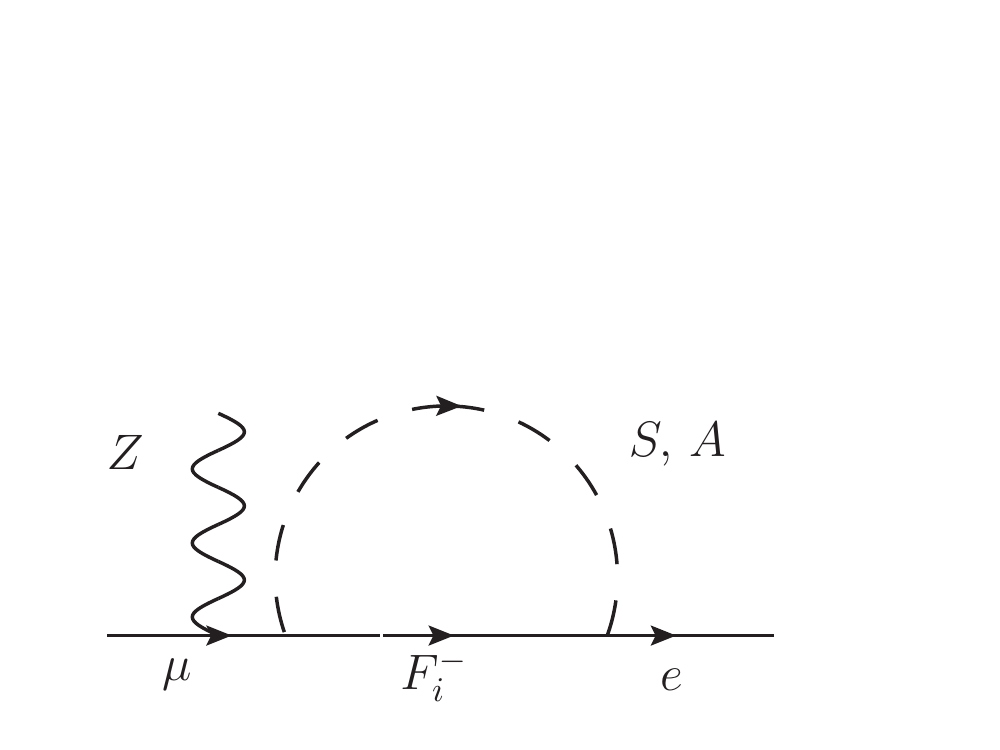}\hspace{-1cm}
 \includegraphics[width=4cm]{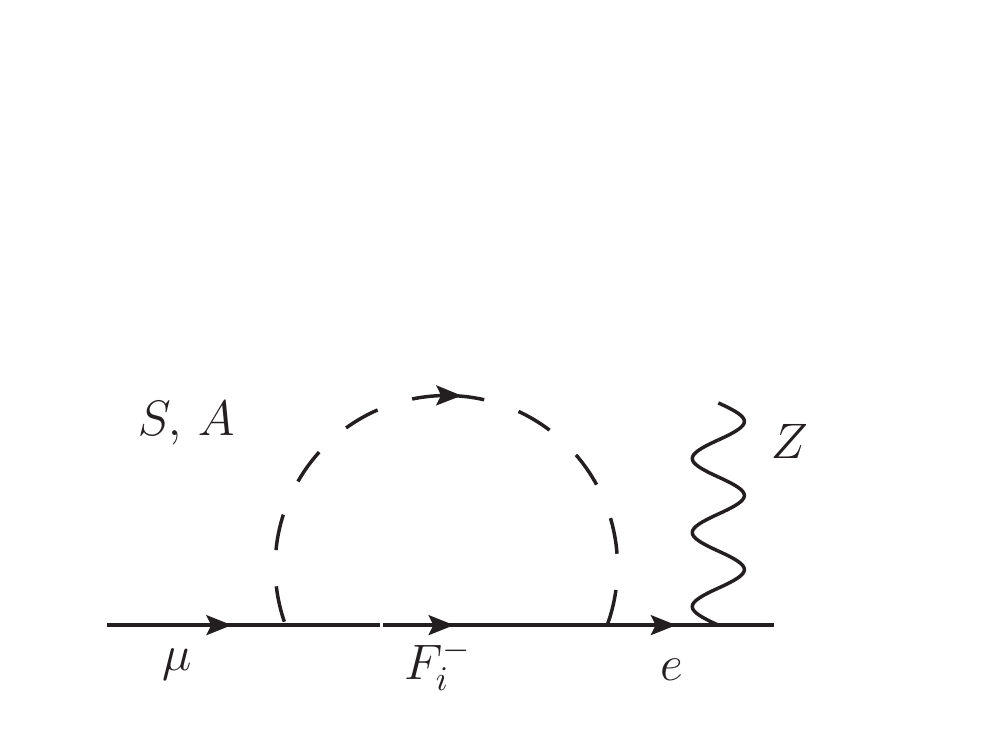}}\vspace{-1cm}
 \end{figure}
\newpage
\subsection{Box diagrams}\label{boxfeyn}
\begin{figure}[h!]
\vspace{-1cm}
 \centerline{
\includegraphics[width=5cm]{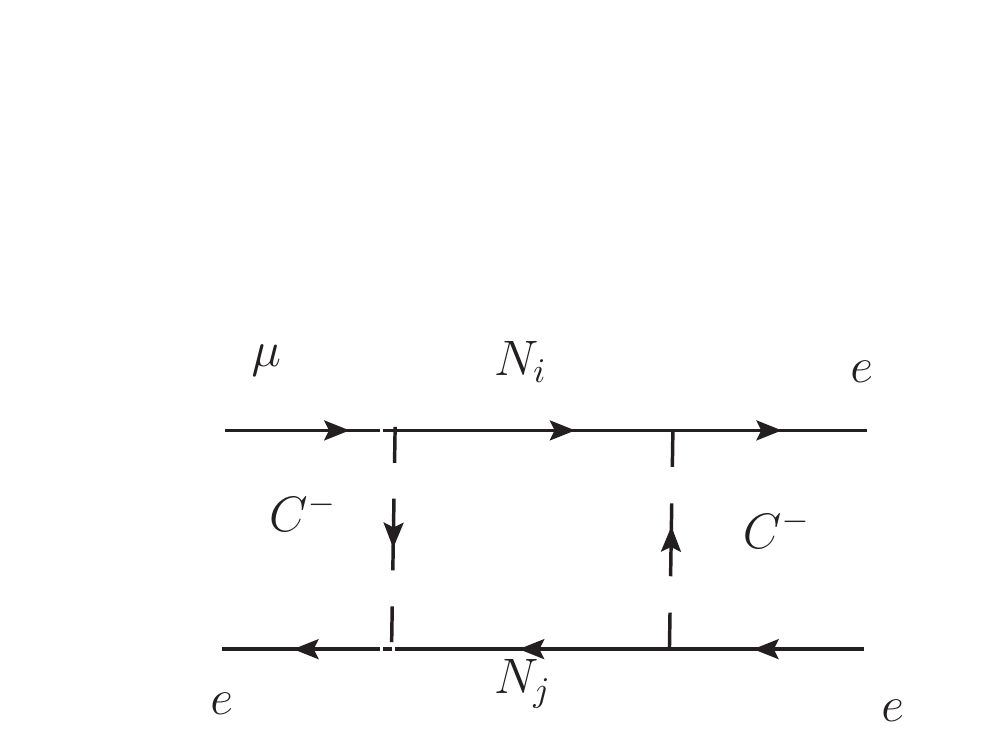}\hspace{0cm}
\includegraphics[width=5cm]{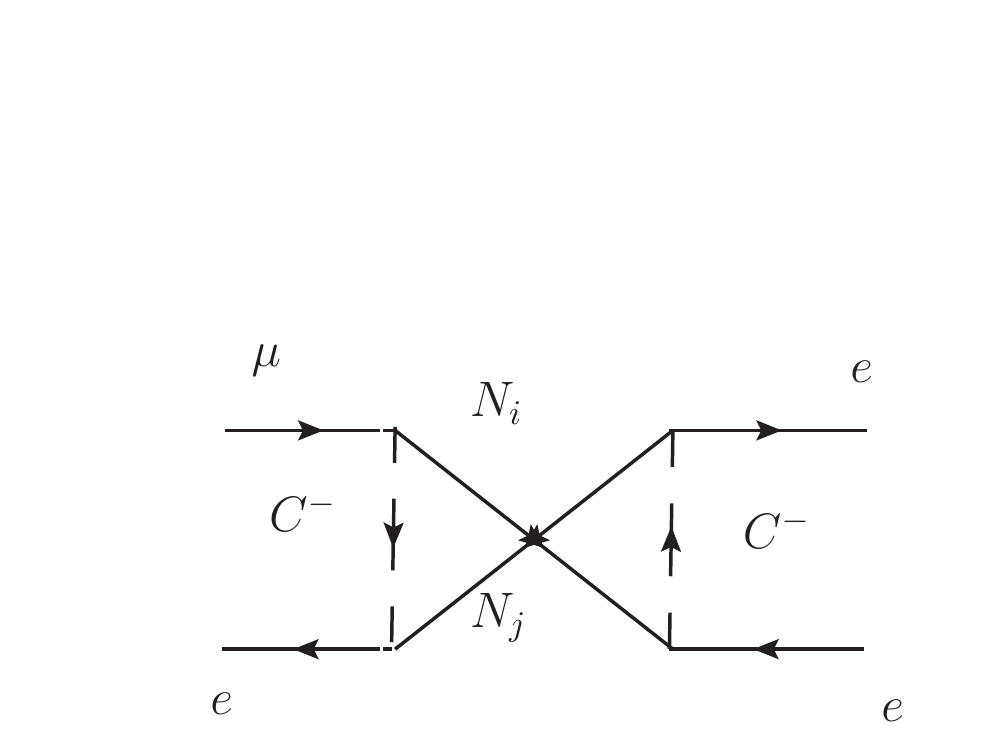}}\vspace{-1.25cm}
\centerline{
\includegraphics[width=5cm]{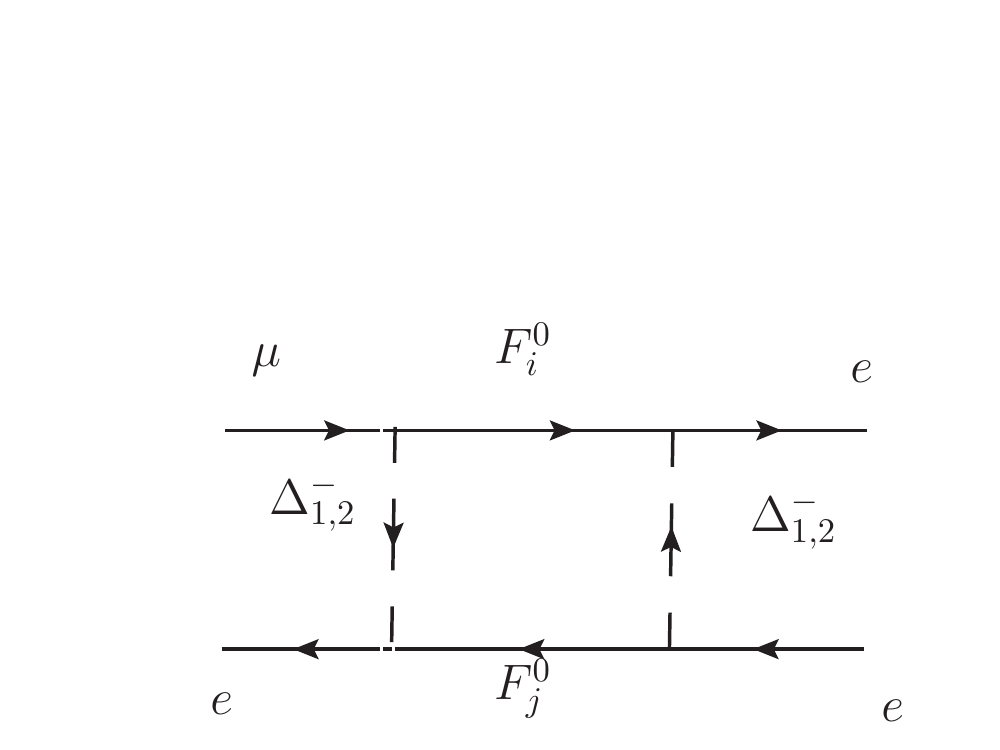}\hspace{0cm}
\includegraphics[width=5cm]{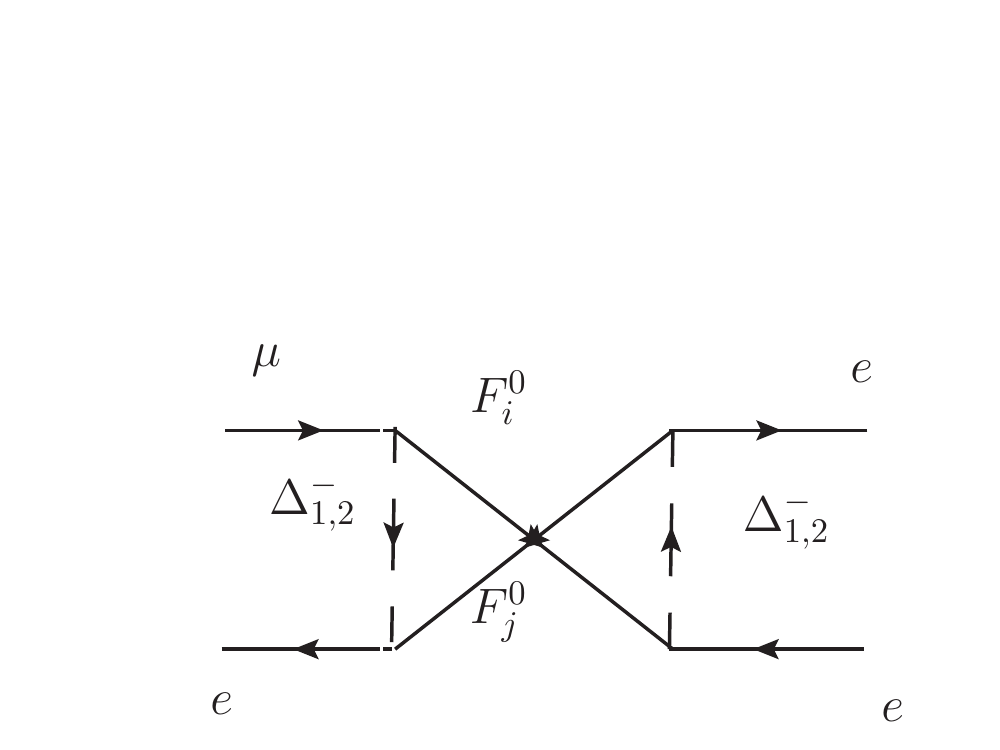}}\vspace{-1.25cm}
\centerline{
\includegraphics[width=5cm]{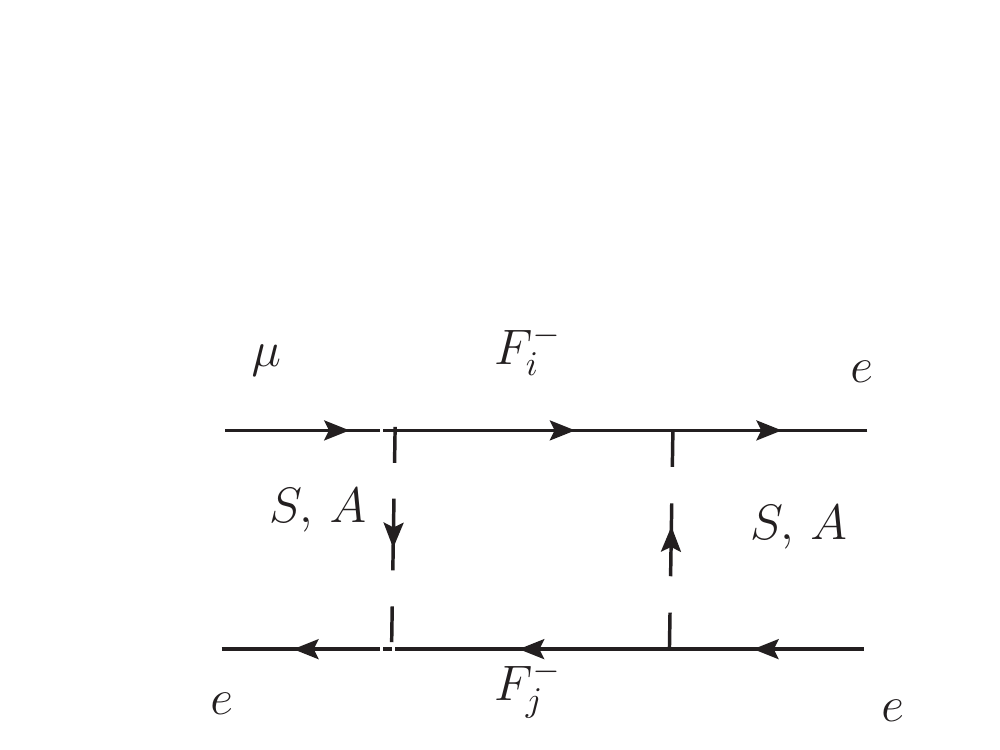}\hspace{0cm}
\includegraphics[width=5cm]{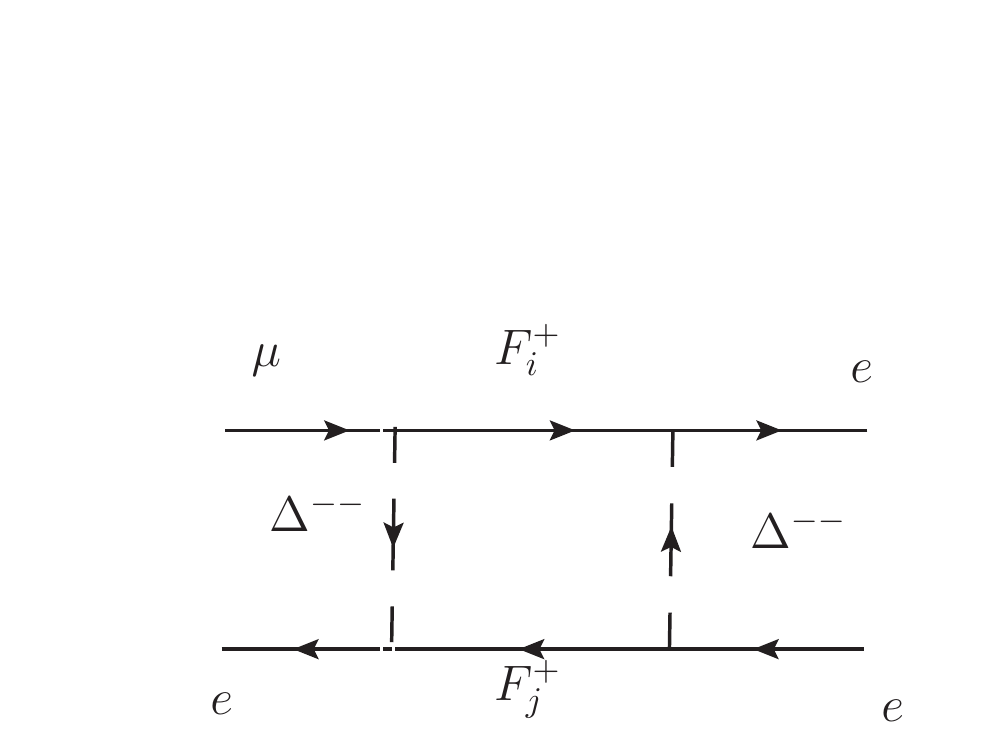}}
\caption{Box diagrams for the doublet (first row) and the quartet (second and third rows).}
\label{boxfigs}\end{figure}

The box diagrams for the doublet and the quartet cases are given in figure \ref{boxfigs},
 \begin{figure}[h!]
\end{figure}


\begin{thebibliography}{99}
 
\bibitem{Ma:2006km} 
  E.~Ma,
  Phys.\ Rev.\ D {\bf 73}, 077301 (2006)
  [hep-ph/0601225].
  
\bibitem{Deshpande:1977rw} 
  N.~G.~Deshpande and E.~Ma,
  Phys.\ Rev.\ D {\bf 18}, 2574 (1978).
  
\bibitem{LopezHonorez:2006gr} 
  L.~Lopez Honorez, E.~Nezri, J.~F.~Oliver and M.~H.~G.~Tytgat,
  JCAP {\bf 0702}, 028 (2007)
  [hep-ph/0612275].
  
\bibitem{Dolle:2009fn} 
  E.~M.~Dolle and S.~Su,
  Phys.\ Rev.\ D {\bf 80}, 055012 (2009)
  [arXiv:0906.1609 [hep-ph]].
  
\bibitem{LopezHonorez:2010tb} 
  L.~Lopez Honorez and C.~E.~Yaguna,
  JCAP {\bf 1101}, 002 (2011)
  [arXiv:1011.1411 [hep-ph]].

\bibitem{Agrawal:2008xz} 
  P.~Agrawal, E.~M.~Dolle and C.~A.~Krenke,
  Phys.\ Rev.\ D {\bf 79}, 015015 (2009)
  [arXiv:0811.1798 [hep-ph]].

\bibitem{Andreas:2009hj} 
  S.~Andreas, M.~H.~G.~Tytgat and Q.~Swillens,
  JCAP {\bf 0904}, 004 (2009)
  [arXiv:0901.1750 [hep-ph]].

\bibitem{Nezri:2009jd} 
  E.~Nezri, M.~H.~G.~Tytgat and G.~Vertongen,
  JCAP {\bf 0904}, 014 (2009)
  [arXiv:0901.2556 [hep-ph]].

\bibitem{Cao:2007rm} 
  Q.~-H.~Cao, E.~Ma and G.~Rajasekaran,
  Phys.\ Rev.\ D {\bf 76}, 095011 (2007)
  [arXiv:0708.2939 [hep-ph]].
  
\bibitem{Martinez:2011ua}
  H.~Martinez, A.~Melfo, F.~Nesti, G.~Senjanovi\'c,
  Phys.\ Rev.\ Lett.\  {\bf 106}, 191802 (2011).
  [arXiv:1101.3796 [hep-ph]]. 

\bibitem{Melfo:2011ie}
  A.~Melfo, M.~Nemev\v sek, F.~Nesti, G.~Senjanovi\' c, Y.~Zhang,
  Phys.\ Rev.\  {\bf D84 } (2011)  034009.
  [arXiv:1105.4611 [hep-ph]].
  	
  
\bibitem{Chowdhury:2011ga} 
  T.~A.~Chowdhury, M.~Nemev\v sek, G.~Senjanovi\'c and Y.~Zhang,
  JCAP {\bf 1202}, 029 (2012)
  [arXiv:1110.5334 [hep-ph]].

\bibitem{Borah:2012pu} 
  D.~Borah and J.~M.~Cline,
  Phys.\ Rev.\ D {\bf 86}, 055001 (2012)
  [arXiv:1204.4722 [hep-ph]].

\bibitem{Gil:2012ya} 
  G.~Gil, P.~Chankowski and M.~Krawczyk,
  Phys.\ Lett.\ B {\bf 717}, 396 (2012)
  [arXiv:1207.0084 [hep-ph]].
	
\bibitem{Cline:2013bln} 
  J.~M.~Cline and K.~Kainulainen,
  Phys.\ Rev.\ D {\bf 87}, 071701 (2013)
  [arXiv:1302.2614 [hep-ph]].
  
\bibitem{Ahriche:2015mea}
  A.~Ahriche, G.~Faisel, S.~Y.~Ho, S.~Nasri and J.~Tandean,
  Phys.\ Rev.\ D {\bf 92} (2015) 3,  035020
  [arXiv:1501.06605 [hep-ph]].
  
\bibitem{Lundstrom:2008ai} 
  E.~Lundstrom, M.~Gustafsson and J.~Edsjo,
  Phys.\ Rev.\ D {\bf 79}, 035013 (2009)
  [arXiv:0810.3924 [hep-ph]].

\bibitem{Dolle:2009ft} 
  E.~Dolle, X.~Miao, S.~Su and B.~Thomas,
  Phys.\ Rev.\ D {\bf 81}, 035003 (2010)
  [arXiv:0909.3094 [hep-ph]].
  
\bibitem{Gustafsson:2012aj} 
  M.~Gustafsson, S.~Rydbeck, L.~Lopez-Honorez and E.~Lundstrom,
  Phys.\ Rev.\ D {\bf 86}, 075019 (2012)
  [arXiv:1206.6316 [hep-ph]].
  
\bibitem{Aoki:2013lhm} 
  M.~Aoki, S.~Kanemura and H.~Yokoya,
  Phys.\ Lett.\ B {\bf 725}, 302 (2013)
  [arXiv:1303.6191 [hep-ph]].
  
\bibitem{Belanger:2015kga}
  G.~Belanger, B.~Dumont, A.~Goudelis, B.~Herrmann, S.~Kraml and D.~Sengupta,
  arXiv:1503.07367 [hep-ph].
  

\bibitem{AbdusSalam:2013eya} 
  S.~S.~AbdusSalam and T.~A.~Chowdhury,
  JCAP {\bf 1405}, 026 (2014)
  [arXiv:1310.8152 [hep-ph]].
 
  
\bibitem{Kubo:2006yx}
  J.~Kubo, E.~Ma and D.~Suematsu,
  Phys.\ Lett.\ B {\bf 642} (2006) 18
  [hep-ph/0604114].
  
\bibitem{Sierra:2008wj}
  D.~Aristizabal Sierra, J.~Kubo, D.~Restrepo, D.~Suematsu and O.~Zapata,
  Phys.\ Rev.\ D {\bf 79} (2009) 013011
  [arXiv:0808.3340 [hep-ph]].
  
\bibitem{Suematsu:2009ww}
  D.~Suematsu, T.~Toma and T.~Yoshida,
  Phys.\ Rev.\ D {\bf 79} (2009) 093004
  [arXiv:0903.0287 [hep-ph]].
  
\bibitem{Adulpravitchai:2009gi}
  A.~Adulpravitchai, M.~Lindner and A.~Merle,
  Phys.\ Rev.\ D {\bf 80} (2009) 055031
  [arXiv:0907.2147 [hep-ph]].
 
\bibitem{Toma:2013zsa}
  T.~Toma and A.~Vicente,
  JHEP {\bf 1401} (2014) 160
  [arXiv:1312.2840, arXiv:1312.2840 [hep-ph]].
  
\bibitem{Vicente:2014wga}
  A.~Vicente and C.~E.~Yaguna,
  JHEP {\bf 1502} (2015) 144
  [arXiv:1412.2545 [hep-ph]].
  
\bibitem{Ma:2008cu}
  E.~Ma and D.~Suematsu,
  Mod.\ Phys.\ Lett.\ A {\bf 24} (2009) 583
  [arXiv:0809.0942 [hep-ph]].
  
\bibitem{Law:2013saa}
  S.~S.~C.~Law and K.~L.~McDonald,
  JHEP {\bf 1309} (2013) 092
  [arXiv:1305.6467 [hep-ph]].
  
\bibitem{Ren:2011mh}
  B.~Ren, K.~Tsumura and X.~G.~He,
  Phys.\ Rev.\ D {\bf 84} (2011) 073004
  [arXiv:1107.5879 [hep-ph]].
  
  
\bibitem{Ahriche:2015wha}
  A.~Ahriche, K.~L.~McDonald, S.~Nasri and T.~Toma,
  Phys.\ Lett.\ B {\bf 746} (2015) 430
  [arXiv:1504.05755 [hep-ph]].
  
  
\bibitem{Adam:2011ch}
  J.~Adam {\it et al.}  [MEG Collaboration],
  Phys.\ Rev.\ Lett.\  {\bf 107} (2011) 171801
  [arXiv:1107.5547 [hep-ex]].
  
\bibitem{Adam:2013mnn}
  J.~Adam {\it et al.}  [MEG Collaboration],
  Phys.\ Rev.\ Lett.\  {\bf 110} (2013) 201801
  [arXiv:1303.0754 [hep-ex]].
  
\bibitem{Baldini:2013ke}
  A.~M.~Baldini, F.~Cei, C.~Cerri, S.~Dussoni, L.~Galli, M.~Grassi, D.~Nicolo and F.~Raffaelli {\it et al.},
  arXiv:1301.7225 [physics.ins-det].
  
\bibitem{Bellgardt:1987du}
  U.~Bellgardt {\it et al.}  [SINDRUM Collaboration],
  Nucl.\ Phys.\ B {\bf 299} (1988) 1.
  
\bibitem{Blondel:2013ia}
  A.~Blondel, A.~Bravar, M.~Pohl, S.~Bachmann, N.~Berger, M.~Kiehn, A.~Schoning and D.~Wiedner {\it et al.},
  arXiv:1301.6113 [physics.ins-det].
  
\bibitem{Bertl:2006up}
  W.~H.~Bertl {\it et al.}  [SINDRUM II Collaboration],
  Eur.\ Phys.\ J.\ C {\bf 47} (2006) 337.
  
\bibitem{Dohmen:1993mp}
  C.~Dohmen {\it et al.}  [SINDRUM II Collaboration],
  Phys.\ Lett.\ B {\bf 317} (1993) 631.
  
\bibitem{Glenzinski:2010zz}
  D.~Glenzinski [Mu2e Collaboration],
  AIP Conf.\ Proc.\  {\bf 1222} (2010) 383.
  
\bibitem{Bartoszek:2014mya}
  L.~Bartoszek {\it et al.}  [Mu2e Collaboration],
  arXiv:1501.05241 [physics.ins-det].
  
\bibitem{Natori:2014yba}
  H.~Natori [DeeMe Collaboration],
  Nucl.\ Phys.\ Proc.\ Suppl.\  {\bf 248-250} (2014) 52.
  
\bibitem{Kuno:2013mha}
  Y.~Kuno [COMET Collaboration],
  PTEP {\bf 2013} (2013) 022C01.
  
\bibitem{Kuno:2005mm}
  Y.~Kuno,
  Nucl.\ Phys.\ Proc.\ Suppl.\  {\bf 149} (2005) 376.
  
\bibitem{Barlow:2011zza}
  R.~J.~Barlow,
  Nucl.\ Phys.\ Proc.\ Suppl.\  {\bf 218} (2011) 44.
  
\bibitem{Cirelli:2005uq}
  M.~Cirelli, N.~Fornengo and A.~Strumia,
  Nucl.\ Phys.\ B {\bf 753} (2006) 178
  [hep-ph/0512090].
  
\bibitem{Cirelli:2007xd} 
  M.~Cirelli, A.~Strumia and M.~Tamburini,
  Nucl.\ Phys.\ B {\bf 787}, 152 (2007)
  [arXiv:0706.4071 [hep-ph]].
  
\bibitem{Cirelli:2009uv} 
  M.~Cirelli and A.~Strumia,
  New J.\ Phys.\  {\bf 11}, 105005 (2009)
  [arXiv:0903.3381 [hep-ph]].
  
\bibitem{Casas:2001sr}
  J.~A.~Casas and A.~Ibarra,
  Nucl.\ Phys.\ B {\bf 618} (2001) 171
  [hep-ph/0103065].
  
\bibitem{Casas:2010wm}
  J.~A.~Casas, J.~M.~Moreno, N.~Rius, R.~Ruiz de Austri and B.~Zaldivar,
  JHEP {\bf 1103} (2011) 034
  [arXiv:1010.5751 [hep-ph]].
  
\bibitem{Heeck:2012fw} 
  J.~Heeck,
  Phys.\ Rev.\ D {\bf 86}, 093023 (2012)
  [arXiv:1207.5521 [hep-ph]].
  
  
\bibitem{Cheng:1976uq}
  T.~P.~Cheng and L.~F.~Li,
  Phys.\ Rev.\ Lett.\  {\bf 38} (1977) 381.
  
\bibitem{Cheng:1977nv}
  T.~P.~Cheng and L.~F.~Li,
  Phys.\ Rev.\ D {\bf 16} (1977) 1425.
  
\bibitem{Cheng:1980tp}
  T.~P.~Cheng and L.~F.~Li,
  Phys.\ Rev.\ Lett.\  {\bf 45} (1980) 1908.
  
\bibitem{Ma:1980gm}
  E.~Ma and A.~Pramudita,
  Phys.\ Rev.\ D {\bf 24} (1981) 1410.
  
\bibitem{Lim:1981kv}
  C.~S.~Lim and T.~Inami,
  Prog.\ Theor.\ Phys.\  {\bf 67} (1982) 1569.
  
\bibitem{Ilakovac:1994kj}
  A.~Ilakovac and A.~Pilaftsis,
  Nucl.\ Phys.\ B {\bf 437} (1995) 491
  [hep-ph/9403398].
  
\bibitem{Blum:2007he}
  A.~Blum and A.~Merle,
  Phys.\ Rev.\ D {\bf 77} (2008) 076005
  [arXiv:0709.3294 [hep-ph]].
  
\bibitem{Hisano:1995cp}
  J.~Hisano, T.~Moroi, K.~Tobe and M.~Yamaguchi,
  Phys.\ Rev.\ D {\bf 53} (1996) 2442
  [hep-ph/9510309].
  
\bibitem{Arganda:2005ji}
  E.~Arganda and M.~J.~Herrero,
  Phys.\ Rev.\ D {\bf 73} (2006) 055003
  [hep-ph/0510405].
  
\bibitem{Krauss:2013gya} 
  M.~E.~Krauss, W.~Porod, F.~Staub, A.~Abada, A.~Vicente and C.~Weiland,
  Phys.\ Rev.\ D {\bf 90}, no. 1, 013008 (2014)
  [arXiv:1312.5318 [hep-ph]].
    
\bibitem{Abada:2014kba}
  A.~Abada, M.~E.~Krauss, W.~Porod, F.~Staub, A.~Vicente and C.~Weiland,
  JHEP {\bf 1411} (2014) 048
  [arXiv:1408.0138 [hep-ph]].
  
\bibitem{Arganda:2014lya}
  E.~Arganda and M.~J.~Herrero,
  arXiv:1403.6161 [hep-ph].
  
\bibitem{Kitano:2002mt}
  R.~Kitano, M.~Koike and Y.~Okada,
  Phys.\ Rev.\ D {\bf 66} (2002) 096002
   [Phys.\ Rev.\ D {\bf 76} (2007) 059902]
  [hep-ph/0203110].
  
\bibitem{Arganda:2007jw}
  E.~Arganda, M.~J.~Herrero and A.~M.~Teixeira,
  JHEP {\bf 0710} (2007) 104
  [arXiv:0707.2955 [hep-ph]].
  
\bibitem{Crivellin:2014cta}
  A.~Crivellin, M.~Hoferichter and M.~Procura,
  Phys.\ Rev.\ D {\bf 89} (2014) 093024
  [arXiv:1404.7134 [hep-ph]].
  
\bibitem{Beringer:1900zz} 
  J.~Beringer {\it et al.}  [Particle Data Group Collaboration],
  Phys.\ Rev.\ D {\bf 86}, 010001 (2012).
  
\bibitem{Aad:2013yna}
  G.~Aad {\it et al.} [ATLAS Collaboration],
  Phys.\ Rev.\ D {\bf 88} (2013) 11,  112006
  [arXiv:1310.3675 [hep-ex]].
  
\bibitem{Cirelli:2014dsa}
  M.~Cirelli, F.~Sala and M.~Taoso,
  JHEP {\bf 1410} (2014) 033
   [JHEP {\bf 1501} (2015) 041]
  [arXiv:1407.7058 [hep-ph]].
  
\bibitem{Ade:2013zuv} 
  P.~A.~R.~Ade {\it et al.}  [Planck Collaboration],
  Astron.\ Astrophys.\  {\bf 571}, A16 (2014)
  [arXiv:1303.5076 [astro-ph.CO]].
  
\bibitem{Hambye:2009pw}
  T.~Hambye, F.-S.~Ling, L.~Lopez Honorez and J.~Rocher,
  JHEP {\bf 0907} (2009) 090
   [JHEP {\bf 1005} (2010) 066]
  [arXiv:0903.4010 [hep-ph]].
  
\bibitem{Alloul:2013bka}
  A.~Alloul, N.~D.~Christensen, C.~Degrande, C.~Duhr and B.~Fuks,
  Comput.\ Phys.\ Commun.\  {\bf 185} (2014) 2250
  [arXiv:1310.1921 [hep-ph]].
  
\bibitem{Belanger:2013oya}
  G.~Belanger, F.~Boudjema, A.~Pukhov and A.~Semenov,
  Comput.\ Phys.\ Commun.\  {\bf 185} (2014) 960
  [arXiv:1305.0237 [hep-ph]].
  
\bibitem{Akerib:2013tjd}
  D.~S.~Akerib {\it et al.} [LUX Collaboration],
  Phys.\ Rev.\ Lett.\  {\bf 112} (2014) 091303
  [arXiv:1310.8214 [astro-ph.CO]].
  
\bibitem{sommerfeldref}
  A.~Sommerfeld,
  Ann.\ Phys.\ {\bf 11}, 257 (1931)
  
\bibitem{Hisano:2003ec} 
  J.~Hisano, S.~Matsumoto and M.~M.~Nojiri,
  Phys.\ Rev.\ Lett.\  {\bf 92}, 031303 (2004)
  [hep-ph/0307216].
  
\bibitem{Hisano:2004ds} 
  J.~Hisano, S.~Matsumoto, M.~M.~Nojiri and O.~Saito,
  Phys.\ Rev.\ D {\bf 71}, 063528 (2005)
  [hep-ph/0412403].
  
\bibitem{Hisano:2005ec} 
  J.~Hisano, S.~Matsumoto, O.~Saito and M.~Senami,
  Phys.\ Rev.\ D {\bf 73}, 055004 (2006)
  [hep-ph/0511118].
  
\bibitem{ArkaniHamed:2008qn}
  N.~Arkani-Hamed, D.~P.~Finkbeiner, T.~R.~Slatyer and N.~Weiner,
  Phys.\ Rev.\ D {\bf 79} (2009) 015014
  [arXiv:0810.0713 [hep-ph]].
  
\bibitem{Slatyer:2009vg}
  T.~R.~Slatyer,
  JCAP {\bf 1002} (2010) 028
  [arXiv:0910.5713 [hep-ph]].
  
\bibitem{Fan:2013faa} 
  J.~Fan and M.~Reece,
  JHEP {\bf 1310}, 124 (2013)
  [arXiv:1307.4400 [hep-ph]].
  
\bibitem{Cohen:2013ama} 
  T.~Cohen, M.~Lisanti, A.~Pierce and T.~R.~Slatyer,
  JCAP {\bf 1310}, 061 (2013)
  [arXiv:1307.4082].
  
\bibitem{Cirelli:2015bda} 
  M.~Cirelli, T.~Hambye, P.~Panci, F.~Sala and M.~Taoso,
  JCAP {\bf 1510}, no. 10, 026 (2015)
  [arXiv:1507.05519 [hep-ph]].
  
\bibitem{Garcia-Cely:2015dda} 
  C.~Garcia-Cely, A.~Ibarra, A.~S.~Lamperstorfer and M.~H.~G.~Tytgat,
  JCAP {\bf 1510}, no. 10, 058 (2015)
  [arXiv:1507.05536 [hep-ph]].
  
\bibitem{Aoki:2015nza} 
  M.~Aoki, T.~Toma and A.~Vicente,
  JCAP {\bf 1509}, 063 (2015)
  [arXiv:1507.01591 [hep-ph]].
  
\bibitem{Quiros:1999jp} 
  M.~Quiros,
  hep-ph/9901312.
  
\bibitem{Morrissey:2012db} 
  D.~E.~Morrissey and M.~J.~Ramsey-Musolf,
  New J.\ Phys.\  {\bf 14}, 125003 (2012)
  [arXiv:1206.2942 [hep-ph]].
  
   
\bibitem{DiLuzio:2015oha}
  L.~Di Luzio, R.~Grober, J.~F.~Kamenik and M.~Nardecchia,
  arXiv:1504.00359 [hep-ph].
  
  
\bibitem{Hamada:2015bra}
  Y.~Hamada, K.~Kawana and K.~Tsumura,
  arXiv:1505.01721 [hep-ph].
  
\bibitem{futureDMcase}
T.~A.~Chowdhury and S.~Nasri,
\it{in preparation.}
 
 
\end{thebibliography}
\end{document}